\documentclass[twoside,twocolumn,12pt]{article}
\usepackage{epsfig}
\usepackage{amsmath}
\usepackage{amsfonts}
\usepackage{ amssymb }

\newcommand{\be}{\begin{equation}}
\newcommand{\ee}{\end{equation}}
\newcommand{\bea}{\begin{eqnarray}}
\newcommand{\eea}{\end{eqnarray}}

\begin{document}

\title{ \vspace{1cm} Neutrino oscillations}

\author{G.\ Bellini$^1$, L.\ Ludhova$^1$,  G.\ Ranucci$^1$, F.L.\ Villante$^2$ \\
\\
{\small $^1$Dip. di Fisica, Universit\`a degli Studi and INFN, 20133 Milano, Italy}\\
{\small $^2$Dip. di Scienze Fisiche e Chimiche, Universit\`a degli Studi and INFN, 67100 L'Aquila, Italy}
}
\maketitle

\begin{abstract} 

In the last decades, a very important breakthrough has been brought in the elementary particle physics by the discovery of the phenomenon of the neutrino oscillations, which has shown neutrino properties beyond the Standard Model. But a full understanding of the various aspects of the neutrino oscillations is far to be achieved.
In this paper the theoretical background of the neutrino oscillation phenomenon is described, referring in particular to the paradigmatic models. Then the various techniques and detectors which studied neutrinos from different sources are discussed, starting from the pioneering ones up to the detectors still in operation and to those in preparation. 
The physics results are finally presented adopting the same research path which has crossed this long saga. The problems not yet fixed in this field are discussed, together with the perspectives of their solutions in the near future.

\end{abstract}

\section{Introduction}
\label{Sec:Intro}

Neutrino studies brought us to some of the most relevant breakthroughs in particle physics of last decades. In spite of that, the neutrino properties are still far to be completely understood. 

The discovery of the oscillation phenomenon produced quite a revolution in the Standard Model of elementary particles, especially through the direct evidence of a non-zero neutrino mass. The first idea of neutrino oscillations was considered by Pontecorvo in 1957~\cite{pontecorvo}, before any experimental indication of this phenomenon. After several-decades lasting saga of experimental and theoretical research, many questions are still open around the interpretation of this phenomenon and on the correlated aspects, on the oscillation parameters, on the neutrino masses, on the mass hierarchy, on CP violation in the leptonic sector, and on a possible existence of a fourth, sterile neutrino.  
	 
The generally accepted MSW model~\cite{Wolfenstein,MSW} to interpret solar neutrino oscillations is presently validated for the oscillation in vacuum and in matter, but not yet in the vacuum-matter transition region. The shape of this transition could be influenced in a relevant way, as suggested by various theories going beyond the Standard Model as for example, the Non-Standard neutrino Interactions and a possible existence of a very light sterile neutrino. For this reason, the transition region deserves further and refined experimental studies.
	
Checks on the neutrino oscillations are under way through several experiments in data-taking phase,  while few others are in preparation or even construction. These projects exploit various approaches, as for example neutrino-flavor disappearance and appearance,  short and long source-to-detector baselines and measure neutrinos and/or antineutrinos of various origin, as the solar, atmospheric, accelerator, geo, and reactor (anti)neutrinos. 

Neutrinos interact with matter only through weak interactions and thus, they can bring to the observer almost undistorted information about their source. For example, by studying solar neutrinos and geo-neutrinos, we gather information not only about the character of neutrino itself but also about the Sun's and the Earth's interior. 

This paper consists of five sections.  In the following Section~\ref{Sec:theory}, the theoretical aspects of neutrino oscillations are reviewed\footnote{For a more detailed discussion, see e.g. \cite{GiuntiBook,Giunti, Vissani}.}. The principles and the structures of the detectors employed in neutrino-physics experiments are discussed in Section~\ref{Sec:detectors}. The most important milestones and the results of neutrino-physics experiments are summarized in Section~\ref{Sec:results}.  We briefly discuss the opened problems of neutrino physics and what can be expected in the near-future of this exciting research field in Section~\ref{Sec:future}.

\section{Neutrino Oscillations}
\label{Sec:theory}

 In the Standard Model (SM) neutrinos are neutral, massless fermions. They only interact with other particles via weak 
interactions, which are described by the charged current (CC) and neutral current (NC) interaction Lagrangians:
\begin{eqnarray}
{\mathcal L}_{\rm CC} &=&
-\frac{g}{2\sqrt{2}} \, j^{\rm CC}_{\rho}\, W^{\rho} + {\rm h.c.}
\label{CClagrangian}
\\
{\mathcal L}_{\rm NC} &=&
-\frac{g}{2 \cos\theta_{\rm W}} \, j^{\rm NC}_{\rho}\, Z^{\rho} 
\label{NClagrangian}
\end{eqnarray} 
In the above relation, $g$ is the $SU(2)_{L}$ gauge coupling constant, $\theta_{W}$ is the weak angle and the 
charged and neutral currents $ j^{\rm CC}_{\rho}$ and $ j^{\rm NC}_{\rho}$ are given by:
\begin{eqnarray}
\label{CC}
 j^{\rm CC}_{\rho} &=& 2 \sum_{\ell=e,\mu,\tau} \overline{\nu}_{\ell{\rm L}} \,\gamma_{\rho} \, \ell_{\rm L} + \dots , \\
\label{NC}
j^{\rm NC}_{\rho} &=& \sum_{\ell=e,\mu,\tau} \overline{\nu}_{\ell{\rm L}}\, \gamma_{\rho} \, \nu_{\ell{\rm L}} + \dots,
\end{eqnarray} 
where $\ell$ are the charged lepton fields and we have written only the terms containing the
neutrino fields $\nu_{\ell}$.

 If neutrino have non-zero masses, the left handed components  $\nu_{\alpha{\rm L}}$ 
of the neutrino fields with definite flavor $\alpha$ (that enter in the CC current definition) can be a superposition 
of the left handed components $\nu_{i {\rm L}}$  of the neutrino fields with definite masses $m_{i}$.
Assuming that neutrinos are ultra-relativistic, we have\footnote{In this section, we use Greek letters ($\alpha$ and $\beta$) 
to refer to neutrino flavors and Latin letters ($i$ and $j$) to refer to neutrino masses.}
\begin{equation}
\nu_{\alpha{\rm L}} = \sum^N_{i=1} U_{\alpha i} \, \nu_{i {\rm L}} 
\end{equation}
where $U$  is an unitary matrix.
By considering that a field operator creates antiparticles, this implies that a {\em flavor eigenstate} $|\nu_\alpha \rangle$
is a superposition of different {\em mass eigenstates} $|\nu_i \rangle$, according to:
\begin{equation}
|\nu_{\alpha}\rangle = \sum^N_{i=1} U^*_{\alpha i} \, |\nu_{i}\rangle 
\end{equation}
For antineutrinos, we obtain correspondingly:
\begin{equation}
|\overline{\nu}_{\alpha}\rangle = \sum^N_{i=1} U_{\alpha i} \, |\overline{\nu}_{i}\rangle 
\end{equation}
In principle, the number $N$ of massive neutrinos can be larger than three. In this case, however, 
we must assume that there are {\em sterile} neutrinos, i.e. light fermions that do not take part to standard weak interactions
(\ref{CClagrangian}) and (\ref{NClagrangian}) and thus are not excluded by LEP results according to which the number of 
{\em active} neutrinos coupled with the $W^{\pm}$ and $Z$ boson is $N_\nu=2.984\pm 0.008$ \cite{ran12}.

In the assumption of 3 massive neutrinos, the {\em neutrino mixing matrix} $U$ can be expressed in terms
of three mixing angles $\theta_{12}$, $\theta_{23}$ and $\theta_{13}$ and one Dirac-type CP phase $\delta$
according to\footnote{We are considering here the assumption 
that neutrinos are Dirac particles. In the case of Majorana (or Dirac-Majorana) mass terms, 
the most general form of the mixing matrix contains two additional phases and it is obtained by $U\rightarrow U \cdot U_{\rm M}$
where $U_{\rm M}={\rm diag}\left( 1, e^{i\phi_1}, e^{i\phi_2}\right)$, see e.g.~\cite{GiuntiBook,Giunti}. 
The Majorana phases $\phi_1$ and $\phi_2$, however, have no observable effects on neutrino oscillations~\cite{Bilenky}.}
\begin{equation}
U=R_{23}\left(\theta_{23}\right) \Gamma\left(\delta\right) R_{13}\left(\theta_{13}\right) \Gamma^\dag\left(\delta\right) R_{12}\left(\theta_{12}\right)
\label{MixingMatrix}
\end{equation}
where $R_{ij}\left(\theta_{ij}\right)$ represents an Euler rotation by $\theta_{ij}$ in the $ij$ plane, and:
\begin{equation}
\Gamma\left(\delta\right)=  {\rm diag}\left( 1,1, e^{i\delta}\right)~.
\end{equation}
In components, the mixing matrix $U$ is expressed as:
\begin{eqnarray}
\nonumber
\footnotesize{U = } & & \\
\nonumber
 && 
\hspace{-1.5cm}
\left(
\footnotesize{\begin{matrix}
\vspace{0.2cm}
\hspace{-0.1cm}  c_{12}c_{13} & \hspace{-1.5cm} s_{12}c_{13}  &  s_{13}  e^{-i\delta}\\
\vspace{0.2cm}
\hspace{-0.1cm}  - s_{12}c_{23}- c_{12}s_{23}s_{13}e^{i\delta} & \hspace{-0.1cm} c_{12}c_{23}- s_{12}s_{23}s_{13}e^{i\delta} & \hspace{-0.1cm} s_{23}c_{13}\\
\vspace{0.2cm}
\hspace{-0.1cm}  s_{12}s_{23} - c_{12}c_{23}s_{13}e^{i\delta}  & \hspace{-0.1cm} - c_{12}s_{23}- s_{12}c_{23}s_{13}e^{i\delta} &\hspace{-0.1cm}  c_{23}c_{13}
 \end{matrix}}
\right)\\
&&
\end{eqnarray}
where $c_{ij}=\cos{\theta_{ij}}$ and $s_{ij}=\sin{\theta_{ij}}$. We indicate with $\Delta m^2_{ij} \equiv m^2_{i} - m^2_{j}$.
As it is usually done, we order the neutrino masses such that $\Delta m^2_{21} > 0$
and $\Delta m^2_{21} \ll \left| \Delta m^2_{31}\right| $. With this choice, the ranges of mixing parameters 
are determined by:
\begin{equation}
0\le \theta_{12},\, \theta_{23},\, \theta_{13} \le \pi/2 \hspace{1.0cm} 0\le\delta\le 2\pi
\end{equation}
The sign of $\Delta m^2_{31}$ determines the neutrino mass hierarchy, being 
$\Delta m^2_{31} >0$ for normal hierarchy (NH) and $\Delta m^2_{31}<0$ for inverted hierarchy (IH).

\subsection{Neutrino evolution equation}
\label{nuEvolEquation}

The evolution of a generic neutrino state $|\nu(t)\rangle$ is described by a Schr\"odinger-like equation:
\begin{equation}
i \frac{d |\nu(t)\rangle}{d t} = H\, |\nu(t)\rangle 
\end{equation}
where $H$ represents the Hamiltonian operator. The above equation can be expressed in the 
flavor eigenstate basis $\left\{|\nu_\alpha\rangle\right\}$. We obtain:
\begin{equation}
i \frac{d \nu^{\rm (f)}(t)}{d t} = H^{\rm (f)} \, \nu^{\rm (f)}(t) 
\end{equation} 
where $\nu^{\rm (f)}(t)$ is the vector describing the flavor content of the neutrino state $|\nu(t)\rangle$
given by:
\begin{equation}
\nu^{\rm (f)}(t) = \left(a_e(t),\,a_\mu(t),\,a_\tau(t), \dots \right)^{T}
\end{equation}
with $a_{\alpha}(t) = \langle \nu_\alpha| \nu(t) \rangle$, and the matrix $H_{f}$ is given by:
\begin{equation}
H^{(\rm f)}_{\alpha\beta} =  \langle \nu_\alpha| H | \nu_\beta \rangle ~.
\end{equation}

 In vacuum, the neutrino Hamiltonian $H_{\rm vac}$ is determined
in terms of neutrino masses and mixing parameters. We have, in fact:
\begin{equation}
H_{\rm vac}^{\rm (f)} = U H_{\rm vac}^{\rm (m)} U^\dag
\label{vachamil}
\end{equation}
where $H_{\rm vac}^{\rm (m)}$ is the representation of the vacuum Hamiltonian in the mass
eigenstate basis, given by:
\begin{eqnarray}
\nonumber
H_{\rm vac}^{\rm (m)} &=& {\rm diag} \left( \sqrt{\vec{p}\,^2+m^2_{1}} ,  \dots, \sqrt{\vec{p}\,^2+m^2_{N}} \right)  \\
&\approx&  
|\vec{p}\,| + \frac{1}{2 |\vec{p}\,|} {\rm diag}\left(m^2_{1} , \dots, m^2_{N} \right) ~.
\end{eqnarray}
In the last equality, we adopted the ultra-relativistic approximation
$E\approx |\vec{p}\,| + m^2/2 |\vec{p}\,|$ and we implicitly assumed that
the neutrino state $|\nu(t)\rangle$ can be described as a superposition of
states with fixed momentum $\vec{p}$. This corresponds to the so-called 
{\em plane-wave approximation} which is adequate to
describe neutrino evolution when coherence of the different
components of the neutrino wave packet is not lost in the detection
and/or propagation processes\footnote{For a wave-packet description of neutrino
oscillations see e.g.~\cite{GiuntiBook}}.

The presence of a matter can affect neutrino propagation 
in a non trivial way. In fact, as it was first realized by~\cite{Wolfenstein},
when a neutrino propagates through a medium, its dispersion relation
(i.e. its energy-momentum relation) is modified by coherent interactions 
with background particles. This phenomenon, that in optics is accounted for 
by introducing a refractive index, can be described by adding an 
effective potential $V$ in the evolution equation, so that:
\begin{equation}
i\frac{d |\nu (t)\rangle }{dt} = \left( H_{\rm vac} + V \right) \, |\nu_{f}(t)\rangle ~.
\label{nuevolution}
\end{equation}
In the SM, the effective potential is diagonal in the flavor basis.
We thus have:
\begin{equation}
V^{\rm (f)} = {\rm diag}\left(V_{e},V_\mu,V_\tau, 0, \dots\right)
\end{equation} 
 where we have taken into account that sterile states do not interact with the
medium. At low energies, the potentials can be evaluated 
by taking the average $\langle{\mathcal H}_{\rm eff}\rangle$ 
of the effective four fermion Hamiltonian due to exchange of 
$W$ and $Z$ bosons over  the state describing the background medium. 
We have:
\begin{equation}
{\mathcal H}_{\rm eff} = {\mathcal H}_{\rm CC} + {\mathcal H}_{\rm NC}
\end{equation}
with:
\begin{eqnarray}
\nonumber
 {\mathcal H}_{\rm CC} &= &\frac{G_{F}}{\sqrt{2} }
 \left[ \overline{\nu}_{e}\, \gamma^{\rho} (1-\gamma^5) \, \nu_e \right] \\
& &\hspace{0.3cm}\times \left[ \overline{e} \, \gamma_{\rho} (1-\gamma^5)\, e \right]
\\
\nonumber
 {\mathcal H}_{\rm NC} &=& \frac{G_{F}}{\sqrt{2}} \;
\sum_{\ell=e,\mu,\tau} \left[ \overline{\nu}_{\ell}\, \gamma^{\rho} (1-\gamma^5) \, \nu_\ell \right] \\  
& & \hspace{0.5cm}\times 
\sum_{b=e, p, n}  \left[ \overline{b} \, \gamma_{\rho} (g^b_{V}-g^b_{A}\gamma^5)\, b \right]
\end{eqnarray}
where $G_{F}$ is the Fermi constant, $g^b_{V}$ and $g^b_{A}$ are the vector and 
axial vector coupling constants of the various background particles
and we have performed a Fierz reshuffling of the fields, see \cite{GiuntiBook,Smirnov}. 
In the above equations, it is taken into account that normal matter does not contain
muons and taus and, consequently,  the CC interactions with the medium only affect 
electron neutrino propagation. For non relativistic unpolarized medium, 
one obtains:
\begin{equation}
V_{\alpha} = A_{\rm CC} \, \delta_{\alpha e } + A_{{\rm NC}}
\end{equation}
where the CC contribution:
\begin{equation}
A_{\rm CC} = \sqrt{2} G_{\rm F} \left(n_{e}-n_{\overline{e}}\right)
\end{equation}
is proportional to difference between the number densities of electrons and positrons.
The NC contribution $A_{\rm NC}$ is equal for all active neutrino flavors and it is given by:
\begin{eqnarray}
\nonumber
\hspace{-0.5cm}
A_{\rm NC} &=& 
\frac{G_{\rm F}}{\sqrt{2}} (1 - 4\, s^2_{W})
\left[
\left(n_{p}-n_{\overline{p}}\right) -
\left(n_{e}-n_{\overline{e}}\right) 
\right] \\
&-& \frac{G_{\rm F}}{\sqrt{2}}
\left(n_{n}-n_{\overline{n}}\right) 
\end{eqnarray}
where $s_{W}\equiv \sin \theta_W$ while $n_{p}$ and $n_{n}$ ($n_{\overline{p}}$ and $n_{\overline{n}}$) are the number densities
of protons and neutrons (anti-protons and anti-neutrons), respectively. 
In neutral matter, it necessarily holds $\left(n_{e}-n_{\overline{e}}\right) = 
\left(n_{p}-n_{\overline{p}}\right)$ that implies that the first term in the r.h.s of the above 
equation vanishes. Moreover, in the absence of sterile neutrinos, the neutral current 
contribution to the total Hamiltonian is proportional to the identity matrix. As a consequence, 
it only introduces  an overall unobservable phase factor in the evolution of $\nu^{\rm (f)}(t)$
and, thus, can be neglected.

Finally, the evolution equation for antineutrinos
is obtained by replacing $U\to U^*$ in Eq.~(\ref{vachamil}) and $V \to -V $ 
in Eq.~(\ref{nuevolution}).
We, thus, understand that CP-violating effects are absent in neutrino oscillations,
if the mixing matrix is real (i.e. $U=U^*$) and 
neutrinos propagate in vacuum or in a CP-symmetric medium (i.e. $V=0$).
 
\subsection{Oscillations in vacuum and in matter}

In formal terms, neutrino oscillations are easily described. 
Let us assume that a neutrino flavor $\nu_{\alpha}$ is created at a time $t_0 =0$. In the flavor 
eigenstate basis, this state is represented by a vector:
\begin{equation}
\nu^{(\rm f)}(0)  = \left(  a_e(0),\,a_\mu(0),\,a_\tau(0), \dots    \right)^T 
\end{equation}
with components $a_{\beta}(0)=\delta_{\beta\alpha}$. After a time interval 
$t$, the neutrino propagated to a distance 
$x\approx t$ and its flavor content has evolved 
according to:
\begin{equation}
\nu^{\rm (f)}(x) = S^{\rm (f)}(x) \, \nu^{\rm (f)}(0)
\end{equation}
where the evolution operator is given by: 
\begin{equation}
S^{\rm (f)}(x) = T\left[\exp{\left(-i \int^x_0 dx'\;  H^{\rm (f)}(x')\right)} \right]
\end{equation}
and $T$ represents the time-ordering of the exponential.
In the presence of neutrino mixing and if neutrino
masses are not degenerate,  the Hamiltonian $H^{\rm (f)}$ is not diagonal. 
Thus, flavor is not conserved and components $\beta\neq\alpha$ can
appear as a result of the evolution.
The probability to detect a neutrino flavor $\nu_{\beta}$
at a distance $L$ from the neutrino production point is given by:
\begin{equation}
P\left(\nu_\alpha\to\nu_{\beta}\right)
=\left| a_{\beta}(L)\right|^2 = \left| S^{\rm (f)}_{\beta\alpha}(L) \right|^2
\end{equation}
In the following, we discuss the expectations for $P\left(\nu_\alpha\to\nu_{\beta}\right)$ in few relevant cases.
\par

\vspace{0.5 cm}
\noindent
{\em Vacuum neutrino oscillations.}\\
In vacuum, the neutrino Hamiltonian $H$ is constant. The evolution 
operator can be explicitly calculated as:
\begin{equation}
S^{\rm(f)} = U S^{\rm (m)} U^\dag
\end{equation}
where $S^{\rm(m)}$ is the evolution operator in the mass eigenstate basis
given by:
\begin{equation}
S^{\rm (m)} = {\rm diag} \left(\exp(i\phi_1), \dots, \exp(i\phi_N)\right)
\end{equation}
with $\phi_i= - m_i^2 \, x  / 2|\vec{p}|$. The probability to observe the 
oscillation $\nu_\alpha\to\nu_{\beta}$ over a distance $L$ is thus given by:
\begin{equation}
P\left(\nu_\alpha\to\nu_{\beta}\right) = 
\sum_{i,j} \left[ U_{\beta i}\,U^*_{\alpha i}\, U^*_{\beta j}\, U_{\alpha j} \right] \exp\left( i \phi_{ij} \right)
\end{equation}
where $\phi_{ij}= \left[ \left( m^2_{j} -m^2_{i} \right) \, L \right]/2E$ and we considered 
that for a relativistic particle $E\approx |\vec{p}|$.

The above expression can be recast in few alternative forms that are useful to discuss the property 
of neutrino oscillations. We obtain, e.g.:
\begin{eqnarray}
\nonumber
P\left(\nu_\alpha\to\nu_{\beta}\right)
&=& \sum_i \left| U_{\beta i } \right|^2 \left| U_{\alpha i } \right|^2 + \\
& & 
\hspace{-2.6cm}
+ \; 2{\rm Re}\left[
\sum_{i>j} 
U_{\beta i}\,U^*_{\alpha i}\, U^*_{\beta j}\, U_{\alpha j} 
\exp(i\phi_{ij}) 
\right]
\end{eqnarray}
that gives the oscillation probability as the sum of a constant and an 
oscillating term. The oscillating part averages to zero if the
phases $\phi_{ij}$ vary over  ranges $\Delta\phi_{ij}\gg1$, as it can be
due e.g. to a spread of the neutrino energy $E$ and/or the neutrino baseline 
$L$. The constant part represents the "classical"' limit
that is obtained by neglecting interference among the different components 
of the neutrino wave-packet and by combining probabilities, rather than amplitudes,
to derive $P\left(\nu_\alpha\to\nu_{\beta}\right)$. 

Alternatively, we can write:
\begin{eqnarray}
\nonumber
P\left(\nu_\alpha\to\nu_{\beta}\right)
&=& \delta_{\alpha\beta} \\
& & 
\nonumber
\hspace{-2.0cm}
- 4
\sum_{i>j} {\rm Re}\left[U_{\beta i}\,U^*_{\alpha i}\, U^*_{\beta j}\, U_{\alpha j}\right] \sin^2(\phi_{ij}/2) \\
& &
\hspace{-2.0cm}
- 2 
\sum_{i>j} {\rm Im}\left[U_{\beta i}\,U^*_{\alpha i}\, U^*_{\beta j}\, U_{\alpha j} \right]\sin(\phi_{ij}) 
\end{eqnarray}
The first two terms in the r.h.s. of the above equation do not change for $U \to U^*$ 
and describe the CP-conserving part of the neutrino oscillation probability. 
The last part, instead, changes sign introducing a difference between neutrino and antineutrino 
oscillation probabilities that can be quantified as:
\begin{eqnarray}
\nonumber
 P\left(\overline{\nu}_\alpha\to\overline{\nu}_{\beta}\right) - P\left(\nu_\alpha\to\nu_{\beta}\right) &=&\\  
& & 
\hspace{-5cm}
4 \sum_{i>j} {\rm Im}\left[U_{\beta i}\,U^*_{\alpha i}\, U^*_{\beta j}\, U_{\alpha j} \right]\sin(\phi_{ij}) 
\end{eqnarray}
For $\alpha = \beta$, this term vanishes showing that CP asymmetry can be measured only in 
transitions between different neutrino flavors.

If we assume two neutrino mixing, i.e. we take only one non-vanishing mixing angle $\theta_{ij}$
in Eq.~(\ref{MixingMatrix}), the oscillation probability reduces to the well known expression: 
\begin{equation}
P\left(\nu_\alpha\to\nu_{\beta}\right) = \sin^2\left(2\theta_{ij}\right) \sin^2\left( \frac{\Delta m_{ij}^2 \, L}{4E}\right)
\label{2nuOsc}
\end{equation}
where $\alpha \neq \beta$ and the involved flavors depends on the mixing angle $\theta_{ij}$.\footnote{In the assumption of two neutrino mixing,
an angle $\theta_{12}\neq 0$ induces $\nu_e\to \nu_\mu$ oscillations; $\theta_{13}\neq 0$ induces  $\nu_e\to \nu_\tau$ oscillations; 
$\theta_{23}\neq 0$ induces $\nu_\mu\to \nu_\tau$ oscillations.}
The survival probability for the case $\alpha=\beta$ can be simply deduced by considering that,
due to unitarity of the mixing matrix, it always holds $\sum_{\beta} P\left(\nu_\alpha\to\nu_{\beta}\right) \equiv 1$
that, in this specific case, gives:
\begin{equation}
P\left(\nu_\alpha\to\nu_{\alpha}\right) = 1 -  \sin^2\left(2\theta_{ij}\right) \sin^2\left( \frac{\Delta m_{ij}^2 \, L}{4E}\right)  ~.
\label{2nuOscSurv}
\end{equation}
Eq.~(\ref{2nuOsc}) describes an oscillating function of $L$. The amplitude of the oscillation is determined by 
$\sin^2 (2\theta_{ij})$ while the oscillation length is given by:
\begin{equation}
L_{ij}=\frac{4\pi E}{|\Delta m^2_{ij}|} = 2.48\frac{E [{\rm MeV}] }{|\Delta m^2_{ij} [{\rm eV^2}]|}{\rm m}
\end{equation}
The oscillation probabilities are unchanged when $\Delta m^{2}_{ij}\to -\Delta m^{2}_{ij}$ or 
$\theta_{ij}\to \pi/2-\theta_{ij}$ showing that two neutrino oscillations in vacuum do not probe 
the hierarchy of the masses $m_i$ and $m_j$ (i.e. the state $\nu_{i}$ and $\nu_{j}$ can be interchanged
with no effect on Eqs.~(\ref{2nuOsc},\ref{2nuOscSurv})).

In the three neutrino case, useful expressions can be derived in the approximation of one-dominant mass scale
(i.e., $\Delta m^2_{21} \ll | \Delta m^2_{31}|\approx \,|\Delta m^2_{32}|$) that is motivated by the fact that the mass difference
 required to explain the atmospheric neutrino anomaly 
is much larger than that required to solve the solar neutrino problem. 
In this assumption, one obtains\footnote{Note that, due to CPT-invariance, it holds
$P\left(\nu_{\beta}\to\nu_{\alpha}\right) = P\left(\overline{\nu}_{\alpha}\to\overline{\nu}_{\beta}\right)$}:
\begin{eqnarray}
\nonumber
P\left(\nu_e\to\nu_{\mu}\right)
&=& s^2_{23}\sin^2{2\theta_{13}}\,S_{23}
\label{Pemu3nuVac}\\
& &\hspace{-0.5cm} + \; c^2_{23}\sin^2{2\theta_{12}}\, S_{12}  - P_{\rm CP}\\
\nonumber
P\left(\nu_e\to\nu_{\tau}\right)
&=& c^2_{23}\sin^2{2\theta_{13}}\,S_{23}\\
& &\hspace{-0.5cm} + \;  s^2_{23}\sin^2{2\theta_{12}}\, S_{12} + P_{\rm CP}\\
\nonumber
P\left(\nu_\mu\to\nu_{\tau}\right)
&=& c^4_{13}\sin^2{2\theta_{23}}\,S_{23}\\
& &\hspace{-0.5cm} - \;  s^2_{23}c^2_{23}\sin^2{2\theta_{12}}\, S_{12} - P_{\rm CP}
\end{eqnarray}
where, following \cite{Vissani}, we adopted the notation $S_{23}=\sin^2(\Delta m^2_{32} \, L/4E)$ and 
$S_{12}=\sin^2(\Delta m^2_{21}\, L/4E)$ and we set $\theta_{13}=0$  in the coefficients of the $S_{12}$ terms.
The CP-violating part $P_{\rm CP}$, that enters with opposite sign in the corresponding expressions
for antineutrinos, is given by:
\begin{equation}
P_{\rm CP} =  8 J \sin \left(\frac{\Delta m^2_{21}\,L}{4E} \right)\sin^2 \left( \frac{\Delta m^2_{31} \,L}{4E}\right)
\end{equation}
where:
\begin{equation}
J= \frac{1}{8} \sin(2\theta_{12}) \sin(2\theta_{23}) \sin(2\theta_{13}) \cos(\theta_{13}) \sin(\delta)
\end{equation}
showing that CP-violation is observed in neutrino oscillations only if all the angles and 
all the mass differences are non vanishing. The magnitude of CP-violating effects depends of the
phase $\delta$, being maximal for $\delta = \pi/2$ and $\delta = 3\pi/2$.

The survival probabilities 
$P\left(\nu_\alpha\to\nu_{\alpha}\right) = P\left(\overline{\nu}_\alpha\to\overline{\nu}_{\alpha}\right)$ 
are given by \cite{Vissani}:
\begin{eqnarray}
\nonumber
P\left(\nu_{e}\to\nu_{e}\right) &=& 1 -\sin^2{2\theta_{13}}\, S_{23}\\
& &\hspace{0.5cm} - \;  c^4_{13} \sin^2{2\theta_{12}}\, S_{12}
\label{Pee3nuVac}\\
\nonumber
P\left(\nu_\mu\to\nu_{\mu}\right) &=& 1-4c^2_{13}s^2_{23}\left(1-c_{13}^2s^2_{23}\right)\, S_{23}\\
& &\hspace{0.5cm}  - c^4_{23}\sin^2{2\theta_{12}}\, S_{12}\\
\nonumber
P\left(\nu_\tau\to\nu_{\tau}\right) &=&1-4c^2_{13}c^2_{23}\left(1-c_{13}^2c^2_{23}\right)\, S_{23}\\
& &\hspace{0.5cm} -\; s^4_{23}\sin^2{2\theta_{12}}\, S_{12}
\end{eqnarray}

We note that, at this level of approximation, there is no sensitivity to neutrino hierarchy since
the oscillation probabilities do not depends on the sign of $\Delta m^{2}_{31}$. Moreover, in the limit $\theta_{13}\to 0$, 
the "atmospheric" mass scale $\Delta m^2_{32}$ does not produce observable 
effects on electron neutrino oscillations that can be regarded as two neutrino 
oscillations, driven by the "solar" mass difference $\Delta m^2_{21}$, between 
$\nu_e$ and the mixed state $\nu_{\mu\tau} = c_{23} \, \nu_{\mu} - s_{23}\, \nu_\tau$. 
This conclusion also holds in presence of matter.

\par
\vspace{1.0 cm}
\noindent
{\em Neutrino oscillations in matter}\\
The evolution of neutrinos in matter is complicated by the fact that the properties of the medium can
change along the neutrino trajectory, thus giving a non-constant Hamiltonian. The evolution
equation reads:
\begin{equation}
i\frac{d \nu^{\rm (f)}(x)}{dx} = \left[ H_{\rm vac}^{\rm (f)} + V^{\rm (f)}(x) \right] \, \nu^{\rm (f)}(x) 
\end{equation}
where, if we neglect sterile neutrinos, the only non-vanishing entry of the matrix $V^{(\rm f)}(x)$ is\footnote{
We omit the NC contribution to matter potential that, in the absence of sterile neutrinos, is proportional to the identity matrix. 
We also assume that the number density of positrons is negligible. See sect.\ref{nuEvolEquation} for details.}:
\begin{equation} 
(V^{(\rm f)})_{ee} = \pm \sqrt{2} \, G_{\rm F} \, n_{e}(x).
\end{equation}
Here, the "+" sign refers to neutrinos while the "-" sign refers to antineutrinos.

It is convenient to diagonalize the Hamiltonian at each point of the space
and discuss the evolution in the basis of the {\em local mass eigenstates} 
defined by the relation:
\begin{equation}
\nu^{\rm (f)}(x) \equiv \tilde{U}(x) \, \nu^{\rm ({\tilde m})}(x)
\end{equation}
where $\tilde{U}(x)$ is the unitary matrix that gives:
\begin{equation}
H_{\rm vac}^{\rm (f)} + V^{\rm (f)}(x) = \frac{1}{2E} \tilde{U}(x) {\tilde{M^2}}(x) \tilde{U}^\dag(x)
\end{equation}
with
\begin{equation}
{\tilde{M^2}}(x) = {\rm diag}\left(\tilde{m}^2_1(x), \dots, \tilde{m}^2_N(x)\right)
\end{equation}
In this basis, the evolution equation becomes:
\begin{equation}
i\frac{d \nu^{\rm (\tilde{m})}(x)}{dx} = \left[\frac{\tilde{M}^2}{2E} - i \tilde{U}^\dag(x) \frac{d \tilde{U} (x)}{dx}\right]\nu^{\rm (\tilde{m})}(x)
\label{MatterEvol}
\end{equation}
We see that the non diagonal entries, that may cause the transitions between the local mass eigenstates, 
are proportional to the derivative of $\tilde{U}(x)$ whose magnitude is essentially determined by the rate of change
of the electrons number density in the background medium.

 This observation can be used to introduce the so-called {\em adiabatic} approximation
that applies with good accuracy to the case of solar neutrino oscillations.
Let us indicate with ${\tilde L}_{ij} \approx 4\pi E/|\Delta {\tilde m}^2_{ij}|$ the length scale over which the components
of the neutrino wave packet with mass ${\tilde m}_{i}$ and ${\tilde m}_{j}$ acquire a phase difference $\Delta\Phi_{ij} = 2\pi$.
If we assume that the various ${\tilde L}_{ij}$ are much smaller than 
the distance over which the medium change its properties $D \equiv (d\ln n_{e}(x)/ dx)^{-1}$, 
the second term in the r.h.s of Eq.~(\ref{MatterEvol}) can be neglected. 
Thus, the components of the vector $\nu^{\rm ({\tilde m})}(x)$ remain constant (in magnitude)
during the evolution, 
even if the decomposition of $\nu^{\rm ({\tilde m})}(x)$ in the flavor basis changes along the 
neutrino trajectory as results of the variations of $n_{e}$. 
If the length scales ${\tilde L}_{ij}$ are also much smaller than the baseline $L$
over which neutrinos propagate, the oscillation probabilities $P\left(\nu_\alpha\to\nu_{\beta}\right)$
only depends on the properties of $\tilde{U}(x)$ at the production point $x_{\rm p}$ and 
at the detection point $x_{\rm d}$. They can be, in fact, deduced by combining incoherently probabilities 
of production and detection, obtaining:
\begin{equation}
P\left(\nu_\alpha\to\nu_{\beta}\right) = \sum_i \left| \tilde{U}_{\alpha i }( x_{\rm p}) \right|^2 \left| \tilde{U}_{\beta i }(x_{\rm d}) \right|^2 ~.
\label{AdiabaticGeneral}
\end{equation}

We now consider the specific case of $\nu_e$ produced by nuclear reactions occurring 
at the center of the Sun. Let us calculate the electron neutrino survival probability, by first considering a two neutrino 
scenario in which only $\theta_{12}\neq 0$. 
The effective mixing angle in matter ${\tilde \theta}_{12}$ can be calculated as:
\begin{equation}
\sin (2\tilde{\theta}_{12} ) =\frac{\sin(2\theta_{12})}{\sqrt{\sin^2(2\theta_{12}) + C^2}}  
\end{equation}
while the difference between the effective neutrino masses is given by:
\begin{equation}
\Delta {\tilde m}^2_{21} \equiv  {\tilde m}^2_2 -  {\tilde m}^2_1 = \Delta  m^2_{21} \sqrt{\sin^2({2\theta_{12}})+C^2}
\end{equation}
with 
\begin{equation}
C(x) =  \cos({2\theta_{12}})  -  \frac{2\sqrt{2} G_{\rm F}\, n_{e}(x) E}{\Delta m^2_{21}}
\label{Cmatt}
\end{equation}
Matter effects break the degeneracies $\Delta m^{2}_{21} \to - \Delta m^{2}_{21}$ 
and $\theta_{12}\to \pi/2 -\theta_{12}$
probing the hierarchy in the 1-2 neutrino sector. In particular,  when 
$\Delta m^{2}_{21}>0$  and $\theta_{12} < \pi/4$ the system has a {\em resonance}.
It exists, in fact, a value of the electron number density, defined by the condition:
\begin{equation}
\Delta m^2_{21} \cos(2\theta_{12})  = 2\sqrt{2} G_{\rm F} \, n_{e} E 
\end{equation}
for which the local mixing  is maximal (i.e. ${\tilde \theta}_{12}=\pi/4$)
while  the mass difference $\Delta {\tilde m}^2_{12}$ reaches 
the minimal value $\Delta {\tilde m}^2_{12} = \Delta m^2_{12} \sin(2\theta_{12})$.
As it was discussed by \cite{MSW}, if the resonance region is sufficiently wide,  
it is possible to achieve a total conversion of $\nu_{e}$ into neutrinos of different flavors. 
This mechanism is called the {\em MSW effect}.
Considering that the electron density in the Sun is $n_{e}\lesssim 10^{26}\, {\rm cm}^{-3}$ and the typical solar neutrino energies are $E\approx 1\,{\rm MeV}$,
the resonance condition requires $\Delta m^2_{21} \cos(2\theta_{12}) \lesssim 10^{-5} \, {\rm eV}^2$.

The evolution equation in the local mass eigenstate basis becomes
\begin{eqnarray}
\hspace{0.5cm}
\nonumber
 i\frac{d \nu^{\rm (\tilde{m})}(x)}{dx} &=& \\
& &
\nonumber
 \hspace{-3.2cm} =
\left[
\frac{1}{2E} \left(
\small{\begin{matrix}
 \tilde{m}^2 _1& 0 \\
0 & \tilde{m}^2 _2
\end{matrix}}
\right) + 
i\left(
\small{\begin{matrix}
 0& \hspace{-0.5cm} -d{\tilde \theta}_{12} / dx \\
d{\tilde \theta}_{12}/dx & 0
\end{matrix}}
\right)\right]\, \nu^{\rm (\tilde{m})}(x)\\
&&
\label{2numatter}
\end{eqnarray}
and the adiabaticity condition can be explicitly expressed as:
\begin{equation}
\gamma(x) \gg 1
\label{AdCondition}
\end{equation}
where the {\em adiabaticity parameter} $\gamma$ is given by the ratio 
between the differences of diagonal elements and off-diagonal elements of Eq.~(\ref{2numatter}):
\begin{equation}
\gamma = \left| \frac{ \Delta {\tilde m}^2_{21}  / 4E}
{ \; d {\tilde \theta}_{12}/dx } \right|
\end{equation}
If condition (\ref{AdCondition}) is fulfilled, the electron neutrino survival probability
can be calculated through Eq.~(\ref{AdiabaticGeneral}) obtaining:
\begin{eqnarray}
\nonumber
P(\nu_e\to\nu_e) &=& \\
& &
\hspace{-3.0cm}
= \frac{1}{2}+\frac{1}{2}\cos(2\tilde{\theta}_{12}) \cos(2\theta_{12})
\label{Pee2nuMatter}
\end{eqnarray}
where $\tilde{\theta}_{12}$ indicates the mixing angle at neutrino production 
point and we assumed that neutrinos are detected in vacuum.

In order to understand the specific features of $P(\nu_e\to\nu_e)$, it is useful to
define a transition energy $E^{*}$, given by:
\begin{equation}
E^* = \frac{\Delta m^2_{21} \cos{2 \theta_{12}}}{2 \sqrt{2} G_{\rm F} \, n_{e,\odot}}
\end{equation}
where $n_{e,\odot}$ is the electron number density at the center of the sun.
For $E\ll E^*$, matter effects are negligible and Eq.~(\ref{Pee2nuMatter}) reduces to:
\begin{equation}
P(\nu_e\to\nu_e) = 1-\frac{1}{2}\sin^2(2\theta_{12})
\end{equation}
that, in fact, corresponds to vacuum averaged neutrino oscillations.
For $E\gg E^*$, matter potential becomes dominant so that "heaviest" mass 
eigenstates in the center of the sun coincides with $\nu_e$. 
As a consequence, we obtain $\cos(2\tilde{\theta}_{12}) = -1 $ and:
\begin{equation}
P(\nu_e\to\nu_e) = \sin^2(\theta_{12})
\end{equation}
For the value of $\theta_{12}$ and $\Delta m^2_{21}$ currently favored by neutrino
oscillation analysis (see sect.\ref{Sec:results}), the transition energy $E^*$ 
is approximately $E^*\approx 1.2\,{\rm MeV}$.

 The violations of adiabaticity can be taken into account by introducing the 
crossing probability $P_{\rm C}$ that represents the probability of a transition
between the local mass eigenstates during the neutrino evolution. 
If $P_{c}\neq0$, the electron neutrino survival probability becomes: 
\begin{eqnarray}
\nonumber
P(\nu_e\to\nu_e) &=& \\
& &
\hspace{-3.0cm}
= \frac{1}{2}+\left(\frac{1}{2}-P_{\rm C}\right)\cos(2\tilde{\theta}_{12}) \cos(2\theta_{12})
\label{Pee2nuMatterPc}
\end{eqnarray}
There are different approaches to calculate $P_{\rm C}$. 
For several cases of interest, the following expression holds (see e.g. \cite{Kuo, Kuo2, Vissani} and references therein): 
\begin{equation}
P_{\rm C} = \frac{\exp{\left(-\frac{\pi}{2}\tilde{\gamma} F \right)}-\exp{\left(-\frac{\pi}{2}\tilde{\gamma} \frac{F}{\sin^2\theta_{12} } \right)}}{1-\exp{\left(-\frac{\pi}{2}\tilde{\gamma} \frac{F}{\sin^2\theta_{12} } \right)} }
\end{equation}
where $\tilde{\gamma}$ is the minimal value of $\gamma(x)$ along the neutrino trajectory
\footnote{In the presence of a resonance, one can often approximate $\tilde{\gamma}\approx 
\gamma_{\rm res}$ where $\gamma_{\rm res}$ is the value of $\gamma(x)$ at the resonance point.
See e.g. \cite{GiuntiBook,Vissani} for discussion.}
and the parameter $F$ depends on the adopted electron density profile. 
In particular, for an exponential density profile $n_{e}\propto \exp(-x)$, 
which is a good approximation for solar neutrinos, one has  $F= 1 - \tan^2\theta_{12}$.

 In the case of three mixed neutrinos, the above picture has to be 
modified to take into account the possibility that $\theta_{23}\neq 0$ and $\theta_{13}\neq 0$. 
Since matter potentials are equal for muon and tau neutrinos, the rotation $R(\theta_{23})$
in Eq.~(\ref{MixingMatrix}) can be re-absorbed in the "mixed" basis $\left\{|\nu_e \rangle, \, 
c_{23} |\nu_\mu  \rangle -s_{23} |\nu_\tau \rangle, \,
s_{23}|\nu_\mu  \rangle + c_{23} |\nu_\tau \rangle  \right\}$. 
This shows that, when $\theta_{13}=0$,
electron neutrinos experience two-neutrino oscillations 
to a mixed state $|\nu_{\mu\tau}\rangle = c_{23} \, |\nu_{\mu}\rangle - s_{23}\, |\nu_\tau\rangle$
and, thus, the electron neutrino survival probability is unchanged.
In presence of $\theta_{13}\neq 0$, we have  instead non trivial modifications
due to the fact that the state $|\nu_e\rangle$  mixes with the state $|\nu_3\rangle$ 
being in fact $|\langle \nu_e |\nu_3\rangle| = s_{13}$.
By repeating the previous calculations, one obtains:
\begin{eqnarray}
\nonumber
P\left(\nu_e\rightarrow \nu_e\right) &=& \sin^4\left(\theta_{13}\right)  \\
& &
\nonumber
\hspace{-3.0cm}
+ \cos^4\left(\theta_{13}\right) \left[
\frac{1}{2}+\left(\frac{1}{2}-P_{\rm C}\right)\cos(2\tilde{\theta}_{12}) \cos(2\theta_{12}) \right]\\
& &
\label{Pee3nuMat}
\end{eqnarray}
where it is assumed that matter effects negligibly modify the $\theta_{13}$ mixing angle (i.e. $\theta_{13}\approx\tilde{\theta}_{13}$).

We finally remark that the above expression applies to solar neutrino detected
during the day, since these neutrinos do not cross the Earth to reach the detector. 
Matter effects due to propagation across the Earth can modify Eq.~(\ref{Pee3nuMat}) by 
introducing a day-night modulation whose magnitude depends on the specific values of mass 
and mixing parameters. 

\section{Neutrino Detectors}
\label{Sec:detectors}

The successful series of solar, atmospheric, reactor and accelerator experiments which led to firmly establish the standard three-flavor neutrino oscillation paradigm involved the realization of sophisticated detectors based on a plurality of techniques. In this paragraph we briefly review their main features, which undoubtedly played a key role in the incredible success of this field.

\subsection{Radiochemical detectors}
\label{Sec:radiochemical}

The emerging hint of the so called Solar Neutrino Problem at the beginning of the 70's from the first results of the pioneering Chlorine experiment (whose final findings are summarized in~\cite{Homestake}), carried out by Ray Davies in the Homestake mine, signaled the experimental beginning of the neutrino oscillation saga. The problem, consisting in a sizable discrepancy between the data and the prediction of the Standard Solar Model, persisted for more than 30 years before being explained as a manifestation of the neutrino oscillation phenomenon. A beautiful account of the early stage of this field can be found in the seminal book of John Bahcall~\cite{Bahcall1989}, where all the steps which brought to shape unambiguously the existence of the experimental puzzle are vividly and clearly explained. In the 90's additional evidence of the existence of the Solar Neutrino Problem came from other two radiochemical experiments, GALLEX/GNO~\cite{Gallex} at Gran Sasso and SAGE~\cite{Sage} at Baksan.
The principle of the radiochemical technique is very simple and elegant: the detection medium is a material which, upon absorption of a neutrino, is converted into a radioactive element whose decay is afterwards revealed and counted.
The Homestake experiment used a chlorine solution as a target for inverse $\beta$-interactions, 
\begin{equation}
\nu_e + ^{37}\rm{Cl} \rightarrow ^{37}\rm{Ar} + e^–
\label{Eq:ClAr}
\end{equation}
characterized by a threshold of 0.814\,MeV. It is worth to remind that such a technique was proposed independently by two giants of modern physics, Bruno Pontecorvo and Louis Alvarez.
The other two experiments, instead, adopted gallium as target, which allows neutrino interaction via 
\begin{equation}
\nu_e + ^{71}\rm{Ga} \rightarrow ^{71}\rm{Ge} + e^–
\label{Eq:GaGe}
\end{equation}
The threshold of this reaction is 233\,keV, low enough to 
essentially probe the entire solar neutrino spectrum (see sect.\ref{Subsec:source} for details)
which on the contrary cannot be revealed with the chlorine reaction due to the higher threshold.
Due to the similarity of the methodology in both cases of chlorine and gallium, in the following its description is focused to the specific gallium implementation.
In GALLEX/GNO the target consisted of 101\,tons of a GaCl$_3$ solution in water and HCl, containing 30.3\,tons of natural gallium; this amount corresponds to about $10^{29}$ $^{71}$Ga nuclei. The solution was contained in a large tank hosted in the Hall A of the underground Gran Sasso Laboratory.

$^{71}$Ge produced by neutrinos is radioactive and decay back by electron capture into $^{71}$Ga. The mean life of a $^{71}$Ge nucleus is about 16\,days: thus the $^{71}$Ge accumulates in the solution, asymptotically reaching equilibrium when the number of $^{71}$Ge atoms produced by neutrino interactions is just the same as the number of the decaying ones. In this equilibrium condition, about a dozen $^{71}$Ge atoms would be present inside the whole gallium chloride solution. Since the exposure time is in practice limited to four weeks, the actual number of $^{71}$Ge atoms is less than the equilibrium value, but still perfectly predictable. Therefore, the solar neutrino flux above threshold is deduced from the number of $^{71}$Ge produced atoms, using the theoretically calculated cross section. The challenging experimental task is thus to identify the feeble amount of $^{71}$Ge atoms.
This is accomplished through a complex procedure which contemplates several steps:
\begin{enumerate}
\item{the solution is exposed to solar neutrinos for about four weeks;}
\item{the $^{71}$Ge atoms present at the end of the four week period in the solution are in the form of volatile GeCl$_4$, which is extracted into water by pumping about 3000\,m$^3$ of nitrogen through the solution;}
\item{the extracted $^{71}$Ge is converted into gaseous GeH$_4$ and introduced into miniaturized proportional counters mixed with Xenon as counting gas. At the end of the process, a quantity variable between 95 and 98\% of the $^{71}$Ge present in the solution at the time of extraction is in the counter; extraction and conversion efficiencies are under constant control using non-radioactive germanium isotopes as carriers;} 
\item{decays and interactions in the counter are observed for a period of 6 months, allowing the complete decay of $^{71}$Ge and a good determination of the counter background. The charge pulses produced in the counters by decays are recorded by means of fast transient digitizers;}
\item{the data, after application of suitable cuts, are then analyzed with a maximum likelihood algorithm to obtain the most probable number of $^{71}$Ge introduced in the counter, with some final corrections applied to take into account the so called "side reaction", i.e. interactions in the solution generated by high energy muons from cosmic rays and by natural radioactivity}. 
\end{enumerate}

The key issue in the overall procedure is the minimization of the possible sources of backgrounds. This is performed through a triple strategy, whose first element is the rigorous application of low-level radioactivity technology in the design and construction of the counters; the second element is the use in the analysis of sophisticated pattern recognition techniques able to perform energy and shape discrimination of the signal and background events; the third and final element is the precise calibration of the counters via an external Gd/Ce X-ray source, to enhance the accuracy of the signal/background discrimination ensured by the pattern recognition method. 

Thanks to the effective methodology adopted, the radiochemical experiments were able to provide very important results in the studies of solar neutrinos, demonstrating unambiguously the discrepancy between the measured and predicted solar neutrino flux and triggering the subsequent vast theoretical and experimental investigations culminated in the proof of the oscillation effect in the solar neutrino sector. 
Such fundamental outputs were achieved despite the incredible challenge of the measurement, that can be well appreciated by considering the smallness of the detected signal.
In about two decades of operation, Homestake and SAGE detected 860 and 870 decays respectively, 
as reported in~\cite{Homestake} and \cite{Sage} (GALLEX/GNO did not publish this number). 

In this respect, it is worth to mention another important ingredient of the radiochemical solar neutrino program, i.e. the source calibration efforts which were performed to prove unambiguously the validity of the entire neutrino detection concept implied by this technique. In particular, GALLEX and SAGE underwent twice through the calibration procedure. GALLEX exploited in both cases a $^{51}$Cr source~\cite{Gallex2010}, while SAGE adopted two different isotopes, $^{51}$Cr in the first instance~\cite{SageSourceCr} and $^{37}$Ar in the second test~\cite{SageSourceAr}. The outcome of the source tests was the definitive validation of the radiochemical approach as an effective method to detect neutrinos. However, the ratio $R$ between the detected and predicted neutrino flux is significantly less than~1: 
taking the four tests together, the global result is $R = 0.86 \pm 0.05$. 
This anomaly can be interpreted as a possible indication of $\nu_e$ disappearance, see e.g. \cite{GiuntiSter}, within models with additional sterile neutrino states (see Sec.~\ref{subsec:sterile}).

\subsection{\v{C}erenkov detectors}
\label{subsec:cerenkov}

The widespread diffusion of the \v{C}erenkov technique in the field of neutrino physics can be appreciated by considering the many experimental set-ups based on this method which have been employed to investigate the entire neutrino spectrum, from the lowest to the highest energies. 

The \v{C}erenkov radiation is produced in a material with refractive index n by a charged particle if its velocity is greater than the local phase velocity of the light. The charged particle polarizes the atoms along its trajectory, generating time dependent dipoles which in turn generate electromagnetic radiation. If $v < c/n$ the dipole distribution is symmetric around the particle position, and the sum of all dipoles vanishes. If $v>c/n$ the distribution is asymmetric and the total time dependent dipole is different from zero, and thus radiates.

The resulting light wavefront is conical, characterized by an opening angle whose cosine is equal to $1/(\beta n)$; the spectrum of the radiation is ultraviolet-divergent, being proportional to $1/\lambda^2$. The propagation properties of the \v{C}erenkov light are therefore fully equivalent to that of the acoustic Mach cone.

\subsubsection{SNO}
\label{subsec:SNO}

The SNO experiment~\cite{SNOarXiv} is a paradigmatic example of how the \v{C}erenkov light can be used as basis to build a very effective neutrino detector\,\footnote{
Since SNO encompasses more experimental features than the other important detector of this kind, the Japanese Super-Kamiokande described in the next Sec.~\ref{subsec:supeK},
we find it convenient, for illustrative purposes, to reverse  the historical order (the data taking of Super-Kamiokande started before SNO).}.
 Located underground, in the Inco mine at Sudbury (Canada), this detector employed heavy water, which acted both as target medium for the neutrinos and as light generating material. The basic idea beyond the choice of heavy water is to perform two independent solar neutrino measurements based on the deuterium target: the first is aimed to detect specifically the electron neutrino component, while the second is sensitive to the all flavor flux. Thus, the comparison of the two results can permit to unambiguously discern if neutrinos, generated only as electron neutrinos in the core of the Sun, undergo flavor conversion during the path Sun-Earth. 

Heavy water makes this possible providing both flavor-specific and flavor-independent neutrino reactions. The first, flavor-specific reaction is the charged current (CC) reaction
\begin{equation}
\nu_e + d \rightarrow p + p + e^–
\label{Eq:CC}
\end{equation}
sensitive only to electron neutrinos.
%
Due to the large energy of the incident neutrinos, the produced electron will be so energetic that it will be ejected at light speed, which is actually faster than the speed of light in water, therefore creating a burst of \v{C}erenkov photons; after traveling throughout the water volume, they are revealed by the spherical array of photomultipliers instrumenting the detector. The amount of light is proportional to the incident neutrino energy, which can be inferred from the number of hits on the PMTs. From the hit pattern, also the angle of propagation of the light can be determined. 

The second flavor-independent reaction is the so called neutral current (NC) reaction 
\begin{equation}
\nu_x + d \rightarrow p + n + \nu_x
\label{Eq:NC}
\end{equation}
whose net effect is just to break apart the deuterium nucleus; the liberated neutron is then thermalized in the heavy water as it scatters around. The reaction can eventually be observed due to gamma rays which are emitted when the neutron is finally captured by another nucleus. The gamma rays will scatter electrons, which produce detectable light via the \v{C}erenkov process, in the same manner as discussed before.

The neutral current reaction is equally sensitive to all neutrino types; the detection efficiency depends on the neutron capture efficiency and the resulting gamma cascades. Neutrons can be captured directly on deuterium $^2$H$(n,\gamma)^3$H, but this is not very efficient. For this reason SNO has employed two separate systems to enhance the detection of NC-interactions. In the so called second SNO phase, $^{35}$Cl has been added to the heavy water in form of 2\,tons of NaCl and neutrons were detected through $^{35}$Cl$(n,\gamma)^{36}$Cl interaction. In the third SNO phase, the 36 proportional $^3$He counters have been deployed in the core of the detector which enabled the neutron detection based on $^3$He$(n,p)^3$H interaction.

There is also a third reaction occurring in the detector, flavor-independent as well, which is the electron scattering (ES) 
\begin{equation}
\nu_x + e \rightarrow \nu_x + e
\label{Eq:EE}
\end{equation}

This reaction is not unique to heavy water, being instead the primary mechanism in other light water detectors, like Kamiokande/Super-Kamiokande (see next sub-paragraph). Although this reaction is sensitive to all neutrino flavors, due to the different cross sections involved the electron neutrino dominates by a factor of six. The final state energy is shared between the electron and the neutrino, thus there is very little spectral information from this reaction. On the other hand, good directional information can be obtained.

The general drawback affecting the \v{C}erenkov technique is that, due to the feeble amount of light produced by the \v{C}erenkov mechanism, the effective neutrino threshold is around 4-5\,MeV, thus allowing the detection only of the high energy component of the solar flux, essentially the $^8$B neutrinos.

\begin{figure}[tb]
\begin{center}
\centering{\epsfig{file=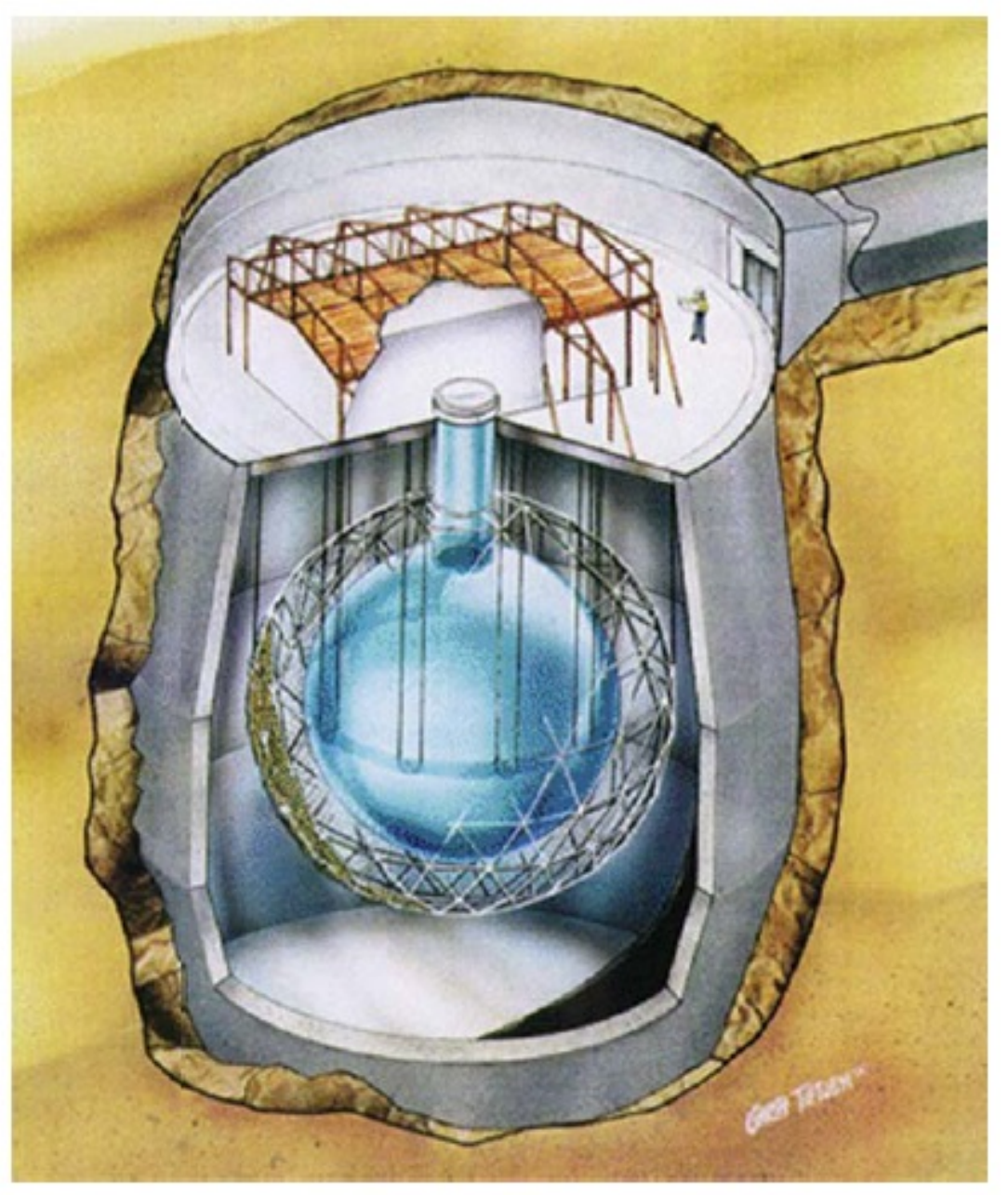,width=0.46\textwidth}}
\caption{Conceptual architectural scheme of the SNO detector.}
\label{Fig:SNO}
\end{center}
\end{figure}

The SNO experiment is now over; its architectural scheme was very simple (see Fig.~\ref{Fig:SNO}), aimed to get the most from the \v{C}erenkov technique: 1000\,tons of heavy water were contained in a thick transparent acrylic vessel, surrounded by an external layer of light water as shielding from the gammas from the radioactivity in the rock. A spherical array of 10000 8''phototubes detected the light from both volumes of water. A key issue for the success of the experiment was the long standing effort throughout the construction and the operation phases to reduce the natural radioactivity in the target volume, not only uranium and thorium, but also in particular the ubiquitous radon gas.

As a result of this experimental effort, the multiple, clean and almost background-free CC, NC, and elastic scattering detection of solar neutrinos provided the unambiguous and model independent proof that neutrinos from the Sun undergo flavor conversion. The specific "smoking gun" indication of the flavor-conversion process was obtained from the comparison of the depleted $\nu_e$-only flux of the CC measurement with the all-flavor flux evaluated through the NC reaction. The first publication of this result in 2002~\cite{SNOPhase1} nailed down definitively the explanation of the Solar Neutrino Problem.

\subsubsection{Super-Kamiokande}
\label{subsec:supeK}
 
As anticipated before, Super-Kamiokande~\cite{ran7}, like its predecessor Kamiokande~\cite{ran8}, is conceptually very similar to SNO, the major difference being the use of normal water instead of heavy water. Hence the neutrino detection occurs only via the scattering reaction off the electrons; the afterwards mechanism of \v{C}erenkov light production and detection via an array of PMT's is equal to that already described for SNO. 

Another major difference is the quantity of water employed, in total 50\,ktons (observed by almost 13000 20'' PMTs), which makes this detector the most massive among the neutrino oscillation experiments built so far. The sufficiently high statistics implied by this huge volume has made possible a fairly precise reconstruction of the spectrum of the scattered electrons, which plays an important role in the subsequent analysis for the interpretation of the data. With its huge mass Super-Kamiokande clearly outperforms the findings of the old Kamiokande (containing only 3000\,tons of water), obtained in the data taking period from 1983 to 1994; however Kamiokande maintains a crucial historical role in the fields of neutrino oscillation and of astrophysical neutrinos, in this case with the detection of the neutrinos from the supernova SN1987A, as witnessed by the 2002 Nobel prize. In this context, it is appropriate to mention also another historically important \v{C}erenkov experiment, the IMB (Irvine-Michigan-Brookhaven) detector~\cite{IMB}, realized with 10000\,tons of water, which shared with Kamiokande the success of detecting the SN1987A neutrinos.  

The intrinsic high directionality of the scattering reaction, coupled to the directionality of the \v{C}erenkov light, provides this experiment with a powerful tool to fight the background due to trace impurities of natural radioactivity dissolved in water, by associating the reconstructed direction of the \v{C}erenkov photons with the angular position of the Sun. Clearly, this is done on top of the purification procedure of the light water, which as for SNO was focused generally on the whole natural radioactivity, but with special emphasis on the radon, which is the factor limiting the threshold at low energy.

An additional important analysis tool is the typical feature of the \v{C}erenkov light to generate sharp \v{C}erenkov rings in case of muon particles, while electrons make rings with fuzzy edges.
Contrary to SNO, Super-Kamiokande is still currently taking data. The long history of this detector started in 1996 and evolved through four phases: the first phase lasted until a major PMT incident in November 2001 and produced a very accurate measure of the $^8$B flux via the ES detection reaction. The phase II with reduced number of PMTs, from the end of 2002 to the end 2005, confirmed with larger error the phase I measurement. After the refurbishment of the detector back to the original number of PMTs, the third phase lasted from the middle of 2006 up to the middle of 2008. Later on, an upgrade of the electronics brought the detector into its fourth, current phase. It is important to highlight the evolution of the energy threshold (total electron energy) in all the phases: 5\,MeV in phase I, 7\,MeV in phase II, 4.5\,MeV in phase III and 4\,MeV for phase IV, thanks to the continuously on-going effort to reduce the radon content in water.

Undoubtedly Super-Kamiokande played a central role in the long path which led to unveil the neutrino oscillation phenomenon, since it has been, and still is, a major player in three of the areas of investigation for neutrino oscillation, e.g those based on solar, atmospheric and accelerator neutrinos. Actually, it was Super-Kamiokande that in 1998~\cite{ran9} announced the epochal discovery of neutrino oscillations, which stemmed from the observed anomaly of the number of atmospheric muon neutrino events compared to electron neutrino events, and it was Super-Kamiokande that first confirmed the oscillation process with a beam of artificial (accelerator) muon neutrinos in the dedicated K2K experiment~\cite{K2K2006}, which took place from 1999 to 2004. And nowadays this successful story continues with the T2K~\cite{ran11} experiment, another accelerator neutrino experiment which is the successor of K2K.

In the solar neutrino study the result provided by Super-Kamiokande are equally of great importance, as key ingredient of the joint analysis of all the experiments to ascertain the allowed regions of the oscillation parameters~\cite{ran12}.

\subsection{Scintillation detectors}
\label{subsec:scintil}

Scintillation detectors have a long and established tradition in the area of neutrino physics, starting from the Cowan-Reines's Savannah River experiment~\cite{ReinesCowan}, which performed the first neutrino detection ever. Other pioneer detectors of this kind which deserve to be mentioned for their historical role in the field (but not necessary in the oscillation study) are the Baksan Underground Scintillation Telescope (BUST)~\cite{BUST}, which also detected the SN1987A neutrinos, the Liquid Scintillation Detector (LSD) at Mont Blanc~\cite{LSD}, the Large Volume Detector (LVD) at Gran Sasso~\cite{LVD} devoted to Supernova search, and the Gosgen~\cite{Gosgen} and Bugey~\cite{Bugey} reactor experiments.

In the following we focus our attention on the more recent implementations of this technique, for the realization of experiments which played a fundamental role in nailing-down the neutrino oscillation properties.

\subsubsection{Borexino}
\label{subsec:BX}

In the context of the solar neutrino research, the Borexino project was conceived and designed to detect in real time the low energy component of the solar flux, with special emphasis on the neutrinos coming from the $^7$Be electron capture in the core of the Sun, exploiting as simple and effective mean to reveal the incoming particles their scattering reaction off the electrons of the target medium.

Specifically, Borexino is a scintillator detector~\cite{ran13} which employs as active detection medium a mixture of pseudocumene (PC, 1,2,4--trimethylbenzene) and PPO (2,5-diphenyloxazole, a fluorescent dye) at a concentration of 1.5\,g/l. Because of its intrinsic high luminosity (50\,times more than in the \v{C}erenkov technique) the liquid scintillation technology is extremely suitable for massive calorimetric low energy spectroscopy. The isotropic nature of the scintillation light does not allow inferring the direction of the incoming particles; it is therefore impossible, contrary to what happens in \v{C}erenkov experiments, to distinguish neutrino scattered electrons from electrons due to natural radioactivity by the association with the direction from the Sun. Thus the key requirement in the technology of Borexino is an extremely low radioactive contamination.

\begin{figure}[tb]
\begin{center}
\centering{\epsfig{file=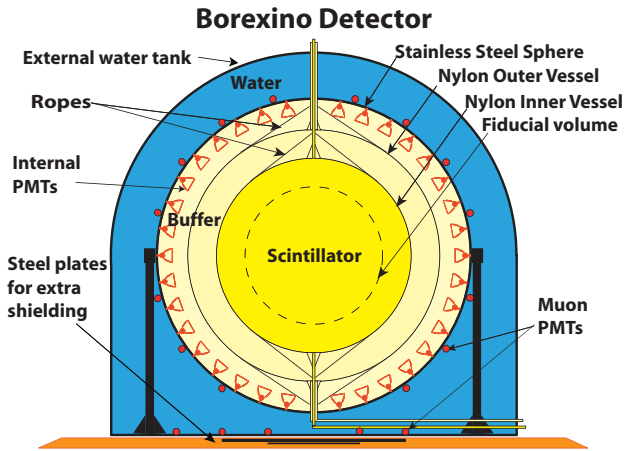,width=0.46\textwidth}}
\caption{Sketch of the Borexino experiment, highlighting its major components arranged according to a graded shielding design.}
\label{Fig:BX}
\end{center}
\end{figure}

To reach ultra-low operating background conditions in the detector, the design of Borexino, as shown in Fig.~\ref{Fig:BX}, is based on the principle of graded shielding, with the inner scintillating core at the center of a set of concentric shells of increasing radiopurity. The scintillator mass (278\,tons) is contained in a 125\,$\mu$m thick nylon Inner Vessel (IV) with a radius of 4.25\,m. Within the IV a fiducial mass is software-defined through the estimated events position, obtained from the PMTs timing data via a time-of-flight algorithm.

A second nylon outer vessel (OV) with radius 5.50\,m surrounds the IV, acting as a barrier against radon and other background contaminations originating from outside. The region between the inner and outer vessels contains a passive shield composed of pseudocumene and 5.0\,g/l (later reduced to 3.0\,g/l) of DMP (dimethylphthalate), a material that quenches the residual scintillation of PC so that spectroscopic signals arise dominantly from the interior of the IV. 

A 6.85\,m radius stainless steel sphere (SSS) encloses the central part of the detector and serves also as a support structure for the PMTs. The region between the OV and the SSS is filled with the same inert buffer fluid (PC plus DMP) which is layered between the inner and outer vessels.

Finally, the entire detector is contained in a tank (radius 9\,m, height 16.9\,m) filled of ultra-pure water. The total liquid passive shielding of the central volume from external radiation (such as that originating from the rock) is thus 5.5\,m of water equivalent (m.w.e). The scintillator material in the IV was less dense than the buffer fluid by about 0.1\% with the original DMP concentration of 5\,g/l; this resulted in a slight upward buoyancy force on the IV, implying the need of thin, low-background ropes made of ultra-high density polyethylene to hold the nylon vessels in place.
This modest buoyancy was further reduced of more than a factor 10 by removing via distillation a fraction of the total DMP content in the buffer: the process ended with a final DMP concentration of 3\,g/l, still perfectly adequate to suppress the buffer scintillation, while at the same time implying less stress applied to the IV.

The scintillation light is viewed by 2212 8'' PMTs uniformly distributed on the inner surface of the SSS. All but 371 photomultipliers are equipped with aluminum light concentrators designed to increase the collection efficiency of the light from the scintillator, and concurrently minimizing the detection of photons not coming from the active scintillating volume. Residual background scintillation and \v{C}erenkov light that escape quenching in the buffer are thus reduced. The PMTs without concentrators can be used to study this background, as well as help identify muons that cross the buffer and not the inner vessel.

Besides being a powerful shield against external backgrounds ($\gamma$'s and neutrons from the rock), the Water Tank (WT) is equipped with 208 PMTs and acts as a \v{C}erenkov muon detector. The muon flux, although reduced by a factor 106 by the 3800 m.w.e. depth of the Gran Sasso Laboratory, is still significant (1.1 muon m$^{-2}$ h$^{-1}$) and an additional reduction (by about 10$^4$) is necessary. Ultra-low radioactive contamination is the distinctive feature of Borexino, achieved through a multiple strategy~\cite{ref3.18} that implied on one hand the careful selection and screening of all the construction materials and components, and on the other the purification of the active scintillator to unprecedented purity levels (see Tab.~\ref{tab:BXpurity}).

\begin{table*}
\begin{center} 
\caption{Radio-purity of the Borexino detector during the phase 1 of the experiment.}
\vspace {2mm}
\label{tab:BXpurity}
\begin{tabular}{l|l|l|l|l}
\hline
Name	  &   Source         &     Typical                       &    Required    & Achieved \\ 
\hline
\hline
$^{14}$C     & intrinsic PC  & $\sim 10^{-12}$ g/g   & $\sim$$10^{-18}$ g/g   & $\sim$$2 \times 10^{-12}$ g/g \\
$^{238}$U   & dust            & $10^{-5} - 10^{-6}$ g/g   & $<  10^{-16}$ g/g   & $ (5.0 \pm 0.9)  \times 10^{-18}$ g/g \\
$^{232}$Th   &            &                                           &                                 & $ (3.0 \pm 1.0)  \times 10^{-18}$ g/g \\
$^{7}$Be   & cosmogenic           & $\sim$$3 \times 10^{-2}$ Bq/ton   & $ <  10^{-6}$ Bq/ton   & not observed \\
$^{40}$K   & dust, PPO           & $\sim$$2 \times 10^{-6}$ g/g (dust)   & $ <  10^{-18}$ g/g   & not observed \\
$^{210}$Po   & surface            &                      & $ < 7$ cpd /ton   & May07: 70 cpd/ton\\
   &     contamination            &                 &                                                                                & May09: 5 cpd/ton \\
$^{222}$Rn   & emanation, rock          &      10 Bq/l (air, water)     & $ < 10$ cpd /100 tons   &  $< 1$ cpd/ 100 tons\\
                      &                                   &  100 - 1000 Bq/kg (rock) &                                    & \\
$^{39}$Ar       & air, cosmogenic  & 17\,mBq/m$^3$ (air)     & $<$1\,cpd/100\,tons    & $<<$ $^{85}$Kr \\
$^{85}$Kr       & air, nuclear weapons &       $\sim$1\,Bq/m$^3$ (air)     &   $<$ 1\,cpd/100\,tons    &   $ 30 \pm 5$\,cpd/100\,tons \\
\hline
\end{tabular}
\end{center}
\end{table*}

Clearly, in this respect key factor are the many liquid purification and handling systems designed and installed to ensure the proper manipulation of the scintillator at the incredible degree of cleanliness demanded by the experiment. The exceptional low-background environment achieved in the core of the liquid scintillator allowed the unprecedented and precise sub-MeV measure of the $^7$Be component of the solar neutrino flux, which to date is the only direct confirmation of the validity in such a low energy range of the MSW mechanism driving the oscillation of neutrinos produced in the core of the Sun.

Borexino has taken data during the so called phase 1 (May 2007-July 2010) and started again to collect data (phase 2), after a further campaign of purification of the scintillator, in October 2011. The purification campaign succeeded to further reduce the residual contamination of the scintillator. 

\subsubsection{Other scintillation experiments} 
\label{subsec:OtherScint}

Other important scintillator based experiments which provided milestone results for the understanding of the neutrino oscillation properties are KamLAND~\cite{KamlPrecision}, and, more recently, Daya Bay~\cite{ran15}, RENO~\cite{RENO} and Double Chooz~\cite{ran17}. While in term of methodology all these experiments are very similar to Borexino, as far as detection criteria, techniques and architectural scheme are considered, their specific characteristics is the measurement target, constituted by antineutrinos from reactors. KamLAND, in particular, is not close to any specific reactor, but rather detects antineutrinos from a number of Japanese power plants located at an average distance of 200\,km, thus performing a long-baseline test. Day Bay, RENO, and Double Chooz, instead, are located a 1\,km from the reactor, acting thus as medium baseline experiments.

While in general the respective technology resembles closely that of Borexino, there are some variations in the type of liquid scintillator, and in the material used for the balloon containing the liquid. The main difference with Borexino stems from the inverse beta reaction which is used to detect antineutrinos (in contrast with the scattering reaction adopted in Borexino to reveal neutrinos)
\begin{equation}
\bar{\nu}_e + p \rightarrow e^+ + n,
\label{Eq:InvBeta}
\end{equation}
After the occurrence of this interaction, the {\it prompt} signal is due to a positron which decelerates and then annihilates producing two 511\,keV $\gamma$ rays. The neutron thermalizes and is captured by a free proton, generating a typical 2.2\,MeV gamma, the so called {\it delayed} signal. The visible energy $E_{vis}$ of the prompt signal is directly correlated with the kinetic energy of the incident antineutrino $E_{\bar{\nu}_e} = E_{vis} + 0.784$ [MeV]. The mean time between the positron production and the neutron capture is about 200-260\,$\mu$s depending on the scintillator type, therefore the tight time coincidence between the respective light signals origins a correlated measurement which ensures a powerful discrimination of a true antineutrino detection with respect to the uncorrelated background events. This kind of signature greatly reduce the requirements for the suppression of the intrinsic radioactivity in the scintillator, marking the major difference between the technology of these reactor experiments and that employed for the solar neutrino detection in Borexino. For example, in KamLAND $^{238}$U has been reduced to $(1.5 \pm 1.8) \times 10^{-19}$\, g/g, $^{232}$Th to $(1.9 \pm 0.2) \times 10^{-17}$\,g/g, $^{40}$K to a limit $< 4.5 \times 10^{-18}$\,g/g, $^{210}$Po to $\sim$2\,mBq/m$^3$,$^{210}$Bi to $<$1\,mBq/m$^3$, and $^{85}$Kr to $\sim$0.1\,mBq/m$^3$.

Historically, the measurement of KamLAND, together with that of SNO, closed the Solar Neutrino Problem, showing unambiguously that also reactor antineutrinos undergo the oscillation phenomenon, while concurrently determining rather precisely the associated mass squared difference parameter $\Delta m^2_{21}$ and, jointly with the outputs of all other solar experiments, the mixing angle $\theta_{12}$.

Daya Bay and RENO (and in future Double Chooz, as well) have the additional characteristics of being equipped with a near detector, so that the far-near arrangement allows determining also the $\theta_{13}$ mixing angle.

As additional remark of this section, it has to be emphasized that also some of the experiments whose outputs are used in the analysis concerning the existence of additional sterile state(s) beyond the established three-neutrino oscillation framework (see Sec.~\ref{subsec:sterile}), are liquid scintillator set-ups. In particular, the LSND~\cite{LSND} and MiniBooNE~\cite{ran19} detectors, at the center of the current hot debate in this area, share essentially all the distinctive features of the other experiments belonging to the same technical "family". Specifically, LSND was a cylindrical tank containing 167\,tons of scintillator viewed by 1220 8\,inch PMTs, while MiniBooNE was based on a spherical detector geometry to contain 800\,tons of scintillator, though still using a similar number of PMTs, 1280. The peculiarity of both set-ups was the exploitation of a special scintillator mixture able to produce a comparable amount of \v{C}erenkov and true scintillation light.

The KARMEN experiment~\cite{KarmenDet} was another player in this debate, but on the other side, since it did not detect the same hints of LSND and MiniBooNE. It was a segmented liquid-scintillator detector; the segmentation, technically the more distinctive feature of the set-up, was realized with 1.5\,mm thick lucite sheets which ensured the transport of the light to the photomultipliers via total internal reflection. The detector was also instrumented with a veto employing plastic scintillator modules.

Finally, the pioneer MACRO detector~\cite{MACRO} at Gran Sasso was, as well, based on segmented liquid-scintillator counters. Actually it comprised three subsystems, being additionally equipped with limited streamer tubes and nuclear track detectors, which altogether provided the experiment with the capability to detect the atmospheric neutrino oscillation phenomenon.

\subsection{Further techniques}
\label{subsec:OtherTechn}

To complete the illustration of the techniques adopted for the neutrino oscillation studies, a brief mention is due to other experiments which have shed light on important aspects of the field, while being not ascribable to any of the methodological categories described so far, starting with MINOS~\cite{ran20} and OPERA~\cite{ran21}.  We briefly describe also the basic features of the near detectors of the already mentioned T2K experiment~\cite{ran11} (that, we remind, uses Super-Kamiokande as the far detector), as well as of the CHORUS~\cite{CHORUS}, NOMAD~\cite{NOMAD}, and ICARUS~\cite{ICARUS} experiments.

The MINOS experiment exploits two detectors to register the neutrino interactions: the near detector at Fermilab characterizing the neutrino beam (NuMI, Neutrinos at the Main Injector, beam) is located about 1\,km from the primary proton beam target, while the far detector performs similar measurements 735\,km downstream. The far detector is located in Soudan, hosted in an inactive iron mine where it is positioned in a cavern excavated on purpose, 705\,m underground (2070\,meters-water-equivalent (m.w.e.)), 210\,m below sea level.

The rationale of the experiment is to make comparisons between event rates, energies and topologies at both detectors, and to infer from those comparisons the relevant "atmospheric" oscillation parameters. The energy spectra and rates are measured separately for $\nu_{\mu}$ and $\nu_e$ charged-current (CC) events, as well as for neutral current (NC) events.

Both the near and far MINOS detectors are steel-scintillator sampling calorimeters, equipped with tracking, energy and topology measurement possibilities. Such a multiple capability is obtained by alternate planes of plastic scintillator strips and 2.54\,cm thick magnetized steel plates.

The 1\,cm thick by 4.1\,cm wide extruded polystyrene scintillator strips are read out using wavelength-shifting fibers coupled to multi-anode photomultiplier tubes. Both detectors ensure equal transverse and longitudinal sampling for fiducial beam-induced events.

The far detector comprises 486 octagonal steel planes, with edge to edge dimension of 8\,m, interleaved with planes of plastic scintillator strips. The total mass is 5400\,tons; the set-up is arranged as two "supermodules" separated by a 1.15\,m distance, individually equipped with an independently controlled magnet coil. 

The near detector, consisting of 282 planes for a total mass of 980\,tons, is located at the extreme of the NuMI beam facility at Fermilab, in a 100\,m deep underground cavern under a 225\,m.w.e. overburden. It exploits the high neutrino flux at this site to identify a relatively small target fiducial volume for selection of events to be employed for the near/far comparison. The upstream part of the detector, i.e. the calorimeter portion, contains the target fiducial volume with all the planes instrumented. The downstream part, the spectrometer section dedicated to the measurement of the momenta of energetic muons, has only one plane every five instrumented with scintillator.

The core of the MINOS detectors’ active system is thus based on the technique of solid scintillator, whose main features are good energy resolution and hermiticity, excellent transverse segmentation, flexibility in readout, fast timing, simple and robust construction, long-term stability, ease of calibration, reliability and, last but not least, low maintenance requirements. Furthermore, the whole set-up met also safety and practicality of construction requirements.

The performances of both detectors rely on some key parameters which are the steel thickness, the width of scintillator strips, and the degree of readout multiplexing, which were carefully studied and optimized during the design phase.

The MINOS detectors represented a significant increase in size from previous fine grained scintillator sampling calorimeters, and therefore the relevant design and construction efforts ended-up with important technical advancements in detector technology of general interest for the field of application of this technique.

This technological effort of the MINOS construction resulted in an impressive scientific success, which brought further evidence to the neutrino oscillation investigation performed by Super-Kamiokande and K2K, sharpening significantly the evaluation of the relevant “atmospheric” oscillation parameters.

The OPERA experiment was designed aiming at the direct observation of $\nu_{\tau}$ appearance stemming from $\nu_{\mu} \rightarrow \nu_{\tau}$ oscillation in a long baseline beam (dubbed CNGS) from CERN to the underground Gran Sasso Laboratory, at a distance of 730\,km, where OPERA is located.

The design of OPERA was specifically tailored to identify the $\tau$ via the topological observation of its decay, reinforced by the kinematic analysis of the event. This goal is pursued through a hybrid apparatus based on two "pillars": real-time detection techniques ("electronic detectors") and the Emulsion Cloud Chamber (ECC) method. A detector based on the ECC approach is made of passive material plates, used as target, alternated with nuclear emulsion films employed as tracking devices, featuring sub-micrometric accuracy.

The sub-micrometric position accuracy, coupled to the adoption of passive material, allows for momentum measurement of charged particles through the detection of multiple Coulomb scattering, as well as for identification and measurement of electromagnetic showers, together with electron/pion separation.

In essence, the main advantage of the ECC technique is the unique property of combining a high accuracy tracker with the capability of performing precise measurements of kinematic variables.

OPERA scaled the ECC technology to an unprecedented size: the basic unit of the experiment is a "brick" realized with 56 plates of lead (1\,mm thick) interleaved with nuclear emulsion films, for a total mass of 8.3\,kg; 150 000 of such target units have been assembled, amounting to an overall mass of 1.25\,kton. The bricks are arranged in 62 vertical structures (walls), orthogonal to the beam direction, interleaved with planes of plastic scintillators.

The detector is made of two identical supermodules, each comprising 31 walls and 31 double layers of scintillator planes followed by a magnetic spectrometer.

The electronic detectors accomplish the twofold task to trigger the data acquisition, identify and measure the trajectory of charged particles and locate the brick where the interaction occurred. 

The momentum of muons is measured by the spectrometers, with their trajectories being traced back through the scintillator planes up to the brick where the track originates. In case of no muons observation, the scintillator signals produced by electrons or hadronic showers are used to predict the location of the brick that contains the primary neutrino interaction vertex. The selected brick is then extracted from the target and afterwards the two interface emulsion films attached on the downstream face of the brick are developed. If tracks related to neutrino interaction are observed in these interface films, the films of the brick are developed, too, following the tracks back by fully automated scanning microscopes until the vertex is located.

The analysis of the event topology at the primary vertex leads to the identification of possible $\tau$ candidates. Topologies of special interest might include one track that shows a clear "kink" due to the decay-in-flight of the $\tau$ (long decays) or an anomalous impact parameter with respect to the primary vertex (short decays) compatible with a decay-in-flight in the first lead plate. Once selected, such topologies are double checked by a kinematic analysis at the primary and decay vertices.

The modular structure of the target ensures to extract only the bricks actually hit by the neutrinos, therefore achieving an efficient analysis strategy of the interaction, while at the same time minimizing the target mass reduction during the run.

In the overall structure of the OPERA detector each brick wall, containing 2912 bricks and supported by a light stainless steel structure, is followed by a double layer of plastic scintillators (Target Trackers, TT) that provide real-time detection of the outgoing charged particles. The instrumented target is further followed by a magnetic spectrometer, consisting of a large iron magnet instrumented with plastic Resistive Plate Chambers (RPC). The bending of charged particles inside the magnetized iron is measured by six stations of drift tubes (Precision Trackers, PT). Left-right ambiguities in the reconstruction of particle trajectories inside the PT are removed by means of additional RPC, with readout strips rotated by $\pm$45$^{\circ}$ with respect to the horizontal plane and positioned near the first two PT stations.
What defined before as a supermodule is actually an instrumented target together with its spectrometer.
Finally, two glass RPC planes mounted in front of the first target allow rejecting charged particles originating from outside the target fiducial region, coming from neutrino interactions in the surrounding materials.

As conclusive remark, OPERA is the first very large scale emulsion experiment: the 150000 ECC bricks include about 110000 m$^2$ emulsion films and 105000 m$^2$ lead plates; the scanning of the events is performed with more than 30 fully automated microscopes. The success of this impressive machine is witnessed by the unambiguous detection of 3 $\tau$ events, so far.

In an arrangement similar to MINOS, T2K~\cite{ran11} employs two near detectors located 280\,m from the graphite proton target to measure the properties of the un-oscillated neutrino beam.

The INGRID near detector comprises 16 modules, 14 of which are positioned in a cross configuration centered on the beam axis. They are made of iron and scintillator layers, allowing the measure of the neutrino rate and profile on the beam axis direction.

The ND280 off-axis near detector is located off the beam axis in the same direction as SK, being exploited to measure the properties of the un-oscillated off-axis beam. It consists of several sub-detectors: the so called Pi-Zero detector (P$\O$D) is a plastic scintillator-based detector optimized for $\pi^0$ detection, followed by a tracking detector made of two fine grained scintillator detector units, in turn sandwiched between three time projection chambers. Both the P$\O$D and tracker are surrounded by electromagnetic calorimeters, including a module located immediately downstream of the tracker itself. The whole detector is located in a magnet with a 0.2\,T magnetic field, serving also as mass for a side muon range detector.

Important predecessors of these efforts were two experiments carried out at CERN in the 90's, CHORUS~\cite{CHORUS} and NOMAD~\cite{NOMAD}.

The active target of CHORUS was realized with nuclear emulsions (total mass of 770\,kg). A scintillating fiber tracker was interleaved, both for timing and for extrapolating the tracks back to the emulsions. The set-up comprised also a hexagonal spectrometer magnet for momentum measurement, a high resolution spaghetti calorimeter for measuring hadronic showers, and a muon spectrometer. The scanning of the emulsions was performed with high-speed CCD microscopes.

NOMAD adopted drift chambers as target and tracking medium. The chambers were 44, located in a 0.4\,T magnetic field, for a total a fiducial mass of 2.7\,tons. They were followed by a transition radiation detector (for $e/\pi$ separation), by additional electron identification devices and by an electromagnetic lead glass calorimeter. The detector comprised also a hadronic calorimeter, 10 drift chambers for muon identification and an iron-scintillator calorimeter of about 20\,tons.

Finally, looking ahead to the future, it must be mentioned that a very promising technique potentially very useful for neutrino oscillation investigation is that based on liquid argon, developed through a very long research and development effort for the ICARUS detector~\cite{ICARUS}. Such liquid argon time projection chamber allows calorimetric measurement of particle energy together with three-dimensional track reconstruction from the electrons drifting in the electric field applied to a volume of sufficiently pure liquid argon. The technique, thus, successfully reproduces the extraordinary imaging features of a bubble chamber, but with the advantage of being a full electronic detector, potentially scalable to the huge masses required for the next round of experimental neutrino studies.

\section{Experimental results}
\label{Sec:results} 

The experimental results concerning the neutrino oscillations have been obtained studying neutrinos from several sources: solar and atmospheric neutrinos, reactor antineutrinos, neutrino and antineutrino accelerator beams. The  neutrino experiments make use of a variety of techniques: radiochemical methods, water and heavy water \v{C}erenkov detectors, liquid and plastic scintillators; in some detectors  also streamer chambers and time projection chambers are used, in addition to nuclear emulsions.

The experiments can be classified as disappearance and appearance ones: the first are measuring a reduced flux of neutrinos having the same flavor as at the source, while the second are looking for neutrinos of different flavor  with respect to those emitted by the source.
  
\subsection{Neutrino sources}
\label{Subsec:source}

 The Sun is one important source of neutrinos.
Energy in the Sun is, in fact, produced by chains of nuclear reactions whose 
overall result is the conversion  of hydrogen into helium
\begin{equation}
4p + 2e^{-} \to {}^{4}{\rm He} + 2\nu_e~.
\label{HydrogenFusion}
\end{equation}
Due to lepton number conservation, helium production is accompanied by the 
production of two electron neutrinos. The total energy released in reaction~(\ref{HydrogenFusion})
is $Q=26.73$~MeV and only a small part of it (about 0.6 MeV on average) is carried away 
by the two neutrinos. The total flux of electron neutrinos arriving on Earth (if they do not oscillate) can be then estimated 
from the radiative flux $K$ produced by the Sun on the Earth surface, obtaining:
\begin{equation}
\Phi_{\rm tot} \approx 2 \, \frac{K}{Q} \approx 6 \times 10^{10}\;{\rm cm}^{-2}{\rm s}^{-1} ~.
\end{equation}  
Due to the eccentricity of the Earth's orbit, the solid angle from the Sun to the Earth changes during the year, 
and thus the solar neutrino flux shows a seasonal variation.

\begin{figure}[tb]
\begin{center}
\centering{\epsfig{file=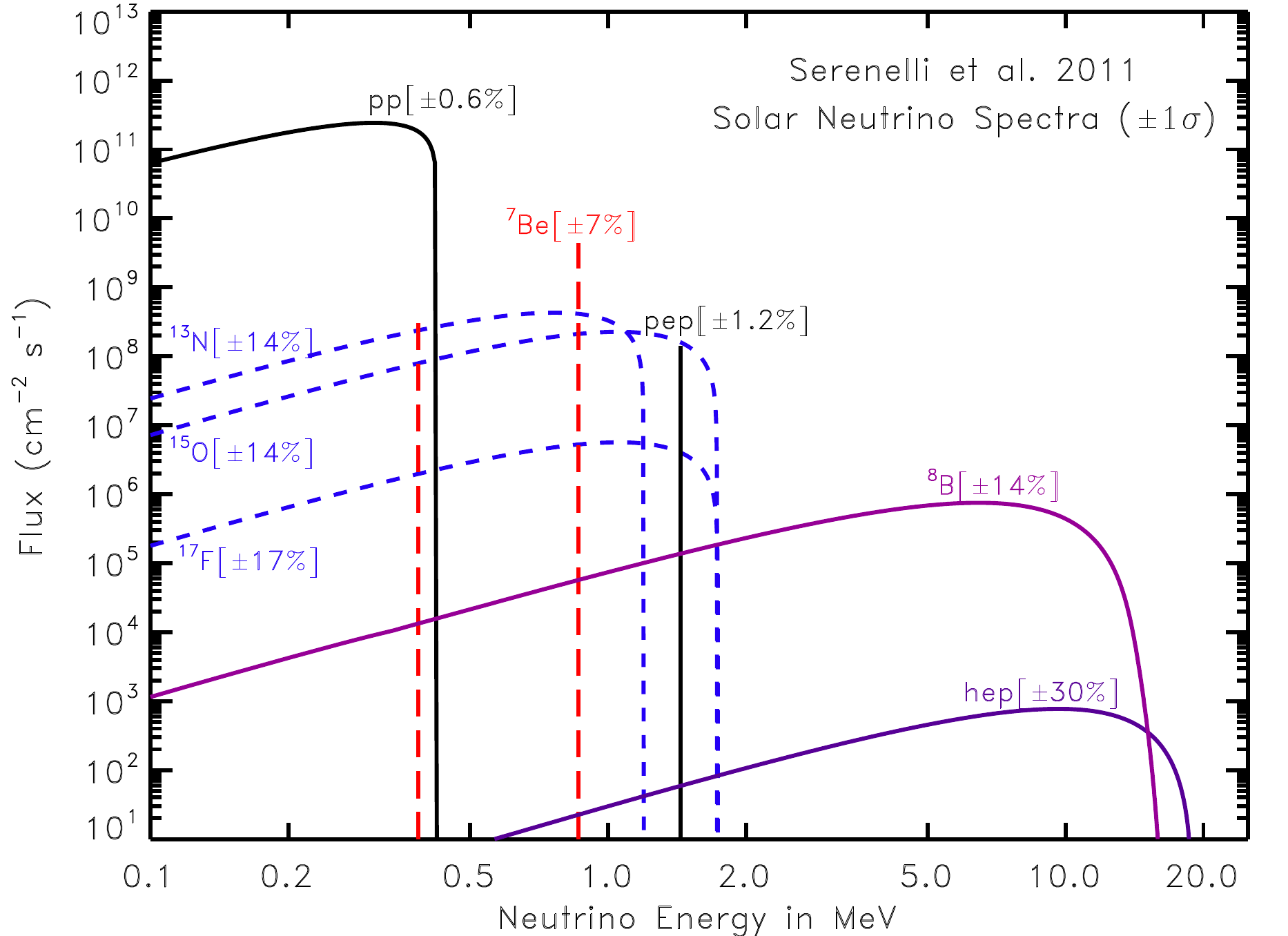,width=0.48\textwidth}}
\caption{The solar neutrino spectrum predicted by 
the SSM calculation of~\cite{Serenelli2011}.}
\label{SolarNuspectrum}
\end{center}
\end{figure}

The interpretation of solar neutrino experiments requires a detailed knowledge
of the solar neutrino spectrum, see Fig.~\ref{SolarNuspectrum}. Hydrogen burning in the Sun proceeds through two chains,
namely the $pp$ chain and the CNO bi-cycle.
At the temperature and density characteristic of the solar interior, hydrogen burns with $\sim$99\% 
probability through the $pp$ chain that is predominantly initiated by the $p+p\to d+e^++\nu_{e}$ reaction. 
This reaction produces the so-called {\em pp neutrinos} which have a continuous spectrum extending up to 
$E=0.42$~MeV and constitutes $\simeq$90\% of the total neutrino flux. 
Alternatively, the $pp$ chain can originate with 0.23\% probability
from the reaction $p+e^-+p\to d+\nu_{e}$ that produces the less abundant 
monochromatic {\em pep neutrinos} with energy $E = 1.445 \,{\rm MeV}$. 

The $pp$ chain has three possible different branches ($pp$-I, $pp$-II and $pp$-III) 
whose relative rates depend on the central temperature of the Sun. 
In the $pp$-II termination, the electron capture reaction 
$e^- +{}^7{\mathrm Be}\to {}^7{\rm Li}+\nu_{e}$ produces 
the monochromatic {\em $^7Be$ neutrinos} with energy $E = 0.863\;{\rm MeV}$\footnote{
This value correspond to transitions to the $^{7}{\rm Li}$ ground state. With $\sim$10\% probability, 
$^{7}{\rm Li}$ is produced  in the first excited states together with a neutrino with energy $E=0.383\;{\rm MeV}$.}.
In the $pp$-III branch, the $\beta$-decay $^{8}{\rm B}\to{}^{8}{\rm Be}^*+e^++\nu_e$
is responsible for the production of the {\em $^8B$ neutrinos}.
The flux of $^8{\rm B}$ neutrinos is extremely low, being approximately equal
to $0.01\%$ of the total flux, but the spectrum extends up to a maximal energy 
$E \approx 15\, {\rm MeV}$.

In the CNO-cycle, the overall conversion of four protons into helium is achieved 
with the aid of C, N and O nuclei present in the Sun. The $\beta$-decays $^{13}{\rm N}\to{}^{13}{\rm C}+e^++\nu_e$,
$^{15}{\rm O}\to{}^{15}{\rm N}+e^++\nu_e$ and, to a minor extent, $^{17}{\rm F}\to{}^{17}{\rm O}+e^++\nu_e$ produce the so-called 
{\em $^{13}N$},  {\em $^{15}O$}, and {\em $^{17}F$  neutrinos}, respectively, all together referred to as {\em CNO neutrinos}.
These three components of the solar neutrino flux have continuous spectra extending up to 
$E\simeq 1.2$, 1.7 and  1.7\,MeV, respectively.

The predictions for each component of the solar neutrino flux are 
obtained by constructing a Standard Solar Model (SSM) which, according to the definition of~\cite{Castellani}, 
is a solution of the stellar structure equations (starting from a chemical homogeneous initial model)
that reproduces,  within uncertainties, the observed properties of the present Sun, by
adopting physical and chemical inputs chosen within their range
of uncertainties. In Tab.~\ref{tabSSM}, we report the neutrino fluxes predicted by two recent SSM calculations
that adopt two different assumptions for the admixture of heavy elements in the 
Sun. Namely, the model labeled GS98 is obtained by using the "old''  composition
from \cite{GS98}, while the model labeled AGSS09 adopts the "new'' admixture
of~\cite{AGSS09}. The reason to consider these two calculations is that, in recent 
years, a new solar problem, often referred to as {\it solar metallicity puzzle}, has emerged. The most recent determinations of the solar photospheric heavy-element abundances (among which~\cite{AGSS09}) have indicated, in fact, that the solar metallicity is lower by 30 to 40$\%$ than previous measurements~\cite{GS98}. However, the internal structure of SSMs calibrated against the newly determined solar surface metallicity do not reproduce the helioseismic 
constraints, see e.g.~\cite{Basu}. The experimental determination of the solar neutrino fluxes, 
beside providing crucial information for flavor neutrino oscillations, may help to shed light 
on the origin of these discrepancies.

\begin{table}
\begin{center}
\caption{The predictions of SSMs implementing GS98 \cite{GS98} and AGSS09 \cite{AGSS09} admixtures.
See~\cite{Serenelli2011} for details.}
\label{tabSSM} 
\vspace{2mm}
\begin{tabular}{l|l|l}
\hline
 & AGSS09 &  GS98   \\
\hline
\hline
$pp$ & $6.03\,(1\pm0.006)$ & $5.98\,(1\pm0.006)$  \\
$pep$ & $1.47\,(1\pm0.012)$ & $1.44\,(1\pm0.012)$ \\
$hep$ & $8.31\,(1\pm0.30)$ & $8.04\,(1\pm0.30)$  \\
$^7{\rm Be}$ & $4.56\,(1\pm0.07)$ & $5.00 \,(1\pm0.07)$  \\
$^8{\rm B}$ & $4.59\,(1\pm 0.14)$ & $5.58\,(1\pm0.14)$ \\
$^{13}{\rm N}$ & $2.17\,(1\pm0.14)$ & $2.96\,(1\pm0.14)$   \\
$^{15}{\rm O}$ &$1.56\,(1\pm0.15)$ & $2.23\,(1\pm0.15)$  \\
$^{17}{\rm F}$ &$3.40\,(1\pm0.17)$ & $5.52\,(1\pm0.17)$ \\
\hline
\end{tabular}
\end{center}
{\small Note: The neutrino fluxes are given in units of
$10^{10}$ ($pp$),
$10^{9}$ ($^7{\rm Be}$),
$10^{8}$ ($pep$, $^{13}{\rm N}$, $^{15}{\rm O}$), 
$10^{6}$ ($^{8}{\rm B}$, $^{17}{\rm F}$),
and $10^{3}$ ($hep$) $\,{\rm cm}^{-2}{\rm s}^{-1}$.}
\vspace{0.5cm}
\end{table}

The atmospheric neutrinos are produced by cosmic rays, which collide with the atmosphere at its most external regions. In these collisions triggered mostly by the cosmic protons (plus a 5\% of He and some minor contributions of heavier nuclei), pions and, at a much smaller rate, kaons are produced~\cite{Honda1, Honda2, Barr, Agrawal, Gaisser}.
The main sources of the atmospheric neutrinos are the following reactions:
\begin{equation}
\pi^+ \rightarrow \mu^+ + \nu_{\mu},~~~~ \mu^+ \rightarrow e^+ + \nu_e + \bar{\nu}_{\mu}
\label{Eq:pi+}
\end{equation}
\begin{equation}
\pi^- \rightarrow \mu^- + \bar{\nu}_{\mu},~~~~ \mu^- \rightarrow e^- + \bar{\nu}_e + \nu_{\mu}
\label{Eq:pi-}
\end{equation}
As a consequence, the produced fluxes are approximately:
%
\begin{equation}
\Phi(\nu_{\mu} + \bar{\nu}_{\mu} )= 2 \Phi (\nu_e + \bar{\nu}_e)
\label{Eq:mue}
\end{equation}
\begin{equation}
\Phi(\nu_{\mu}) \sim \Phi(\bar{\nu}_{\mu})
\label{Eq:muantimu}
\end{equation}
Moreover, due to the cosmic ray isotropy and the sphericity of the Earth, the {\it up} and {\it down} neutrino fluxes (i.e. 
having the Zenith angle  $\theta$ corresponding to $\cos \theta < 0$ and  $\cos \theta > 0$, respectively) 
are expected to have the same magnitude:
\begin{equation}
\Phi (E_{\nu_x},\cos \theta) \sim \Phi (E_{\nu_x},-\cos \theta)
\label{Eq:updown}
\end{equation}
where $E_{\nu_x}$ is the energy of neutrino with flavor $x$.

The atmospheric neutrino flux can be evaluated with an uncertainty $<$10$\%$ at $1<E<10$\, GeV, while at $E<1$\,GeV the error is larger. 
At $E<10$\, GeV, the relation~(\ref{Eq:mue}) is valid within 2-3\% errors. The accuracy worsen at  larger energies due to kaon production. 
Eq.~(\ref{Eq:muantimu}) is confirmed at 1\% at $E< 1$\,GeV, and has an uncertainty $<$1\% at 1\,GeV.

\begin{figure}[tb]
\begin{center}
\centering{\epsfig{file=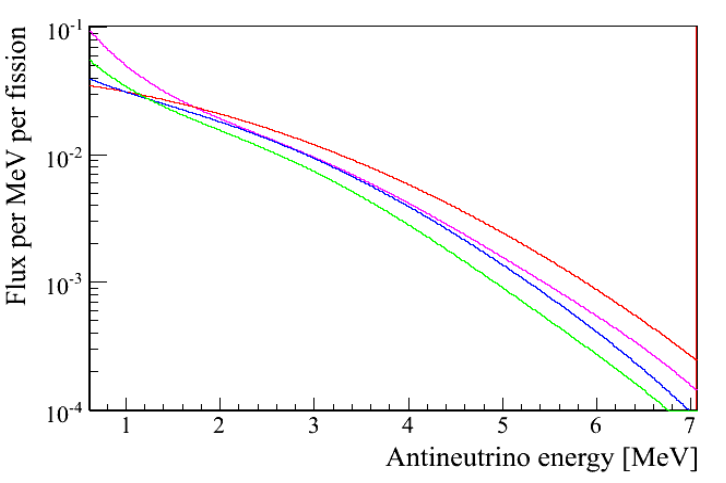,width=0.46\textwidth}}
\caption{Energy spectra of antineutrinos released in the fission of the main isotopes ($^{235}$U (cyan), $^{238}$U (red), $^{239}$Pu (green), and $^{241}$Pu (blue)) used as the fuel in the cores of nuclear power plants.}
\label{Fig:ReactorFlux}
\end{center}
\end{figure}

Nuclear reactors are a source of electron antineutrinos. The energy spectra of antineutrinos released in the fission of the main isotopes used as the fuel in reactor cores ($^{235}$U, $^{238}$U, $^{239}$Pu, and $^{241}$Pu) are shown in Fig.~\ref{Fig:ReactorFlux}. The reactor antineutrino flux is different from site-to-site and strongly depends on the presence of reactors in the neighborhoods. Its evaluation~\cite{Vogel, Huber, Mueller, Huber2} has to take into account different reactor characteristics, some of them time-dependent, as their thermal power and the power fractions of fuel isotopes. The reactor-detector distance has a strong influence on the shape of the oscillated, electron antineutrino energy spectrum. The mean energy of reactor antineutrinos which can be detected by the inverse beta decay reaction given in Eq.~\ref{Eq:InvBeta} is about 4\,MeV.


Supernova explosions represent another possible source of neutrinos and antineutrinos of all flavors. 
The observation of neutrinos produced by a galactic Supernova could bring important information to comprehend the explosion mechanism and to study neutrino propagation 
in the dense Supernova environment.  Supernova neutrino oscillations have a complex and interesting phenomenology;
their potential in neutrino oscillation studies may be affected by the large uncertainties of the astrophysical Supernova models.

Finally, neutrinos and antineutrinos of various energies can be produced by accelerators. At CERN, FNAL, KEK, and Los Alamos Neutron Science Center, neutrino and antineutrino beams are produced for short and long baseline neutrino experiments.

\subsection{The neutrino oscillation study}
\label{subsec:ExpOscil}

The experimental study of the neutrino oscillations can be divided in several phases: the solar neutrino problem, the first evidences of the oscillation phenomenon from atmospheric and solar neutrino experiments, precise measurements of the oscillation parameters $\Delta m^2_{21}$ and $\theta_{12}$ by studying the nuclear-reactor antineutrinos, extension of the oscillation analysis to the low-energy neutrinos and the vacuum regime, confirmation of the oscillation phenomenon via disappearance and appearance experiments with accelerator beams, measurements 
of non-zero $\theta_{13}$, and finally indication of a third $\Delta m^2$ and therefore of a possible sterile neutrino.
 
\subsubsection{Evidences of the neutrino oscillation phenomenon}
\label{subsec:EvidencesOscil}

The road towards the first understanding of the neutrino oscillation phenomenon passed several milestones.

 {\it 1) The Solar Neutrino Problem: an apparent deficit in the solar neutrino flux} 

Pioneering experiments used the radiochemical techniques (see Sec.~\ref{Sec:radiochemical}) applied to the observation of solar neutrinos; they are Homestake~\cite{Homestake}, GALLEX~\cite{Gallex}, and SAGE~\cite{Sage}. These experiments are based upon the charged-current interaction of electron-flavor neutrino on a nucleus.
Because the solar $\nu_e$'s oscillate to different flavors, the experiments, which are sensitive only to electron neutrinos, detect a reduced number of events with respect 
to the expectations based on the Standard Solar Model (SSM). 
This lack of signal has been called "Solar Neutrino Problem" and the possible explanations were either a wrong description of the Sun by SSMs or the phenomenon of the "Neutrino Oscillations", an hypothesis introduced by B.~Pontecorvo in 1957~\cite{pontecorvo}.

The radiochemical experiments measure the integrated flux from the detection 
reaction threshold to the upper limit of the solar neutrino energy spectrum. Homestake measurements start from a threshold of $\sim$0.814\,MeV
and, thus, do not probe the {\em pp-}neutrino component of the solar neutrino flux; it observes $2.56 \pm 0.23$\,SNU\footnote {SNU = Solar Neutrino Unit  equals to the neutrino flux producing $10^{-36}$ captures per target atom per second} to be compared with the SSM expectation of $7.7^{+1.2}_{-1.0}$\,SNU~\cite{Homestake}. A deficit of the solar neutrino signal was confirmed later by GALLEX, which, with a threshold at $\sim$0.23\, MeV, found $83 \pm 19 \rm {(stat)} \pm 8 \rm{(syst)}$\,SNU to be compared to the expected $127\pm 7$ SNU~\cite{Gallex}. SAGE is still running and its results agree with the GALLEX's ones. The reduction is higher in the Homestake data ($\sim$67\%) than in GALLEX ($\sim$35\%). This difference is partly, but not completely, explained by the dependence of $\nu_e$ survival probability from the neutrino energy. A recent hypothesis of the existence of a light sterile neutrino~\cite{ref3.5} could explain the Homestake result.

The Solar Neutrino Problem raised by the radiochemical experiments has been confirmed in 1991 by a real-time experiment based on the water \v{C}erenkov technique, Kamiokande, detecting the $\nu - e$ elastic scattering~\cite{ref3.6}.  The $\nu - e$ elastic scattering cross section $\sigma$ is lower for $\mu, \tau$ flavor neutrino than for the electron flavor neutrino (for the muon flavor $\sigma(\nu_{\mu} - e) \sim 1/7 \sigma(\nu_e - e)$). Kamiokande finds the solar neutrino flux reduced by 40\% with respect to what expected by the SSM. The measured neutrino energy range includes only the $^8$B solar neutrinos, because the threshold in Kamiokande is at $\sim$5.0\,MeV of the recoil-electron energy (which corresponds to $\sim$5.2\,MeV for the neutrino energy).

 {\it 2)   The first experimental evidences of neutrino oscillations}

The experimental evidence for the existence of the neutrino oscillation phenomenon has been provided by three \v{C}erenkov experiments (see Sec.~\ref{subsec:cerenkov}), studying the atmospheric neutrinos with water (Kamiokande and Super-Kamiokande) and the solar neutrinos with heavy water (SNO). Here, we demonstrate
the atmospheric-neutrino measurements on the Super-Kamiokande results, since they are fully compatible with those of Kamiokande, but are based on higher statistics.

Super-Kamiokande observed~\cite{ref3.7} an important discrepancy in the atmospheric-$\nu_{\mu}$ {\it up} and {\it down} fluxes, not observed, on the other hand, in the $\nu_e$ rates. The measured ratio of {\it up} and {\it down} $\nu_{\mu}$ fluxes is well different from 1, contrary to what is expected in the absence of neutrino oscillations, see Eq.~(\ref{Eq:updown}). 
 
\begin{figure*}[tb]
\begin{center}
\centering{\epsfig{file=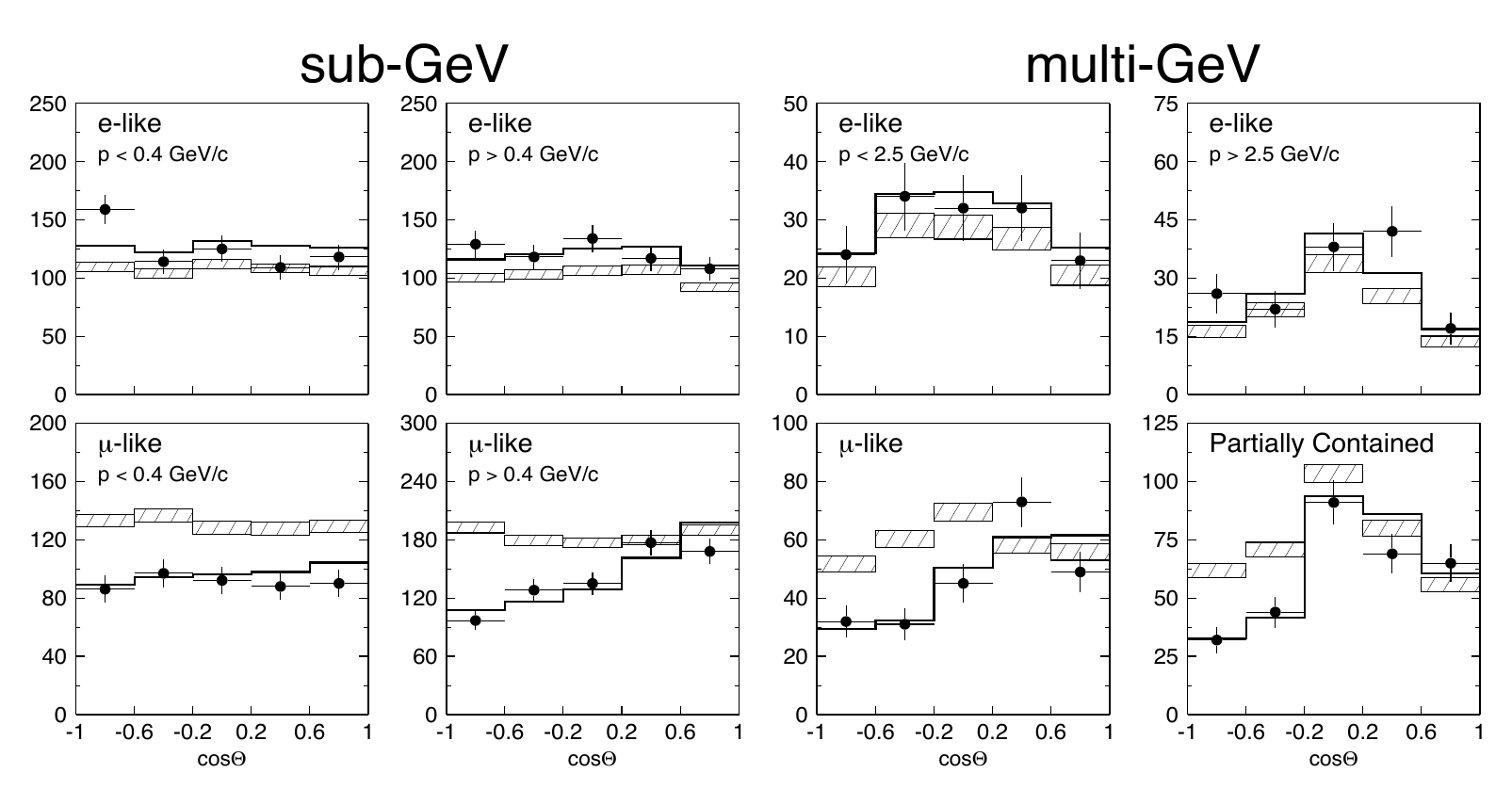,width=0.81\textwidth}}
\caption{Zenith angle distributions of muon and electrons for sub-GeV and multi-GeV data from Super-Kamiokande~\cite{ref3.7}. The hatched regions are the Monte Carlo expectations for no-oscillations. The solid black lines plot the best-fits for $\nu_{\mu} \rightarrow \nu_{\tau}$ oscillations; in the fit the overall flux normalization if left as a free parameter.}
\label{Fig:SuperK}
\end{center}
\end{figure*}

The results are summarized in the Tab.~\ref{tab:superK} and in the Fig.~\ref{Fig:SuperK}. Super-Kamiokande detects the muons produced by $\nu_{\mu}$ and the electrons produced by $\nu_e$: muons and electrons are fast enough to produce a \v{C}erenkov-light cone. The sub-GeV events are fully contained in the detector, while this is not the case for the multi-GeV events.

\begin{table}
\begin{center} 
\caption{The "up/down" asymmetry for muons and electrons observed in Super-Kamiokande~\cite{ref3.7}. Here, "up" refers to incident neutrinos within the zenith angle range $-1 < \cos \theta < -0.2$ and "down" within $0.2 < \cos \theta < 1$.}
\vspace {2mm}
\label{tab:superK}
\begin{tabular}{l|l}
\hline
Source & "up/down" asymmetry  \\ 
\hline \hline
multi-GeV $e$-like & $1.04 \pm 0.03 \pm 0.03$ \\
multi-GeV $\mu$-like & $0.52 \pm 0.05 \pm 0.006$ \\
\hline
sub-GeV $e$-like & $1.09 \pm 0.02 \pm 0.03$ \\
sub-GeV $\mu$-like & $0.65 \pm 0.05 \pm 0.001$ \\
\hline
\end{tabular}
\end{center}
\end{table}     

The observed ``up/down'' asymmetry can be interpreted in terms of  $\nu_{\mu} \rightarrow \nu_{\tau}$ oscillations in vacuum. The best fit values of the oscillation parameters obtained from these data are  $1.9 \times 10^{-3} \rm{eV}^2 < |\Delta m^2_{23} | < 3.0 \times 10^{-3} \rm{eV}^2$ and $\sin 2\theta_{23} > 0.90$ (for Kamiokande data~\cite{ref3.8}, the $|\Delta m^2_{23}|$ allowed region is ranging between $1.3 \times 10^{-2}$ and $2.95 \times 10^{-3} \rm{eV}^2$).

This oscillation effect can be understood on the basis of the oscillation length in vacuum, corresponding to a neutrino energy of $\sim$1\,GeV and to $\Delta m^2_{23} \sim 3 \times 10^{-3} \rm{eV}^2$  (see Sec.~\ref{Sec:theory}):
\begin{equation}
L_0 = 2.48 \rm{[m]} \frac {E \rm{[MeV]}} {\Delta m^2_{23} \rm{[eV^2]}} \sim 1000 ~\rm{km}
\label{Eq:nuen}
\end{equation}

\begin{figure}[tb]
\begin{center}
\centering{\epsfig{file=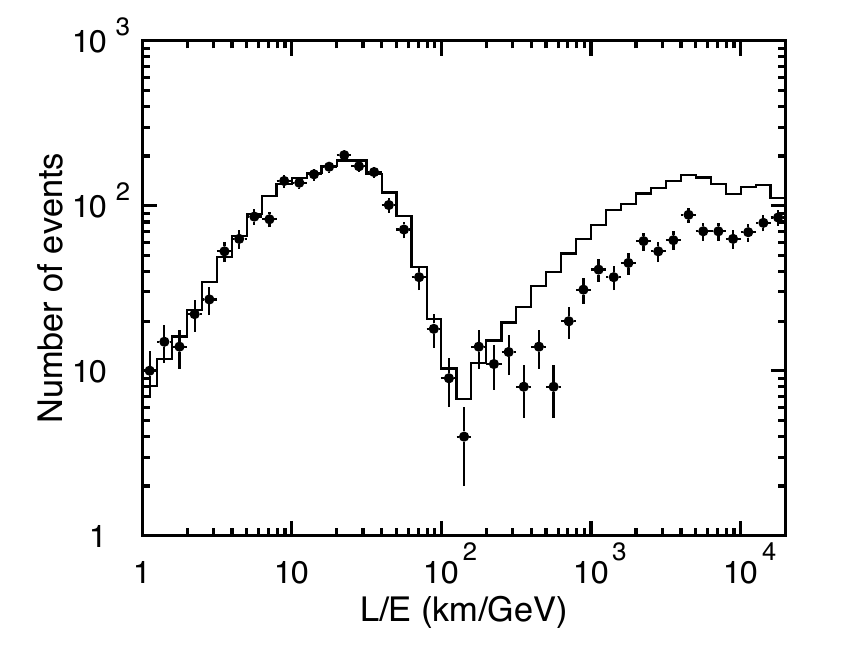,width=0.48\textwidth}}
\caption{Plot of the number of events vs $L/E$ ($L$ is the distance between the neutrino production and the detector and $E$ is the neutrino energy) for the Super-Kamiokande data (points with error bars); the histogram is the result of the Monte Carlo simulation for atmospheric neutrino events without oscillations~\cite{ref3.9}.}
\label{Fig:SuperK2}
\end{center}
\end{figure}

The "down-going" neutrinos  are reaching the detector after $\sim$10\,km from their production, while the "up-going" neutrinos travel on average $\sim$6000\,km. As a consequence, the distance between production and detection for the down-going neutrino is too short to observe a relevant flavor change. In Fig.~\ref{Fig:SuperK2}, the number of events vs $L/E$ ($L$ is the distance between the neutrino production and the detector and $E$ is the neutrino energy) is shown. 

These Super-Kamiokande results have demonstrated for the first time the existence of an oscillation phenomenon on the atmospheric neutrinos. It is essentially model independent and not influenced by any hypotheses assumed in the cosmic ray simulations. 
The results obtained by Kamiokande and Super-Kamiokande have been confirmed by MACRO experiment~\cite{MACRO}  with a smaller statistics.

The SNO detector is a heavy water \v{C}erenkov experiment installed in the Sudbury Inco mine (see Sec.~\ref{subsec:SNO}). 
The use of a deuterium target allowed to study two independent neutrino interactions: Charge Current (CC, Eq.~\ref{Eq:CC}) and Neutral Current (NC, Eq.~\ref{Eq:NC}). 
In addition, the $\nu_x - e$ elastic scattering (Eq.~\ref{Eq:EE}) has been detected. The data have been collected during three phases, characterized by different techniques to capture the neutron emitted in the NC-reactions. The 5\,MeV SNO threshold limits the detectable neutrinos to the $^8$B-component of the solar neutrino flux. Later SNO has repeated the analysis pushing down the threshold to 3.5\,MeV (SNO LETA~\cite{SNO-LETA}). The $^8$B-$\nu$ fluxes measured by SNO are summarized in Tab.~\ref{tab:SNO}.

\begin{table*}
\begin{center} 
\caption{Fluxes of $^8$B solar neutrinos measured by SNO in the three phases of the data taking.}
\vspace {2mm}
\label{tab:SNO}
\begin{tabular}{l|l|l|l}
\hline
Data set & $\Phi_{CC}$  & $\Phi_{ES}$ & $\Phi_{NC}$   \\
              & $\times 10^6 \rm{cm}^{-2} \rm{s}^{-1}$ & $\times 10^6 \rm{cm}^{-2} \rm{s}^{-1}$ & $\times 10^6 \rm{cm}^{-2} \rm{s}^{-1}$\\
\hline \hline
Phase 1 (306 live days)~\cite{SNOPhase1}  & $1.76^{+0.06+0.09}_{-0.05-0.09}$ & $2.39^{+0.24+0.12}_{-0.23-0.12}$ & $5.09^{+0.44+0.46}_{-0.43-0.43}$\\
 Phase 2 (391 live days)~\cite{SNOPhase3} & $1.68^{+0.06+0.08}_{-0.06-0.09}$ & $2.35^{+0.22+0.15}_{-0.22-0.15}$ & $4.94^{+0.21+0.38}_{-0.21-0.34}$\\
 Phase 3 (385 live days)~\cite{SNOPhase3} & $1.68^{+0.05+0.07}_{-0.04-0.08}$ & $1.77^{+0.24+0.09}_{-0.21-0.10}$ & $5.54^{+0.33+0.36}_{-0.31-0.34}$\\
\hline
\end{tabular}
\end{center}
\end{table*}    

The CC interactions are produced only by $\nu_e$, while the NC ones are triggered by all-flavor neutrinos. Therefore, it is clear by comparing the results from Tab.~\ref{tab:SNO} that part of the $\nu_e$ produced in the nuclear reactions in the Sun's core has been transformed to other flavors. The final estimate of the $^8$B neutrino flux from the NC reactions, obtained from a joint analysis of the three phases is $(5.25 \pm 0.16 \rm{(stat)} ^{+0.11}_{-0.13} \rm{(syst)}) \times 10^6$\,cm$^{-2}$ s$^{-1}$~\cite{SNOarXiv}, in a good agreement with the SSM prediction of 
$(4.59 \pm 0.64)\times 10^6$\,cm$^{-2}$ s$^{-1}$.

The SNO results can be interpreted as direct evidence of matter effects on neutrino oscillations,  see Sec.~\ref{Sec:theory}. 
The second term in the r.h.s of Eq.~(\ref{Cmatt}) 
is, in fact, not negligible 
due to the high electron density in the solar interior and to the relatively high energy of the neutrinos 
detected by SNO. Therefore, flavor oscillations are enhanced  due to neutrino propagation through the Sun.
  
Super-Kamiokande also measured the solar neutrinos with a threshold of $\sim$5\,MeV. The $\nu_e - e$ elastic scattering gives a result of $2.32 \pm 0.04 \rm{(stat)} \pm 0.05 \rm{(syst)} \times 10^6$\,cm$^{-2}$s$^{-1}$~\cite{ref3.11} , fully compatible with the SNO measurement.

\begin{figure}[tb]
\begin{center}
\centering{\epsfig{file=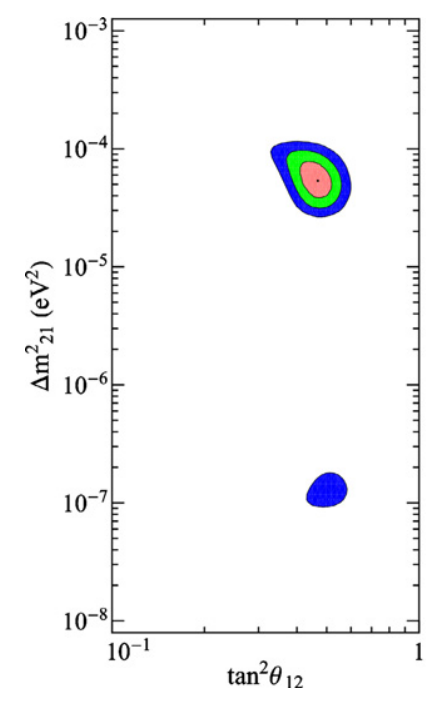,width=0.3\textwidth}}
\caption{Allowed regions (68.27\%, 95.45\%, and 99.73\% C.L.) in the oscillation parameter plane obtained fitting the Homestake + GALLEX + Super-Kamiokande + SNO data in the frame of the MSW model. Note that these results are based on solar neutrinos only, without considering KamLAND antineutrino data. From~\cite{BXDayNight}.}
\label{fig:GlobalSolar}
\end{center}
\end{figure}

In a two neutrino analysis, the allowed regions in the $\Delta m^2_{21}$ versus $\tan ^2 \theta_{12}$ plane, obtained by a global fit of the radiochemical plus \v{C}erenkov experiments are shown in Fig.~\ref{fig:GlobalSolar}. The region at top right is called Large Mixing Angle\footnote{The name of Large Mixing Angle (LMA) has been assigned to that region, in the frame of MSW model, to distinguish from another allowed region, in the $\Delta m^2_{21}$ range of $[10^{-4} - 10^{-5}]$\,eV$^2$, which was called Small Mixing Angle (SMA). The SMA region has been ruled out by more recent solar and KamLAND data.}, the one at bottom right is the LOW region.

 Finally, other experiments were performed to look for neutrino oscillations 
with high-energy artificial neutrino beam at short baseline. 
In particular, the NOMAD~\cite{NOMAD} and CHORUS~\cite{CHORUS} experiments at CERN obtained a null result. 
On the other hand, the LSND experiment (see~\cite{LSND} and Sec.~\ref{subsec:sterile}), which took data at Los Alamos with intermediate energy beam, obtained evidences of neutrino oscillation, even if the result is controversial.

\subsubsection{Checks and refinements of the solar and atmospheric neutrino measurements}
\label{subsec:SolarAtm}

The results obtained by SNO and Super-Kamiokande on solar and atmospheric neutrinos have been confirmed by other experiments. We discuss here: KamLAND, K2K, OPERA, and MINOS. These experiments, despite using using different techniques and different  neutrino sources (reactor $\bar{\nu}_e$, accelerator $\nu_{\mu}$ beams) have $E/L$ ratios, where $E$ is the neutrino energy and $L$ is the neutrino baseline,
that permit to probe the same $\Delta m^2$ region as the "solar" and "atmospheric" data. OPERA is an appearance experiment, KamLAND, K2K, MINOS are disappearance experiments. KamLAND detects the $\bar{\nu}_e$s from the 55 Japanese nuclear reactors; K2K, MINOS, and OPERA study the $\nu_{\mu}$ beam produced by the KEK, Fermilab, and CERN accelerators, respectively (see Sec.~\ref{subsec:OtherTechn}). It has to be recalled that it is possible to consider the $\bar{\nu}$  data in the same framework of the $\nu$ results only if the CPT invariance is assumed.

The techniques used by these experiments are very different, see also Sec.~\ref{Sec:detectors}. K2K is a long baseline experiment: a 1.3\,GeV $\nu_{\mu}$ beam is sent from KEK to Super-Kamiokande, 350\,km apart. The $E/L$ ratio is $\sim$$5 \times 10^{-3}\;{\rm eV^2}$, very close to the atmospheric $\Delta m^2_{23}$ range.

KamLAND studies the ${\overline \nu}_{e}$ conversion by observing the $\overline{\nu}_{e}$ produced by nuclear reactors with an average baseline
$L \sim 200$ km. The E/L ratio falls just in the range of solar neutrinos.

The goal of the OPERA experiment is direct experimental observation of the $\nu_{\tau}$ appearance in the $\nu_{\mu}$ beam via the conversion $\nu_{\mu} \rightarrow \nu_{\tau}$.  The $E/L$ for OPERA is on the average $\sim$$2.4 \times 10^{-2}\,{\rm eV}^2$, partially in the range of atmospheric neutrinos. OPERA is expecting to observe no more than 5 to 8 $\tau$ decays. 

MINOS is a long baseline experiment with a near and a far detector.  The measured energy spectrum in the far detector is compared to the predictions obtained on the basis of the near-detector data. In this way many sources of systematic uncertainty cancel out. The $E/L$ is $\sim$$4 \times 10^{-3}\,{\rm eV}^2$, in the range of atmospheric neutrinos.

\begin{figure}[tb]
\begin{center}
\centering{\epsfig{file=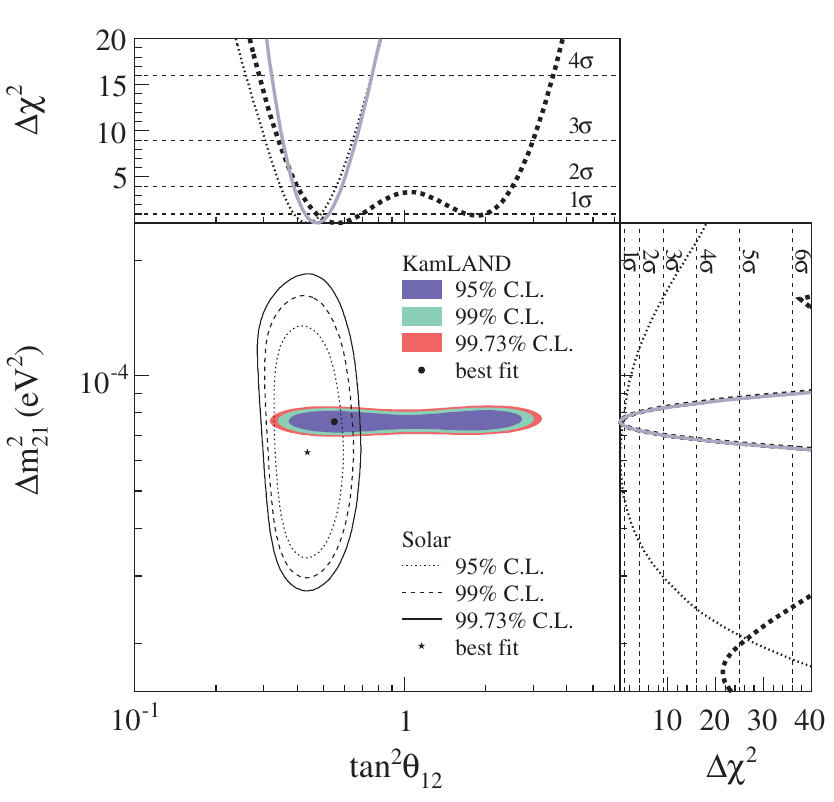,width=0.48\textwidth}}
\caption{Allowed regions~\cite{KamlPrecision} for the oscillation parameters $\Delta m^2_{21}$ and $\tan^2 \theta_{12}$ from solar and KamLAND data. The allowed LMA area is the crossing between the solar and the KamLAND allowed regions.}
\label{Fig:KL_precision}
\end{center}
\end{figure}

The KamLAND experiment had a big impact since it permitted to discriminate among the possible solution of the solar neutrino problem.
Its results~\cite{KL2003}, in fact, ruled out the LOW solution which was still allowed by the solar neutrino data only (see Fig.~\ref{fig:GlobalSolar}) and restricted 
the LMA region. This is demonstrated on Fig.~\ref{Fig:KL_precision} from~\cite{KamlPrecision}.
In the frame of two neutrino approach the electron antineutrino survival probability can be written as  
\begin{equation}
P(\bar{\nu}_e \rightarrow \bar{\nu}_e) = 1 - \sin^2 2\theta_{12} \sin^2 \left ( \frac {\Delta m^2_{21} L} {4E} \right) 
\label{Eq:Pee_antinu}
\end{equation}
In this approximation, the best fit parameters from the KamLAND data only are $\Delta m^2_{21} = (7.58 ^{+0.14}_{-0.13} \rm {(stat)} \pm 0.15 \rm {(syst)} )\times 10^{-5} \rm{eV}^2$ and $\tan^2 \theta_{12} = 0.56^{+0.10}_{-0.07} \rm{(stat)} ^{+0.10}_{-0.06} \rm{(syst)}$~\cite{KamlPrecision}.
Combining with solar neutrino data, the best fit parameters are $\Delta m^2_{21} = 7.59 ^{+0.21}_{-0.21} \times 10^{-5} \rm{eV}^2$ and $\tan^2 \theta_{12} = 0.47^{+0.06}_{-0.05}$ 

K2K~\cite{ref3.13} studied both the $\nu_{\mu}$ disappearance and a possible appearance of $\nu_e$: the first to check the oscillation parameters in the atmospheric $\Delta m^2_{23}$ region, the second to study the $\theta_{13}$ mixing angle. From the study of $\nu_{\mu}$ disappearance, K2K confirmed the results of Super-Kamiokande with atmospheric neutrinos and obtained fully consistent values of the oscillation parameters $1.5 \times 10^{-3} < |\Delta m^2_{23} | < 3.4 \times 10^{-3} \rm {eV}^2$ and $\sin^2 2\theta_{23} > 0.92$. In the search of possible conversion of $\nu_{\mu} \rightarrow \nu_e$ K2K succeeded to extract only an upper limit for $\theta_{13}$.

\begin{figure}[tb]
\begin{center}
\centering{\epsfig{file=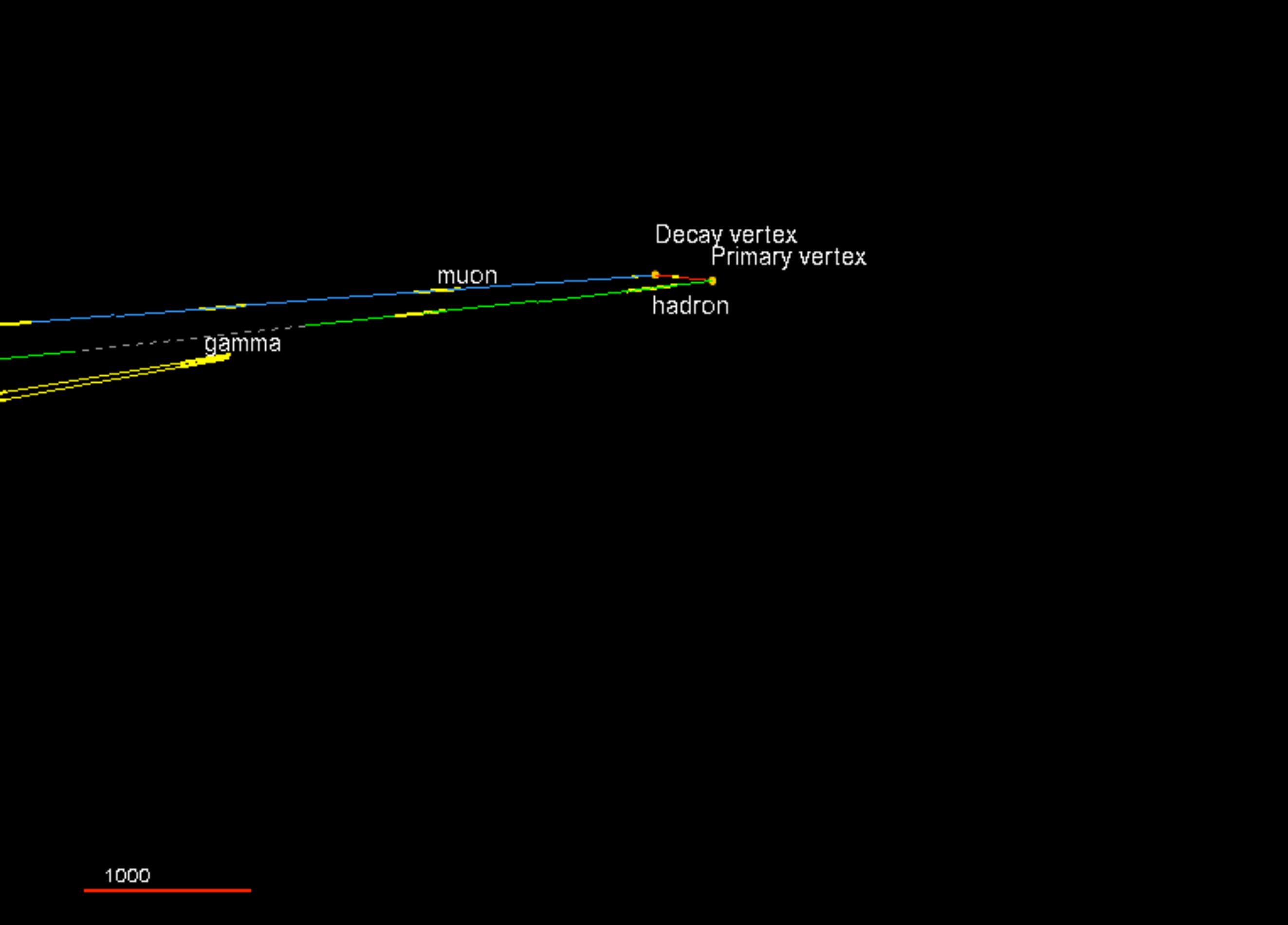,width=0.4\textwidth}}
\caption{One of the OPERA $\nu_{\tau}$ candidates~\cite{opera}. In the primary vertex the $\nu_{\tau}$ interacts producing a $\tau$, which decays after 376\,$\mu$s into a muon plus neutrinos. In addition, the conversion of a $\gamma$, produced in the primary vertex, is visible.}
\label{Fig:tau}
\end{center}
\end{figure}

OPERA has completed its data taking because the neutrino beam has been switched off at CERN, where the activity for the upgrading of the LHC energy has started. The data have been collected during 797 beam days and up to day the 2008-2009 data have been already analyzed, while the analysis of the 2010-1012 events is ongoing.  OPERA has observed up to now three $\nu_{\tau}$ candidates~\cite{ran21}, \cite{opera}, one of them is shown in Fig.~\ref{Fig:tau}. The probability to have observed $\nu_{\tau}$ appearance from $\nu_{\mu}$,
corresponds to $\sim$3.5$\sigma$ C.L.

MINOS measures the muon flavor disappearance with a $\nu_{\mu}$~\cite{ran20} and a $\bar{\nu}_{\mu}$~\cite{ref3.16} enhanced beam. The detected charged current interactions, $\nu_{\mu} (\bar{\nu}_{\mu}) + N_1 \rightarrow \mu^- (\mu^+) + N_2$, give the opportunities to reject the $\bar{\nu}_{\mu} (\nu_{\mu})$ background in the $\nu_{\mu} (\bar{\nu}_{\mu})$ beam, respectively, by analyzing the curvature of the reconstructed muon track in the magnetic calorimeter. 
The statistics of the neutrino data is more than an order of magnitude higher than the antineutrino data. The no-oscillation hypothesis is disfavored, in the case of $\bar{\nu}_{\mu}$ , at 6.3$\sigma$ C.L. 
By fitting the data in the context of two neutrino oscillations and using independent mass and mixing parameters for the neutrino and antineutrino case, the best fit results are:
%
\begin{itemize}
\item{for $\nu_{\mu}$:
$ |\Delta m^2_{23}| = 2.32^{+0.13}_{-0.08} \times 10^{-3} \rm{eV}^2$ and $\sin^2(2\theta_{23}) > 0.90$ (90\% C.L.)}
\item{for $\bar{\nu}_{\mu}$:  $ |\Delta \bar{m}^2_{23} | = (3.36^{+0.46}_{-0.40} \rm{(stat)}  \pm 0.06 \rm{(syst)}) \times 10^{-3} \rm{eV}^2$ and $\sin^2(2\bar{\theta}_{23}) = 0.86^{+0.11}_{-0.12} \rm{(stat)} \pm 0.01 \rm{(syst)}$ . }
\end{itemize}

MINOS has analyzed also the neutral current interactions in order to investigate a possible active to sterile neutrino mixing. Because the neutral current interaction cross sections are the same for the three flavors, an observation of a neutral current event depletion between the near and far detectors could be due to the existence of a fourth sterile neutrino. MINOS  found that the fraction of $\nu_{\mu}$, which may show a transition to a sterile neutrino, is $<$22\% (90\% C.L.)~\cite{ref3.16}.

\subsubsection{Low-energy solar neutrinos and the oscillation in vacuum}
\label{Subsec:lenu}

The analysis of the solar neutrinos by the \v{C}erenkov experiments has been carried out with an energy threshold at $\sim$5\,MeV of detectable energy (for the incident neutrinos this threshold is slightly higher and depends on the reaction). Only SNO tried to push down the threshold to $\sim$3.5\,MeV~\cite{SNO-LETA}, but the obtained results have large uncertainties. Thus, the spectrum analyzed by \v{C}erenkov technique corresponds to $\sim$0.01\% of the total solar spectrum and concerns the matter-enhanced neutrino oscillation.

The reason of such an high-energy threshold is the natural radioactivity, present in the environment and in the materials used to build the detectors. Two main families are present: $^{232}$Th and $^{238}$U; the highest-$Q$ (2.8\,MeV) member is $^{208}$Tl from the $^{232}$Th decay chain. Therefore, in order to safely exclude the natural radioactivity (taking into account also the energy resolution) from the data, a high energy threshold had to be applied.

\begin{figure}[tb]
\begin{center}
\vspace{-7 mm}
\centering{\epsfig{file=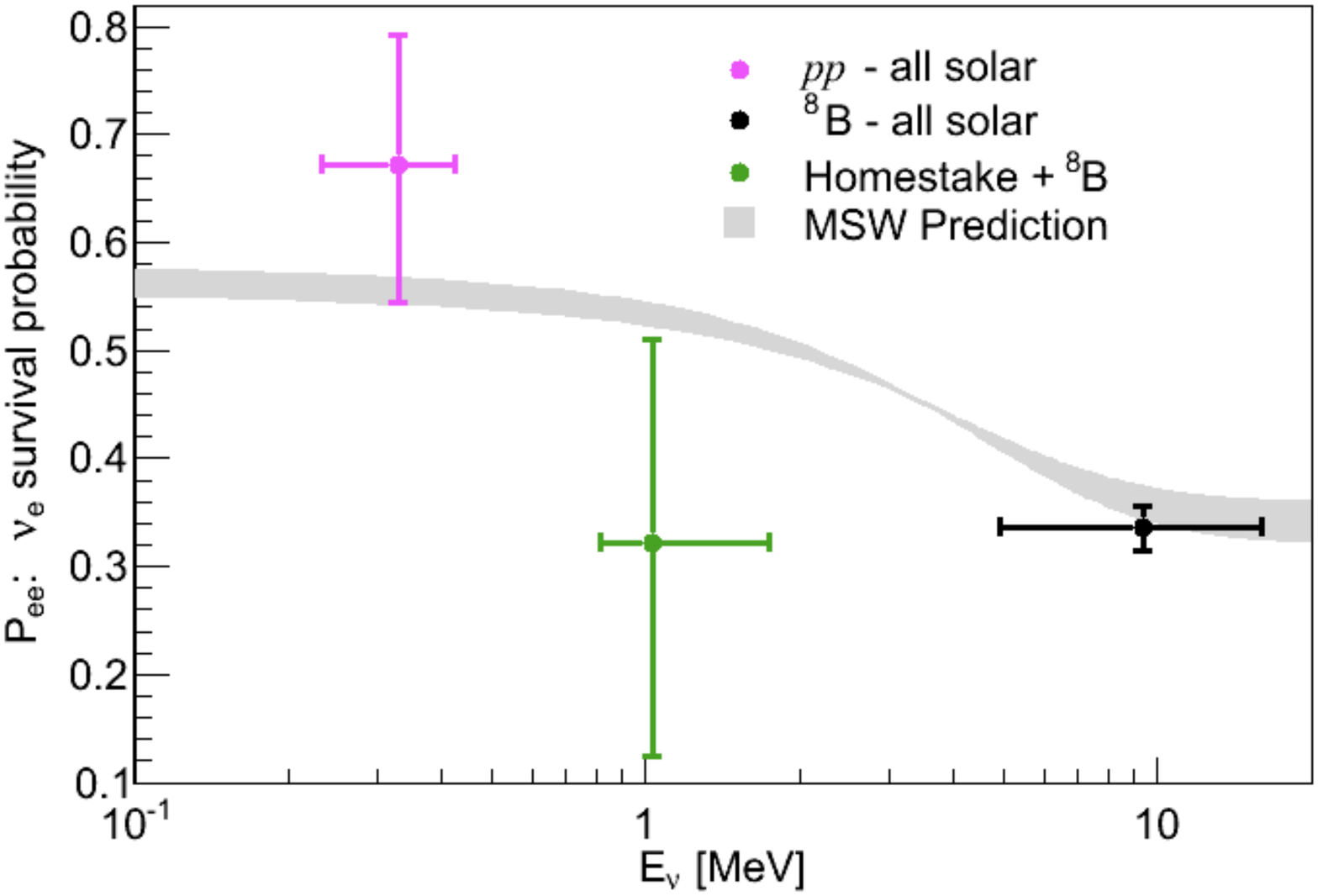,angle=270,width=0.50\textwidth}}
\caption{Solar $\nu_e$ survival probability~\cite{Bellini2012} as a function of neutrino energy. The data points are from the radiochemical and the water \v{C}erenkov experiments. The grey band is the prediction of the LMA solution in the frame of MSW model. }
\label{Fig:Peenobx}
\end{center}
\end{figure}

The understanding of the solar-neutrino oscillations which has been reached thanks to the radiochemical and the \v{C}erenkov experiments can be demonstrated on a plot of the $\nu_e$ survival probability as a function of energy, shown in Fig.~\ref{Fig:Peenobx}. The grey band is the prediction of the LMA solution in the framework of MSW model calculated using the best fit values of the oscillation parameters from a global fit of solar + KamLAND data: the thickness of the band takes into account the uncertainties of the oscillation parameters. 
The two plateaus, at the low and at the high energy regions, correspond to the oscillation in {\it vacuum} and in {\it matter} respectively, as it is explained in Sec.~\ref{Sec:theory}. The intermediate region is called the {\it transition region}. The black experimental point in the high energy region is obtained from a proper average of the SNO + Super-Kamiokande data; the other two data points are from the radiochemical experiments.
As it can be seen, the LMA solution in the frame of MSW model is validated with good accuracy only for the matter-enhanced oscillation regime, while checks of increased precision are needed for the vacuum regime and the transition region.

The study of the low-energy neutrinos, say below 2\,MeV, needs a strong effort and the development of new techniques to strongly suppress the natural radioactivity background, to purify the active part of the detector and the shielding materials. Only one experiment succeeded to solve these problems: Borexino, installed at the Gran Sasso underground laboratory, see Sec.~\ref{subsec:BX}.

During the phase 1 (May 2007- July 2010), Borexino succeeded to measure the $^7$Be~\cite{ref3.19}, $pep$~\cite{ref3.20}, $^8$B (with the lower threshold down to 3.2\,MeV)~\cite{ref3.21} neutrino fluxes and to obtain an upper limit for the CNO neutrino flux~\cite{ref3.20}. 
In Tab.~\ref{tab:BXresults} we summarize the measured rates, while in Tab.~\ref{tab:SolarFlux} we compare the corresponding fluxes, 
calculated by using the best fit oscillation parameters, with the predictions obtained by the "low'"  (AGSS09 \cite{AGSS09})
and the "high'" metallicity (GS98 \cite{GS98}) SSMs. The fluxes measured by Borexino and by the \v{C}erenkov experiments are in good agreement with the SSM predictions, 
but are unable to discriminate between high and low metallicity, mainly due to the experimental errors and 
uncertainties in solar model construction. 

\begin{table}
\begin{center} 
\caption{Solar-neutrino rates as measured by Borexino.}
\vspace {2mm}
\label{tab:BXresults}
\begin{tabular}{l|l}
\hline
Reaction  & Rate \\
 in the Sun & [counts / day / 100 tons] \\ \hline \hline
$^7$Be & $46 \pm 1. 59 \rm{(stat)} ^{+1.6}_{-1.5} \rm{(syst)}$ \\
$pep$ & $3.13 \pm 0.55 \rm{(stat)} \pm 0.23 \rm{(syst)}$ \\
CNO & $<$1.4 \\
$^8$B & $0.22 \pm 0.04 \rm{(stat)} \pm 0.01 \rm{(syst)}$ \\
\hline
\end{tabular}
\end{center}
\end{table}

\begin{table*}
\begin{center} 
\caption{Comparison between the SSM predictions for the solar neutrino fluxes with high (GS98) and low metallicity (AGSS09) and the experimental results.
The CNO flux corresponds to the sum of the $^{13}{\rm N}$, $^{15}{\rm O}$ and $^{17}{\rm F}$ solar neutrino components.}
\vspace {2mm}
\label{tab:SolarFlux}
\begin{tabular}{l|l|l|l}
\hline
          & GS98 & AGSS09 & Experimental result \\
\hline \hline
$pep$  & $1.44 \pm 0.017$ & $1.47  \pm 0.018$ & $1.6 \pm 0.3$  {\small\em (Borexino)} \\
$^7$Be  & $ 5.00 \pm 0.35 $ & $4.56 \pm 0.32$ & $4.87 \pm 0.24$ {\small\em (Borexino)}  \\
 CNO   &   $5.25 \pm 0.79$ & $3.76 \pm 0.56$ &  $<$7.7 (95\% C.L.) {\small\em (Borexino)}\\
$^8$B  & $5.58 \pm 0.78$ & $4.59 \pm 0.64$ & $5.2 \pm 0.3$ {\small\em (SNO + SK + Borexino + KamLAND)} \\
          &   & & $5.25 \pm 0.16 \rm{(stat)} ^{+0.11}_{-0.013} \rm{(syst)} $ {\small\em (SNO-LETA)}  \\
\hline
\end{tabular}
\vspace{-0.3cm}
\end{center}
{\small Note: The neutrino fluxes are given in units of
$10^{9}$ ($^7{\rm Be}$),
$10^{8}$ ($pep$, CNO), 
and $10^{6}$ ($^{8}{\rm B}$) $\,{\rm cm}^{-2}{\rm s}^{-1}$.}
\end{table*}

\begin{figure}[tb]
\begin{center}
\centering{\epsfig{file=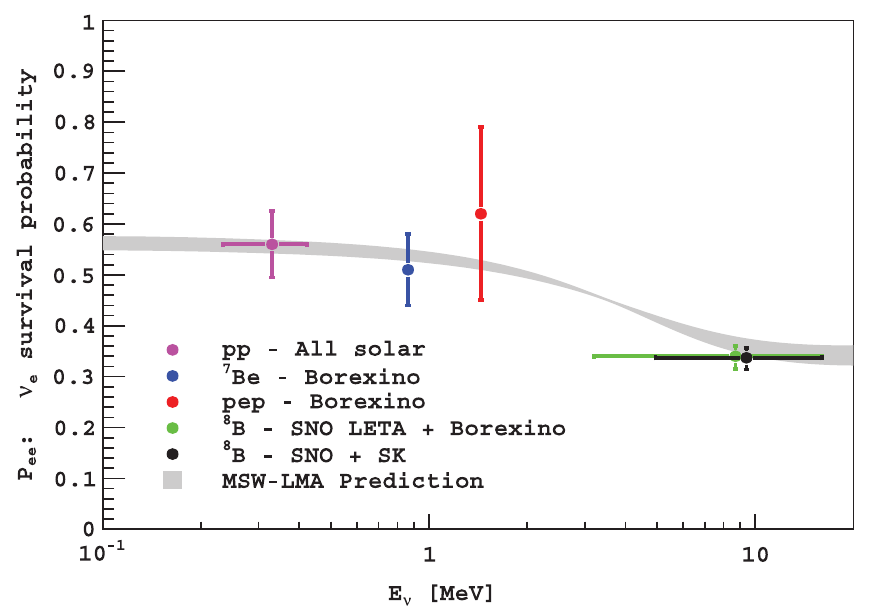,width=0.47\textwidth}}
\caption{Solar $\nu_e$ survival probability~\cite{ref3.20} as a function of neutrino energy including all solar (with Borexino) experimental results. The grey band is the prevision of the LMA solution in the frame of MSW model. }
\label{Fig:Peebx}
\end{center}
\end{figure}

The impact of Borexino results on the determination of the solar $\nu_e$ survival probability, is shown in Fig.~\ref{Fig:Peebx}. In addition of the new measurements of $^7$Be and $pep$ neutrino fluxes, the constraints on the $pp$ flux is much improved following the $^7$Be-$\nu$ flux knowledge. Thus, the plateau corresponding to the vacuum regime is validated and through the $^7$Be and the $pep$ neutrinos (this last even if with large errors) a check of the transition region has started. Finally, the $^8$B analysis is extended to lower energies.

A further result of the Borexino phase 1 is the measurement~\cite{BXDayNight} of the day/night asymmetry $A_{DN}$ defined as:
\begin{equation}
A_{DN} = 2 \frac{R_N - R_D}{R_N + R_D} 
\label{Eq:Be7DN}
\end{equation}
for the rate of the $^7$Be neutrinos, $R_N$ and $R_D$ being the corresponding rates during the day and night. It has been found null at $\sim$1\% error:  $A_{DN} = 0.001 \pm 0.012 \rm{(stat)} \pm 0.007 \rm{(syst)}$. In a global fit with solar results only, this result is able to rule out the LOW region (see Fig.~\ref{fig:GlobalSolar} for the situation before Borexino), at 6.2$\sigma$ C.L.,  and thus, 
without assuming the CPT invariance which is instead implicitly assumed when KamLAND anti-neutrino measurements are taken into account (see Fig.~\ref{Fig:KL_precision}).
                            
All these results will be improved by Borexino during the phase 2 because of the further radio-contaminants reduction and the effort to leave untouched the detector during at least three years. The phase 2 goals are: 1) the reduction of the uncertainties on the $^7$Be and $pep$ neutrino fluxes; 2) the direct measurement of the $pp$ neutrino flux; 3) either an improvement of the upper limit or a direct measurement of the CNO neutrino flux; 4) a measurement of the solar neutrino flux seasonal variation. The physics impact of these goals concern the determination of the shape of the vacuum-to-matter transition region of the solar-$\nu_e$ survival probability, which could be influenced by the Non-Standard neutrino Interactions (NSI)~\cite{ref3.31} or by the existence of an ultra-light sterile neutrino~\cite{ref3.5}, \cite{ref3.32}. In the same direction goes the effort to measure with reduced errors the $^8$B-neutrino flux allowing an experimental point in the range 3-5\,MeV.

\subsubsection{A third $\Delta m^2$ range around 1\,eV$^2$?}
\label{subsec:sterile}          

The possibility of a $\Delta m^2$ with a higher value than the solar and atmospheric ones has been considered in connection with the LSND~\cite{LSND} and MiniBooNE~\cite{ran19} data. Both these experiments are short baseline projects and the short distance between the neutrino source and the detector makes them impossible to observe oscillations driven by "atmospheric'' mass difference $\Delta m^2_{23}$ or by the "solar'' mass difference $\Delta m^2_{21}$.

LSND has taken data during the periods 1993-1995 and 1996-1998 at Los Alamos Neutron Science Center with a $\bar{\nu}_{\mu}$ beam produced by $\pi^+$ and $\mu^+$ decays, most of which at rest, and a $\nu_{\mu}$ beam.
The energy spectrum of both neutrinos and antineutrinos is broad, the maximum of the spectrum is at $\sim$60\,MeV for $\nu_{\mu}$ and at $\sim$45\,MeV for $\bar{\nu}_{\mu}$. The distance between the beam stop and the detector is 30\,m. Therefore $E/L$ is spanning around $1\,{\rm eV}^2$.

Strong effort has been devoted to reject the electron induced reactions, but in any case the electron background in the beam is very limited. In addition, the energy range is restricted within $20<E<200$\, MeV to study the oscillation $\bar{\nu}_{\mu} \rightarrow \bar{\nu}_e$ and within $60<E<200$\, MeV for $\nu_{\mu} \rightarrow \nu_e$, in order to suppress various background sources. The $\bar{\nu}_e$ are identified, as usual,  through the inverse beta decay, see Eq.~\ref{Eq:InvBeta}.

In the runs with $\bar{\nu}_{\mu}$ LSND founds an excess of $117.9 \pm 22.4$ inverse beta-decay interactions. Subtracting from this sample $19.5 \pm 3.9$ events due to $\mu^-$ in the beam  and $20.5 \pm 4.6$ events due to $\bar{\nu}_{\mu} + p \rightarrow \mu^+ + n$, a final sample of $87.9 \pm 22.4 \pm 6.0$ events remain. The maximum likelihood best fit for $\Delta m^2$ falls in the range 0.2 - 2.0\,eV$^2$~\cite{LSND}. LSND does not find any effect with the $\nu_{\mu}$ beam.

A check of this result has been carried out by the experiment KARMEN with a detector and an energy similar to LSND, but with a distance from the neutrino source of 17.5\,m. KARMEN, which is installed at the ISIS facility of the Rutherford Appleton Laboratory, did not find any evidence of $\bar{\nu}_{\mu} \rightarrow \bar{\nu}_e$ oscillation, but its sensitivity is lower than LSND. In any case, KARMEN succeeded to rule out a large part  (but not all) of the $(\Delta m^2, \sin^2 2\theta)$ region~\cite{ref3.25} allowed by LSND.

More recently a new collaboration, which some LSND members have joined, designed and carried out the experiment MiniBooNE~\cite{ran19} at Fermilab, just to  check the LSND results. They use a $\nu_{\mu}$ beam, peaking at 600\,MeV of energy, and a $\bar{\nu}_{\mu}$ beam at 400\,MeV, while the detector is located at 541\,m from the beam target, thus with an $E/L$ in the same range of LSND. The signature of a possible transition $\nu_{\mu} \rightarrow \nu_e$ and $\bar{\nu}_{\mu} \rightarrow \bar{\nu}_e$ is an excess of charged-current quasi-elastic events induced by $\nu_e$ and $\bar{\nu}_e$. 

MiniBooNE finds 480 events passing the $\bar{\nu}_e$ event selections at the neutrino energy range 200 - 3000\,MeV, to be compared with the background expectation of $399.6 \pm 20.0 \rm{(stat)} \pm 20.3 \rm{(syst)}$. Then the excess is $78.4 \pm 28.5$ $(2.8\sigma)$~\cite{ran19}. 

But, contrary to LSND, MimiBooNE observes also an $\nu_e$ excess of $128.8 \pm 43.4$ in the $\nu_{\mu}$ beam, in the range 200 - 475\,MeV , while no excess has been observed above 475 MeV. The best fits for the oscillation parameters for the antineutrino mode, quoted by MiniBooNE in various papers, using a two neutrino approach, vary from some hundredths to some tenths of eV$^2$ for $\Delta m^2$, and from some tenths to few hundredths for  $\sin^2 2 \theta$.

\begin{figure}[!thb]
\begin{center}
\centering{\epsfig{file=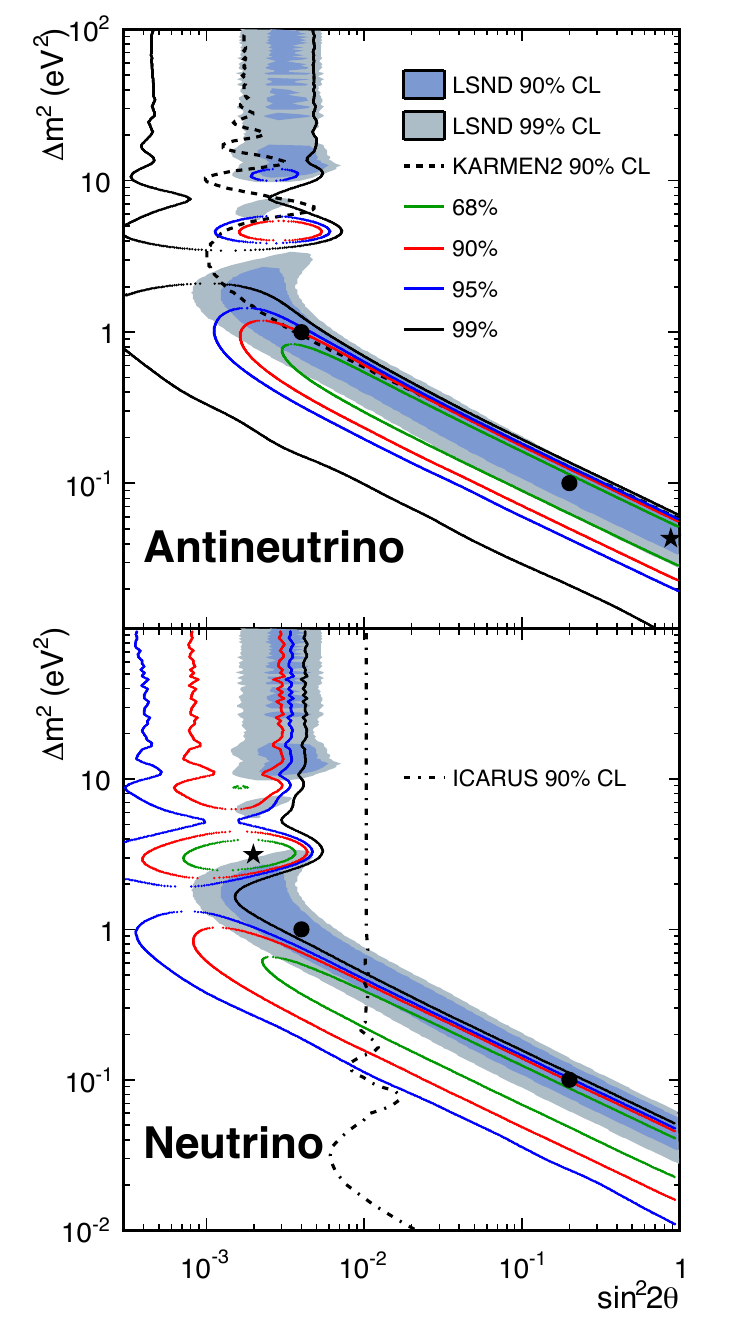,width=0.48\textwidth}}
\caption{The allowed regions for $\bar{\nu}_{\mu} \rightarrow \bar{\nu}_e$  and  $\nu_{\mu} \rightarrow \nu_e$ from the MiniBooNE data~\cite{ran19} are shown. In the case of antineutrinos, they are compared with the LSND allowed region and with the KARMEN exclusion plot (to the right of the dashed line). }
\label{Fig:sterile}
\end{center}
\end{figure}

These oscillation parameters can only be allowed by assuming a fourth sterile (anti)neutrino, which does not interact, but mix with the active (anti)neutrinos. The present status of the art on this topic is summarized in Fig.~\ref{Fig:sterile}, where the allowed and excluded regions in the $(\Delta m^2,\sin^2 2 \theta$) plane by LSND, MiniBooNE, and KARMEN are shown. 

\subsubsection{The mixing angle $\theta _{13}$}
\label{subsec:theta13}

In the framework of three neutrino oscillations, 
if we neglect for simplicity CP-violating effects (i.e. we set the CP phase $\delta =0$, see Sec.~\ref{Sec:theory}), 
there are five parameters, two squared mass differences $\Delta m^2 _{21}, \Delta m^2 _{23}$ and
three angles $\theta_{12}, \theta_{23}, \theta_{13}$. 
These parameters have been measured by the experiments described in the previous paragraphs, except the angle $\theta_{13}$ for which, however, 
an upper limit was obtained.  Therefore, in order to measure $\theta_{13}$, high statistics and very low systematic errors are needed. One important improvement can be reached using in the same experiment two detectors, one close to the neutrino source ({\it near detector}) and another one ({\it far detector}) at a distance of few kilometers (in case of low energy neutrinos) or some hundreds of kilometers (in case of high energy neutrinos). In a set-up with both near and far detectors many sources of systematic cancel out.

Three experiments, which succeeded to achieve a $\theta_{13}$ measurement, are assembled with near and far detectors: Daya Bay, RENO, and T2K. Daya Bay~\cite{ref3.27} consists of 6 detectors, exposed at 6 nuclear reactors at 26 different distances ranging from 362 to 1925 m. RENO~\cite{RENO} also detects the reactor $\bar{\nu}_e$ with the two detectors at 294\,m and 1393\,m from the center of a six reactor array. In T2K~\cite{T2K2012}, which is a second generation follow-up to the K2K, a $\nu_{\mu}$ beam is sent off-axis (2-3$^{\circ}$, in order to reduce the beam energy below 1\,GeV and then to have a proper $E/L$) to Super-Kamiokande detector, 295\,km away, with a near detector 280\,m from the beam target. A further experiment, Double Chooz~\cite{ran17}, has taken data with only a far detector (1050\,m away) exposed to 2 reactors.

All detectors make use of the liquid scintillator technique, with the only exception of the magnetic tracking system of the T2K near detector. Daya Bay, RENO, and Double Chooz are disappearance experiments, while T2K is looking for appearance of $\nu_e$ in a $\nu_{\mu}$ beam. All detectors are installed under some hundreds meters of water equivalent overburden.

The experiments exposed to reactor antineutrinos determine the $\overline{\nu}_e$ survival probability
that in the context of three neutrino oscillations is given by Eq.~(\ref{Pee3nuVac}).
%
%
Neglecting the term proportional to $S_{12}$, this can be written as:
\begin{equation}
P\left(\overline{\nu}_e\to\overline{\nu}_e\right) =  1  - \sin^2 2\theta_{13} \sin^2\left( \frac {1.267 \Delta m^2_{23} L \rm{[m]}} {E \rm{[MeV]} } \right)
\label{Eq:Pee3fapprox}
\end{equation}
The $\theta_{13}$ values that are obtained from the experimental data are summarized in Tab.~\ref{tab:theta13}.

\begin{table*}
\begin{center} 
\caption{Results of the $\theta_{13}$ measurements from the three experiments detecting the $\bar{\nu}_e$ from nuclear reactors.}
\vspace {2mm}
\label{tab:theta13}
\begin{tabular}{l|l|l}
\hline
Experiment & N(measured)/N(expected) events  & $\sin ^2 \theta_{13}$ \\
\hline \hline
Daya Bay & $0.944 \pm 0.007 \rm{(stat)} \pm 0.003 \rm{(syst)} $ & $0.089 \pm 0.010 \rm{(stat)} \pm 0.005 \rm{(syst)}$ \\ 
RENO & $0.920 \pm 0.009 \rm{(stat)} \pm 0.014 \rm{(syst)} $ & $0.113 \pm 0.013 \rm{(stat)} \pm 0.019 \rm{(syst)}$  \\ 
Double Chooz &  & $0.097 \pm 0.034 \rm{(stat)} \pm 0.034 \rm{(syst)}$ \\ 
\hline
\end{tabular}
\end{center}
\end{table*}    

T2K observed 21 $\nu_e$ candidates. The oscillation probability $P\left(\nu_\mu\to \nu_e\right) $ 
can be deduced from Eq.~(\ref{Pemu3nuVac}) obtaining approximately:
\begin{eqnarray}
\nonumber
P\left(\nu_\mu\to\nu_e\right) &=&\\
& & 
\hspace{-3cm}
 \sin^2\theta_{23} \sin^2 2\theta_{13} \sin^2\left( \frac {1.267 \Delta m^2_{23} L \rm{[m]}} {E \rm{[MeV]} } \right)
\end{eqnarray}
From the experimental results, one obtains $\sin^2 \theta_{13} = 0.104 \pm 0.060 \rm{(stat)} \pm 0.045 \rm{(syst)}$.

\subsubsection{A global analysis}
\label{subsec:global}

The most powerful tool to extract information on neutrino parameters is provided by global analysis in which
all experimental neutrino data are fitted simultaneously in the context of three neutrino oscillations.
In fact, this approach provided the evidence (and the magnitude) for  non zero $\theta_{13}$ before 
the direct experimental measurements were performed (see Fogli et al.~\cite{FogliGLobal}). 
We present here the status of global fits by reviewing the results of~\cite{global} 
(for a similar approach see also \cite{FogliLast}).
In this global analysis the following experimental information are included: 
the atmospheric neutrino data (Super-Kamiokande~\cite{SuperK} phase 1-4), the long baseline accelerator experiments (K2K~\cite{K2K2006}, 
disappearance and appearance data of MINOS~\cite{minos2012} and T2K~\cite{T2K2012}), 
the reactor experiments (CHOOZ~\cite{CHOOZ1999}, Palo Verde~\cite{PaloVerde}, Double-Chooz~\cite{DoubleChooz2012}, Daya Bay~\cite{DayaBayTalk}, RENO~\cite{RENO}, KamLAND~\cite{Kamland2011}), solar radiochemical  (Homestake~\cite{Homestake}, GALLEX~\cite{Gallex2010}, SAGE~\cite{Sage}) and solar real-time experiments (Super-Kamiokande~\cite{SuperK2006}, SNO~\cite{SNOarXiv}, Borexino~\cite{ref3.19, ref3.21}). 
The best fit oscillation parameters and the corresponding allowed regions 
are shown in Tab.~\ref{tab:global}.  The two columns correspond to two different
assumptions for the reactor neutrinos. In the first, the fit is carried out allowing for a free normalization 
of the reactor neutrino fluxes and including in the analysis the results of the reactor experiments 
with very short baseline $L \leq 100$\,m (RSBL), as Bugey, ROVNO, Krasnoyarsk, ILL, G\"osgen, and SRP.
The results presented in the second column are, instead, obtained by assuming the recent reactor fluxes 
calculated by Huber et al.~\cite{Huber2} and leaving out RSBL data. These two choices permit
to explore the relevance of the reactor neutrino flux normalization in the extraction of neutrino parameters, 
taking also into account that recent calculations of the reactor fluxes as in Mueller et al.~\cite{Mueller} and Huber et al. \cite{Huber2} 
give a deficit of about 3\% with respect to the previous flux evaluations~\cite{Vogel, Huber}.
The adopted choice slightly affects the $\theta_{13}$ determination, partly due to the 
tension between the new fluxes and the RSBL data (the reactor anomaly): an increase of statistics of the 
Daya Bay and Reno will reduce the uncertainty connected with this choice.

For $\sin ^2 \theta_{23}$, two disconnected 1$\sigma$ intervals are shown, the first-one corresponds to the absolute minimum 
while the second-one represents a secondary local minimum. The global analysis prefers a non-maximal value of $\theta_{23}$; 
this result is mostly driven by the MINOS $\nu_{\mu}$ disappearance results.

We also see that there is a marginal sensitivity to $\delta$ provided by the combination of the MINOS disappearance results, 
the $\nu_{\mu} \rightarrow  \nu_e$ data from long baseline experiments, and the atmospheric data.

Finally, while the sign of the mass difference $\Delta m^2_{21}$ is determined from
matter effects in solar neutrino oscillation, the sign of the ``atmospheric'' mass splitting $\Delta m^2_{31}\simeq \Delta m^2_{32}$
 is not known.  Correspondingly, we have two different options, the normal hierarchy (NH) for  $\Delta m^2_{31} > 0$ and 
the  inverted hierarchy (IH) for  $\Delta m^2_{31} < 0$,  
that provide a fit of very similar quality to the available data set.

\begin{table*}
\begin{center} 
\caption{Neutrino oscillation parameters from a global fit from~\cite{global}. Details in text.}
\vspace {2mm}
\label{tab:global}
\begin{tabular}{l|l|l|l}
\hline 
Parameter & Unit &  (1) & (2)  \\
\hline \hline
$\sin^2 \theta_{12}$ & - & $0.302 ^{+0.013}_{-0.012}$ & $0.311  ^{+0.013}_{-0.013}$ \\
$\theta_{12}$&  $[^{\circ}]$ & $33.36  ^{+0.81}_{-0.78}$ & $33.87 ^{+0.82}_{-0.80}$ \\
$\sin^2 \theta_{23}$ & - & $0.413 ^{+0.037}_{-0.025}$ ${\scriptstyle\bigoplus}$ $0.594^{+0.021}_{-0.022}$ & $0.416  ^{+0.036}_{-0.029}$ ${\scriptstyle\bigoplus}$ $0.600^{+0.019}_{-0.026}$ \\
$\theta_{23}$ & $[^{\circ}]$ & $40.0 ^{+2.1}_{-0.1.5}$ ${\scriptstyle\bigoplus}$ $50.4^{+1.3}_{-1.3}$ & $40.1 ^{+2.1}_{-1.6}$ ${\scriptstyle\bigoplus}$ $50.7^{+1.2}_{-1.5}$ \\
$\sin^2 \theta_{13}$ & - & $0.0227  ^{+0.0023}_{-0.0024}$ & $0.0255  ^{+0.0024}_{-0.0024}$ \\
$\theta_{13}$ & $[^{\circ}]$ & $8.66 ^{+0.44}_{-0.46}$ & $9.2  ^{+0.41}_{-0.45}$ \\
$\delta$ & $[^{\circ}]$ & $300  ^{+66}_{-138}$ & $298 ^{+59}_{-145}$ \\
$\Delta m_{21}^2$&  $[10^{-5} {\mathrm eV}^2]$ &  $7.50 ^{+0.18}_{-0.19}$ & $7.51^{+0.21}_{-0.15}$ \\
$\Delta m_{31}^2$ (NH) &  $[10^{-3} {\mathrm eV}^2]$ &  $2.473 ^{+0.070}_{-0.067}$ & $2.489^{+0.055}_{-0.051}$ \\\
$\Delta m_{32}^2$ (IH) &  $[10^{-3} {\mathrm eV}^2]$ &  $-2.427 ^{+0.042}_{-0.065}$ & $-2.468^{+0.073}_{-0.065}$ \\
\hline
\end{tabular}
\end{center}
\end{table*}

\section{Open problems and future projects}
\label{Sec:future}

In neutrino physics, there are still many open problems.
Concerning neutrino oscillations, the more relevant issues are the determination of 
neutrino mass hierarchy,  the measurement of the CP phase $\delta$, 
and the precise evaluation of $\theta_{23}$ (see also Sec.~\ref{subsec:global}).
Of course, there is a much vaster playground to play, considering such exciting possibilities as the existence of sterile neutrino, of Non Standard neutrino Interactions (NSI), the determination of the origin of neutrino mass (Majorana versus Dirac), the absolute mass of neutrinos, the role of neutrinos in cosmology and possible connections to dark matter.
The phenomenon of neutrino oscillations, to which is dedicated this review, provides still a unique tool to answer several of these questions.
This last Section is dedicated to some future prospects of physics of neutrino oscillations.

The CP violation was observed so far only in the quark sector and the CP violation in the leptonic sector is still a big unknown.
%
%
Among the current and near-future experiments, T2K and NO$\nu$A~\cite{NOVA} have a limited sensitivity to CP violation via studying the $\nu_{\mu} \rightarrow \nu_e$ versus $\bar{\nu}_{\mu} \rightarrow \bar{\nu}_e$ appearance. 
The NO$\nu$A experiment at NuMI beam is finalizing the construction phase at FNAL.
A much improved sensitivity to $\delta$ and a strong discovery potential is expected only for the experiments of not so immediate future.
T2HK  (HK stands for Hyper-Kamiokande) will be a successor of T2K, with upgraded J-PARC beam and with the far detector represented by a next generation underground water \v{C}erenkov detector Hyper-Kamiokande~\cite{HyperK} with about 1\,Mton of fiducial mass. The Hyper-Kamiokande construction is expected to start in 2016.
The DAE$\delta$ALUS~\cite{daedalus}, a phased neutrino physics program using cyclotron decay-at-rest neutrino sources would have a strong discovery potential when combined with Hyper-Kamiokande as the detector. 
The Long Baseline Neutrino Experiment (LBNE)~\cite{LBNE}, planning to use the strong neutrino beam from Fermilab travelling 1300\,m baseline to 34\,kton liquid argon time-projection chamber has been recently approved. 
LAGUNA-LBNO (Large Apparatus for Grand Unification~\cite{laguna} and Neutrino Astrophysics and Long Baseline Neutrino Oscillations~\cite{LBNO}) is a European long baseline project using CERN neutrino beam.
The $\nu_{e} \rightarrow \nu_{\mu}$ versus $\bar{\nu}_e \rightarrow \bar{\nu}_{\mu}$ appearance is a music of far future of $\nu$-factories ($\nu$'s from decays of $\mu$'s from accelerator) and $\beta$-beams ($\nu$'s from $\beta$-decays of light nuclei) which would have a decisive sensitivity to $\delta$ but face a problem of exceeding costs.

The long-baseline projects mentioned before for the CP violation search in the neutrino sector have also the ability to determine the neutrino mass hierarchy.
For example, LBNE, jointly with T2K/NO$\nu$A, can in principle address the mass hierarchy issue with a significance of more than 3$\sigma$ by 2030.
Hyper-Kamiokande can get a similar discovery reach on a comparable timescale through the combination of atmospheric neutrino data with a shorter baseline measurement.
In principle, the LAGUNA-LBNO project can afford a very high sensitivity, more than 5$\sigma$, and with a relatively limited data taking period (few years), thanks to its very long baseline.
Several other experiments have been discussed and proposed which may have the capability to test the neutrino mass hierarchy on a time frame shorter than that of the long baseline projects.
Among these alternatives, the most promising approaches seem to be reactor neutrinos (JUNO~\cite{JUNO} formerly known as Daya Bay II) and atmospheric neutrinos in ice (PINGU~\cite{PINGU} at IceCube) or water (ORCA~\cite{ORCA}).
Through a significant breakthrough in the technology and in the detector performance, JUNO is targeted to a potential sensitivity of more than 3$\sigma$ (4$\sigma$), depending upon present and future uncertainties on $\Delta m^2_{32}$.
The experiment has been already approved in China and an international Collaboration is being formed for its construction and operation. 
PINGU, as well as ORCA, could guarantee extremely good statistical sensitivity to the hierarchy, provided the systematic effects are under control and well understood. 
Actually, the estimates of the sensitivity may vary depending upon the choice of oscillation parameters and hierarchy; in an optimistic configuration, a 4$\sigma$ measurement could be obtained after 3 years of data.

In the solar neutrino physics, the most important open issues are the 
determination of the $\nu_e$ survival probability $P(\nu_e\to\nu_e)$ in the
transition region and the measurement of the CNO-neutrino flux. 

The LMA solution of the solar neutrino problem predicts an upturn in 
$P(\nu_e\to\nu_e)$ at $E\approx {\rm few}\,{\rm MeV}$ that corresponds to 
the transition from matter-enhanced oscillations (high-energy end) to 
vacuum-averaged oscillations (low-energy end), see Sec.~\ref{Sec:theory} and the grey bend in Fig.~\ref{Fig:Peebx}. The observation of this feature
would be the final confirmation of the LMA paradigm. 
However, the experiments that measure solar neutrinos have not observed it yet. 
The statistics in each individual experiment does not allow firm conclusions, but
the effect indicating deviations from the LMA predictions is systematically present in all data. Super-Kamiokande, SNO-LETA, and BOREXINO 
seem to favor a flat distribution. In particular, Super-Kamiokande data disfavor the expected 
upturn at 1.3 to 1.9$\sigma$ C.L.  
In addition, the Homestake result is about 1.5$\sigma$ below the LMA prediction.

The shape of $P(\nu_e\to\nu_e)$ in the transition region could be strongly influenced 
by the possible existence of the Non Standard neutrino Interactions (NSI)~\cite{ref3.31} and/or 
of an ultra-light sterile neutrino~\cite{ref3.5, ref3.32}. 
As an example, in presence of an ultra-light sterile neutrino that mixes very weakly with active neutrinos,
a dip in the $P(\nu_e\to\nu_e)$ is expected in the transition region; its precise position is determined by the sterile neutrino mass 
and its width and depth depend on the mixing angle~\cite{ref3.5}.
The possibility of neutrino NSI with other fermions has been predicted by several extensions of the SM, 
as for instance the left-right symmetric models and supersymmetric models with $R$-parity violation.
The NSI can be described at low energy by effective four fermion interactions:
\begin{equation}
\mathcal{L}_{NSI}  =  -2 \sqrt{2} G_F \epsilon^{e,u,d}_{\alpha,\beta}  \left(\bar{\nu}_{\alpha} \gamma^{\mu} P_L \nu_{\beta}  \right ) \left(\bar{f} \gamma_{\mu} P_C f'  \right )
\label{Eq:NSNI}
\end{equation}
where $G_F$ is the Fermi constant, $\alpha$ and $\beta$ are the neutrino flavors, $f$ and $f'$ are the electron or the light quarks, $P_C$ is the chirality of the operator, $C$ can be $L$ or $R$, 
and finally $\epsilon^{e,u,d}_{\alpha,\beta} $  is a dimensionless number which, coupled with the weak coupling constant, parameterizes the strength of the interaction. 
Also in this case the shape of $P(\nu_e\to\nu_e)$  in the transition region is influenced by the NSI and its study can limit the range of the parameter $\epsilon$. Even more effective is its study in the $\nu - e$ elastic scattering.
In order to constrain the shape of the transition region, it is critical to achieve a measurement as precise as possible of $pep$, CNO, and $^8$B (with a lower energy threshold down to 3\,MeV) solar neutrinos. In addition to this, a measurement of CNO neutrinos is important also for other purposes.

The CNO bi-cycle contributes for $\sim$1\% of the energy emitted by Sun.
Despite being sub-dominant in the Sun, the CNO cycle  has, however, a key role in 
astrophysics, being the prominent source of energy in more massive stars 
and in advanced evolutionary stages of solar-like stars.
The evaluation of CNO efficiency is connected with various interesting problems,
like e.g. the determination of globular clusters age from which we extract
a  lower limit to the age of the Universe.  At the moment, we still miss a direct observational 
evidence for CNO energy generation in the Sun. The detection of CNO solar neutrinos would clearly provide a direct test of the CNO cycle efficiency.
The measurement of the CNO solar neutrino flux can also provide clues to
solve the so called "solar composition problem'' (see Sec.~\ref{Subsec:source}).
The flux is, in fact, directly related to the abundance of 
carbon, nitrogen, and oxygen in the core of the Sun and, so to the admixture of heavy elements. 
Thus, a determination of the CNO flux can help in solving the solar metallicity puzzle.

In the neutrino physics, one important problem is the  possible existence of a fourth sterile neutrino. The LSND~\cite{LSND} and MiniBooNE~\cite{ran19} results (see Sec.~\ref{subsec:sterile}) need independent checks, which could either rule out or confirm a third $\Delta m ^2$ around 1\,eV$^2$. In addition, other two indications exist, which could favor the hypothesis of the fourth sterile neutrino: the reactor anomaly~\cite{Mueller, Huber2} and the calibration of the radiochemical experiments with artificial sources~\cite{Giunti2011}. 
The comparison of the antineutrino flux from nuclear reactors with the results of short baseline reactor experiments shows, in fact, a deficit of $\sim$3.5\%. A deficit has been also evidenced in the GALLEX~\cite{Gallex2010} and SAGE~\cite{SageSourceCr} calibrations campaigns with a $^{51}$Cr source (SAGE also with a $^{37}$Ar source~\cite{SageSourceAr}). Each of these deficit is at $\sim$3$\sigma$ level.
A possible interpretation of these anomalies is connected to oscillations into new light sterile neutrino.

The existence of sterile neutrinos appears in many extensions of the Standard Model: they would be simply gauge singlets of the model.  
The simplest model (3 + 1 scheme) introduces only one sterile neutrino $\nu_s$. 
%
In this scenario the four flavor eigenstates ($\nu_e, \nu_{\mu}, \nu_{\tau}, \nu_s$) mix through the matrix elements ($U_{e4}, U_{\mu 4}, U_{\tau4 }, U_{s4}$) 
with a fourth mass eigenstate $\nu_4$.  
The $\Delta m ^2_{i4}$ ($i$ = 1,2,3) are supposed to be $\simeq$ 1\,eV$^2$ in order to provide explanation of the observed anomalies 
and, thus, they are much larger with respect to solar and atmospheric mass splittings. 

The neutrino community favors new and decisive experimental tests on this matter~\cite{WhiteSterile}.
Four experiments will try to address these problems in the near future: Borexino, Borexino-SOX, MicroBooNE, and Icarus-Nessie. 

Borexino phase 2 has as a possible goal the experimental reproduction of the survival-probability transition region. 

Borexino-SOX~\cite{ref3.36} is a project, in which the Borexino detector is taking data with a $^{51}$Cr $\nu_e$ source installed in a small tunnel below it, at $\sim$8\,m from the detector center. The $E/L$ is in the same range as that of the LSND and MiniBooNE experiments. It will look for the possible existence of a sterile neutrino, and will give an important check on the NSI, studying the $\nu - e$ elastic scattering.

MicroBooNE is an experiment based on a 70\,tons fiducial volume Liquid Argon Time Projection Chamber exposed to the NuMI neutrino beamlines at Fermilab. Its goal is to repeat the same measurements of MiniBooNE with high resolution at low energy, starting below 200\,MeV~\cite{ref3.37}.

Finally, Icarus-Nessie is an experiment using the Icarus Liquid Argon Time Projection Chamber technique coupled to near and far spectrometers exposed to $\nu_{\mu}$ and $\bar{\nu}_{\mu}$ beams with an $E/L \sim 1\,{\rm eV}^2$, again to check the possible sterile-neutrino existence~\cite{ref3.38}.

\section*{Conflict of Interests}

The authors declare that there is no conflict of interests regarding the publication of this article.


\begin{thebibliography}{99}

\itemsep -2pt 
\bibitem{pontecorvo} B.~Pontecorvo, "Mesonium and anti-mesonium", Sov. Phys. JETP, Vol. 6, 429, 1957; B.~Pontecorvo, J. Exp. Theor. Phys., Vol. 33,
549, 1957; B.~Pontecorvo, J. Exp. Theor. Phys., Vol. 34, 247, 1958.
\bibitem{Wolfenstein} L. Wolfenstein, "Neutrino Oscillations in Matter", Phys. Rev. D, Vol. 17, 2369-2374, 1978. 
\bibitem{MSW} S. P. Mikheev and A. Y. Smirnov, "Resonance Amplification of Oscillations in Matter and Spectroscopy of Solar Neutrinos'', Sov. J. Nucl. Phys., Vol. 42, 913-917, 1985 and Nuovo Cim. C, Vol. 9, 17-26, 1986.
\bibitem{GiuntiBook} C. Giunti and C. W. Kim, "Fundamentals of Neutrino Physics and Astrophysics", Oxford University Press.
\bibitem{Giunti} S.~M.~Bilenky, C.~Giunti and W.~Grimus, "Phenomenology of neutrino oscillations'', Prog. Part. Nucl. Phys., Vol. 43, 1, 1999.
\bibitem{Vissani} A. Strumia and F. Vissani, "Neutrino masses, mixings and ...", e-Print: hep-ph/0606054, IFUP-TH-2004-1.
\bibitem{ran12} J. Beringer et al. (Particle Data Group), "The Review of Particle Physics'',  Phys. Rev. D, Vol. 86, 010001, 2012.
\bibitem{Bilenky} S.~M.~Bilenky, J.~Hosek and S.~T.~Petcov, "On Oscillations of Neutrinos with Dirac and Majorana Masses'', Phys. Lett. B, Vol. 94, 495-498, 1980.
\bibitem{Smirnov}  M.~Blennow and A.~Y.~.Smirnov, "Neutrino propagation in matter", Adv. High Energy Phys., Vol. 2013, 972485, 2013.
\bibitem{Kuo} T.~K.~Kuo and J.~T.~Pantaleone,  "Nonadiabatic Neutrino Oscillations in Matter,''  Phys. Rev. D, Vol. 39, 1930-1939, 1989.
\bibitem{Kuo2} T.~K.~Kuo and J.~T.~Pantaleone, ''Neutrino Oscillations in Matter'',  Rev. Mod. Phys.,  Vol. 61, 937-979, 1989.
\bibitem{Homestake} B.T. Cleveland et al. (Homestake Collaboration), "Measurement of the Solar Electron Neutrino Flux with the Homestake Chlorine Detector", Ap. J., Vol. 496, 505-526, 1998.
\bibitem{Bahcall1989} J.N. Bahcall, "Neutrino Astrophysics", Cambridge University, Cambridge, 1989.
\bibitem{Gallex} W. Hampel et. al. (GALLEX Collaboration), "GALLEX solar neutrino observations: results for GALLEX IV", Phys. Lett. B, Vol. 447, 127-133, 1999.
\bibitem{Sage} J. N. Abdurashitov et al. (SAGE Collaboration), "Measurement of the solar neutrino capture rate with gallium metal", Phys. Rev. C, Vol. 60, 055801, 1999; J.N. Abdurashitov et al. (SAGE Collaboration), "Measurement of the solar neutrino capture rate with gallium metal. III. Results for the 2002-2007 data-taking period", Phys. Rev. C, Vol. 80, 015807, 2009.
\bibitem{Gallex2010} F. Kaether et al., "Reanalysis of the GALLEX solar neutrino flux and source experiments", Phys. Lett. B, Vol. 685, 47-54, 2010.
\bibitem{SageSourceCr} J. N. Abdurashitov et al. (SAGE Collaboration), "Measurement of the response of a gallium metal solar neutrino experiment to neutrinos from a $^{51}$Cr source",  Phys. Rev. C, Vol. 59, 2246, 1999.
\bibitem{SageSourceAr} J. N. Abdurashitov et al. (SAGE Collaboration), "Measurement of the response of a Ga solar neutrino experiment to neutrinos from a $^{37}$Ar source", Phys. Rev. C, Vol. 73, 045805, 2006.
\bibitem{GiuntiSter}  M.~A.~Acero, C.~Giunti and M.~Laveder, "Limits on nu(e) and anti-nu(e) disappearance from Gallium and reactor experiments'', Phys.\ Rev.\ D  78, 073009, 2008
\bibitem{SNOarXiv} B. Aharmin et al. (SNO Collaboration), "Combined Analysis of all Three Phases of Solar Neutrino Data from the Sudbury Neutrino Observatory", arXiv:1109.0763, 2011.
\bibitem{SNOPhase1} Q. R. Amhad et al. (SNO Collaboration), "Direct Evidence for Neutrino Flavor Transformation from Neutral-Current Interactions in the Sudbury Neutrino Observatory", Phys. Rev. Lett., Vol. 89, 011301, 2002;
\bibitem{ran7} S. Fukuda et al. (The Super-Kamiokande Collaboration), "The Super-Kamiokande Detector", Nucl. Instr. Meth. A, Vol. 501, 418-462, 2003.
\bibitem{ran8} K. Hirata et al. (Kamiokande Collaboration), "Observation of $^8$B Solar Neutrinos in the Kamiokande-II Detector", Phys. Rev. Lett., Vol. 63, 16-19, 1989.
\bibitem{IMB} ] R. Becker-Szendy et al., “Search for muon neutrino oscillations with the Irvine-Michigan-Brookhaven detector”, Phys. Rev. Lett., Vol. 69, 1010-1013, 1992.
\bibitem{ran9} Y. Fukuda et al. (The Super-Kamiokande Collaboration), "Evidence for Oscillation of Atmospheric Neutrinos", Phys. Rev. Lett., Vol. 81, 1562-1567, 1998.
\bibitem{K2K2006} M. H. Ahn et al, (The K2K Collaboration), "Measurement of neutrino oscillation by the K2K experiment", Phys. Rev. D, Vol. 74, 072003, 2006.
\bibitem{ran11} K. Abe et al., (The T2K Collaboration), "Indication of Electron Neutrino Appearance from an Accelerator-Produced Off-Axis Muon Neutrino Beam", Phys. Rev. Lett., Vol. 107, 041801, 2011.
\bibitem{ReinesCowan} F. Reines and C. L. Cowan, Jr., "Detection of the Free Neutrino", Phys. Rev., Vol. 92, 830-831, 1953.
\bibitem{BUST} E.N. Alekseev et al. (BUST Collaboration), "The Baksan Underground Scintillation Telescope", Phys. Part. Nucl., Vol. 29, 254-256, 1998.
\bibitem{LSD} M. Aglietta et al. (LSD Collaboration), "Results of the Liquid Scintillation Detector of the Mont Blanc Laboratory", Nuovo Cim. C, Vol. 9, 185-195, 1986.
\bibitem{LVD} M. Aglietta et al. (LVD Collaboration), "The Most powerful scintillator supernovae detector: LVD", Nuovo Cim. A, Vol. 105, 1793-1804, 1992.
\bibitem{Gosgen} G. Zacek et al., "Neutrino oscillation experiments at the Gosgen nuclear power reactor", Phys. Rev. D, Vol. 34, 2621-2636, 1986.
\bibitem{Bugey} M. Abbeset al., "The Bugey-3 neutrino detector", Nucl. Instrum. Meth. A, Vol. 374, 164-187, 1996.
\bibitem{ran13} G. Alimonti et al. (Borexino Collaboration), "The Borexino detector at the Laboratori Nazionali delran Sasso", Nucl. Instr. and Meth A, Vol. 600, 568-593, 2009.
\bibitem{ref3.18} G. Alimonti et al. (Borexino Collaboration), "The liquid handling systems for the Borexino solar neutrino detector", Nucl. Instr. and Methods A, Vol. 609, 58-78, 2009.
\bibitem{KamlPrecision} S. Abe et al. (KamLAND Collaboration), "Precision Measurement of Neutrino Oscillation Parameters with KamLAND", Phys. Rev. Let., Vol. 100, 221803, 2008.
\bibitem{ran15} F. P. An et al. (Daya Bay Collaboration), "Observation of Electron-Antineutrino Disappearance at Daya Bay", Phys. Rev. Lett., Vol. 108, 171803, 2012.
\bibitem{RENO} J. K. Ahn et al., (RENO Collaboration), "Observation of Reactor Electron Antineutrinos Disappearance in the RENO Experiment", Phys. Rev. Lett., Vol. 108, 191802, 2012. 
\bibitem{ran17} S. Abe et al. (Double Chooz Collaboration), "First measurement of  $\theta_{13}$ from delayed neutron capture on hydrogen in the Double Chooz experiment", Phys. Lett. B, Vol. 723, 66-70, 2013.
\bibitem{LSND} A. Aguilar et al., (LSND Collaboration), "Evidence for neutrino oscillations from the observation of $\bar{\nu}_e$ appearance in a $\bar{\nu}_{\mu}$ beam", Phys. Rev. D, Vol.  64, 112007, 2001.  
\bibitem{ran19} A. A. Aguilar-Arevalo et al. (MiniBooNE Collaboration), "Improved Search for $\bar{\nu}_{\mu} \rightarrow \bar{\nu}_e$ Oscillations in the MiniBooNE Experiment", Phys. Rev. Lett., Vol.110, 161801, 2013.
\bibitem{KarmenDet} H. Gemmeke et al., "The High resolution neutrino calorimeter KARMEN", Nucl. Instrum. Meth. A, Vol. 289, 490-495, 1990.
\bibitem{MACRO} (MACRO Collaboration) G. Giacomelli and A. Margiotta,"MACRO results on atmospheric neutrinos", Nucl. Phys. Proc. Suppl., 145, 2005.  
\bibitem{ran20} D. G. Michael et al. (MINOS Collaboration), "Observation of Muon Neutrino Disappearance with the MINOS Detectors in the NuMI Neutrino Beam", Phys. Rev. Lett., Vol. 97, 191801, 2006.
\bibitem{ran21} N. Agafonova et al., (OPERA Collaboration), "Search for oscillation $\nu_{\mu} \rightarrow \nu_{\tau}$ with the OPERA experiment in the CNGS beam", New J. Phys., Vol. 14, 033017, 2012.
\bibitem{CHORUS} D. Macina et al. (CHORUS Collaboration), "The CHORUS experiment", Nucl. Phys. Proc. Suppl., Vol. 48, 183-187, 1996.
\bibitem{NOMAD} J. Altegoer et al. (NOMAD Collaboration), "The NOMAD experiment at the CERN SPS", Nucl. Instrum. Meth. A, Vol. 404, 96-128, 1998.
\bibitem{ICARUS} C. Rubbia et al. (ICARUS Collaboration), "Underground operation of the ICARUS T600 LAr-TPC: first results", JINST, Vol. 6, P07011, 2011.

\bibitem{Serenelli2011} A. Serenelli, W. C. Haxton, C. Pena-Garay, "Solar models with accretion. I. Application to the solar abundance problem", Ap. J., Vol. 743, 24, 2011. 
\bibitem{Castellani} V. Castellani et al., "Solar neutrinos: beyond standard solar models", Phys.Rept., Vol. 281, 309-398, 1997.
\bibitem{GS98}  N. Grevesse and A.J. Sauval, "Standard Solar Composition'', Space Science Rev., Vol. 85, 161-174, 1998.
\bibitem{AGSS09} M. Asplund et al., "The chemical composition of the Sun", Ann. Rev. Astr. Astroph., Vol. 47, 481-522, 2009.
\bibitem{Basu} S. Basu and H. M. Antia, "Helioseismology and solar abundances'',  Phys. Rept., Vol. 457, 217-283, 2008.
\bibitem{Honda1} M. Honda et al., "Calculation of the flux of atmospheric neutrinos", Phys. Rev. D, Vol. 52, 4985-5005, 1995.
\bibitem{Honda2} M. Honda et al., "Atmospheric neutrino fluxes", Phys. Lett. B, Vol. 248, 193-198, 1990.
\bibitem{Barr} S. Barr, T.K. Gaisser, and T. Stanev, "Flux of atmospheric neutrinos", Phys. Rev. D, Vol. 39, 3532-3534, 1989.
\bibitem{Agrawal} V. Agrawal et al., "Atmospheric neutrino flux above 1 GeV", Phys. Rev. D, Vol. 53, 1314-1323, 1996.
\bibitem{Gaisser} T.K. Gaisser and T. Stanev, in Proc. 24th Int. Cosmic Ray Conf. (Rome), Vol. 1, 694-697, 1995.
\bibitem{Vogel} P. Vogel et al., "Reactor antineutrino spectra and their application to antineutrino-induced reactions", Phys. Rev. C, Vol. 24, 1543-1553, 1981.
\bibitem{Huber} P. Huber and T. Schwetz, "Precision spectroscopy with reactor antineutrinos", Phys. Rev. D, Vol. 70, 053011, 2004.
\bibitem{Mueller} Th.A. Mueller et al., "Improved predictions of reactor antineutrino spectra", Phys. Rev. C, Vol. 83, 054615, 2011.
\bibitem{Huber2} P. Huber, "Determination of antineutrino spectra from nuclear reactors", Phys. Rev. C, Vol. 84, 024617, 2011.
\bibitem{ref3.5} P.C. de Holanda and A.Yu. Smirnov, "Solar neutrino spectrum, sterile neutrinos, and additional radiation in the Universe", Phys. Rev. D, Vol. 83, 113011, 2011.
\bibitem{ref3.6} K. S. Hirata et al. (Kamiokande Collaboration), "Real-time, directional measurement of $^8$B solar neutrinos in the Kamiokande II detector", Phys. Rev. D, Vol. 44, 2241-2260, 1991.
\bibitem{ref3.7} Y. Fukuda et al. (Super-Kamiokande Collaboration), "Evidence for Oscillation of Atmospheric Neutrinos", Phys. Rev.Lett., Vol. 81, 1562-1567, 1998.
\bibitem{ref3.8} K. S. Hirata et al. (Kamiokande Collaboration), "Experimental study of the atmospheric neutrino flux", Phys. Lett. B, Vol. 205, 416-420, 1988.
\bibitem{ref3.9} Y. Ashie et al. (Super-Kamiokande Collaboration), "Evidence for an oscillatory signature in atmospheric neutrino oscillation", Phys. Rev. Lett., Vol. 93, 101801, 2004.
\bibitem{SNO-LETA} B. Aharmim et al. (SNO Collaboration), "Low-energy-threshold analysis of the Phase I and Phase II data sets of the Sudbury Neutrino Observatory", Phys. Rev. C, Vol. 81, 055504, 2010.
\bibitem{SNOPhase3} B. Aharmim et al. (SNO Collaboration), "Measurement of the νe and total $^8$B solar neutrino fluxes with the Sudbury Neutrino Observatory phase-III data set, Phys. Rev. C, Vol. 87, 015502, 2013. 
\bibitem{ref3.11} K. Abe et al. (Super-Kamiokande Collaboration), " Solar neutrino results in Super-Kamiokande-III", Phys. Rev. D, Vol. 83, 052010, 2011.
\bibitem{BXDayNight} G. Bellini et al., (Borexino Collaboration), "Absence of a day-night asymmetry in the $^7$Be solar neutrino rate in Borexino", Phys. Lett. B, Vol. 707, 22-26, 2012.
\bibitem{KL2003} K. Eguchi et al. (KamLAND Collaboration), "First Results from KamLAND: Evidence for Reactor Anti-Neutrino Disappearance", Phys. Rev. Lett., Vol. 90, 021802, 2003.
\bibitem{ref3.13} E. Aliu et al. (K2K Collaboration), "Evidence for muon neutrino oscillation in an accelerator-based experiment", Phys. Rev. Lett., Vol. 94, 081802, 2005; S. Yamamoto et al. (K2K Collaboration), "An improved search for $\nu_{\mu} \rightarrow \nu_e$ oscillation in a long-baseline accelerator experiment", Phys. Rev. Lett., Vol. 96, 181801, 2006. 
\bibitem{opera} G. DeLellis for OPERA Collaboration, talk at LNGS, 16/5/2013.
\bibitem{ref3.16} P. Adamson et al. (MINOS Collaboration), "First Direct Observation of Muon Antineutrino Disappearance", Phys. Rev. Lett., Vol. 107, 021801, 2011; P. Adamson et al. (MINOS Collaboration), "Improved Search for Muon-Neutrino to Electron-Neutrino Oscillations in MINOS", Phys. Rev. Lett., Vol. 107, 181802, 2011.
\bibitem{Bellini2012} G. Bellini, A. Ianni, and G. Ranucci, "Borexino and solar neutrinos", Rivista del Nuovo Cimento, Vol. 035, 481-537, 2012.
\bibitem{ref3.19} G. Bellini et al. (Borexino Collaboration),  "Precision Measurement of the $^7$Be Solar Neutrino Interaction Rate in Borexino", Phys. Rev. Lett., Vol. 107,  141302, 2011.
\bibitem{ref3.20} G. Bellini et al. (Borexino Collaboration), "First Evidence of $pep$ Solar Neutrinos by Direct Detection in Borexino", Phys. Rev. Lett., Vol. 108, 051302, 2012.
\bibitem{ref3.21} G. Bellini et al. (Borexino Collaboration), "Measurement of the solar $^8$B neutrino rate with a liquid scintillator target and 3 MeV energy threshold in the Borexino detector', Phys. Rev. D, Vol. 82, 033006, 2010.
\bibitem{ref3.31} Y. Minakata and C. Pe\~na-Garay, "Solar Neutrino Observables Sensitive to Matter Effects", Adv. High En. Phys., Vol. 2012, Article ID 349686, 2012.
\bibitem{ref3.32} P. C. de Holanda and A. Yu. Smirnov, "Homestake result, sterile neutrinos, and low energy solar neutrino experiments", Phys. Rev. D, Vol. 69, 113002, 2004. 
\bibitem{ref3.25}  B. Armbruster et al. (KARMEN Collaboration),  "Upper limits for neutrino oscillations $\bar{\nu}_{\mu} \rightarrow \bar{\nu}_e$ from muon decay at rest", Phys. Rev. D, Vol. 65, 112001, 2002. 
\bibitem{ref3.27} F. P. An et al. (Daya Bay Collaboration), "Improved measurement of electron antineutrino disappearance at Daya Bay", Chinese Physics. C, Vol. 37, 011001, 2013.
\bibitem{FogliGLobal} G.~L.~Fogli et. al, "Evidence of $\theta_{13} >0$ from global neutrino data analysis", Phys. Rev. D, Vol, 84, 053007, 2011.  
\bibitem{global} M.C. Gonzalez-Garcia et al., "Global fit to three neutrino mixing: critical look at present precision", JHEP, Vol. 12, 123, 2012. 
\bibitem{FogliLast}   G.~L.~Fogli et al., "Global analysis of neutrino masses, mixings and phases: entering the era of leptonic CP violation searches'', Phys.\ Rev.\ D, Vol. 86, 013012, 2012.
\bibitem{SuperK} R. Wendell et al. (Super-Kamiokande Collaboration), "Atmospheric neutrino oscillation analysis with sub-leading effects in Super-Kamiokande I, II, and III", Phys. Rev. D, Vol. 81, 092004, 2010. 
\bibitem{minos2012} P. Adamson et al. (MINOS Collaboration), Improved measurement of muon antineutrino disappearance in MINOS, Phys. Rev. D., Vol. 108, 191801, 2012.
\bibitem{T2K2012} K. Abe et al. (T2K Collaboration), "First muon-neutrino disappearance study with an off-axis beam", Phys. Rev. D, Vol. 85, 031103, 2012.
\bibitem{CHOOZ1999} M. Apollonio et al. (CHOOZ Collaboration), "Limits on neutrino oscillations from the CHOOZ experiment", Phys. Lett. B, Vol. 466, 415-430, 1999.
\bibitem{PaloVerde} A. Piepke et al. (Palo Verde Collaboration), "Final results from the Palo Verde neutrino oscillation experiment", Progr. Part. Nucl. Phys., Vol. 48, 113–121, 2002.
\bibitem{DoubleChooz2012} Y. Abe et al. (Double Chooz Collaboration), "Reactor electron antineutrino disappearance in the Double Chooz experiment", Phys. Rev. D, Vol. 86, 052008, 2012.
\bibitem{DayaBayTalk} D. Dwyer for Daya Bay Collaboration. Talk given at the XXV International Conference on Neutrino Physics, Kyoto, Japan, June 3-9, 2012.
\bibitem{Kamland2011}  A. Gando et al. (KamLAND Collaboration), "Constraint on $\theta_{13}$ from a three-flavor oscillation analysis of reactor antineutrinos at KamLAND", Phys. Rev. D, Vol. 83, 052002, 2011.
\bibitem{SuperK2006}  J. Hosaka et al. (Super-Kamiokande Collaboration), "Solar neutrino measurements in
Super-Kamiokande-I", Phys. Rev. D, Vol. 73, 112001, 2006.
\bibitem{NOVA} D. Perevalov for NO$\nu$A Collaboration, talk at WIN 2013 conference, Natal, Brasil, 2013.
\bibitem{HyperK} M. Shiozawa for Hyper-Kamiokande Collaboration, "The Hyper-Kamiokande project", Nucl. Phys. B - Proc. Suppl., Vol. 237–238, 289–294, 2013.
\bibitem{LBNE} C. Adams et al. (LBNE Collaboration), "Opportunities with the Long-Baseline Neutrino Experiment", arXiv:1307.7335, 2013.
\bibitem{daedalus} C. Aberle et al. (DAE$\delta$ALUS Collaboration), "Whitepaper on the DAEdALUS Program", arXiv:1307.2949, 2013.
\bibitem{laguna}  D. Angus et al. (LAGUNA Collaboration), "The LAGUNA design study - towards giant liquid based underground detectors for neutrino physics and astrophysics and proton decay searches", arXiv:1001.0077, 2009.
\bibitem{LBNO} M. Ghosh et al., "Synergies between neutrino oscillation experiments: An `adequate' configuration for LBNO", arXiv:1308.5979, 2013.
\bibitem{JUNO} Y.F. Li et al. (JUNO Collaboration), "Unambiguous Determination of the Neutrino Mass Hierarchy Using Reactor Neutrinos", Phys. Rev. D, Vol. 88, 013008, 2013.
\bibitem{PINGU} M. G. Aartsen et al. (IceCube, PINGU collaboration), "PINGU Sensitivity to the Neutrino Mass Hierarchy", arXiv:1306.5846, 2013.
\bibitem{ORCA} P. Coyle et al. (The Km3Net collaboration), contribution to the European Strategy Preparatory Group Symposium, September 2012, Krakow, Poland.
\bibitem{Giunti2011} C. Giunti and M. Laveder, "Statistical significance of the gallium anomaly", Phys. Rev. C, Vol. 83,  065504, 2011.
\bibitem{WhiteSterile} K. N. Abazajian et al., "Light Sterile Neutrinos: A White Paper", arXiv:1204.5379, 2012.
\bibitem{ref3.36} G. Bellini et al. (Borexino Collaboration), "SOX: Short distance neutrino Oscillations with BoreXino", JHEP, Vol. 2013, Article ID 38, 2013.
\bibitem{ref3.37} H. Chen et al., "A proposal for a new experiment using the Booster and NuMI neutrino beamlines: MicrooBooNE". Fermilab.
\bibitem{ref3.38} Icarus-Nessie Collaboration, "Search for anomalies in the neutrino sector with muon spectrometers and large LArTPC imaging detectors at CERN", European Strategy for Particle Physics, Krakow, 2012.




\end{thebibliography}
\end{document}